\documentclass[a4paper,11pt]{article}
\usepackage{jheppub} 
\usepackage{lineno}

\usepackage{amssymb}
\usepackage{amsmath}
\usepackage{mathtools}
\usepackage{epsfig}
\usepackage{hyperref}
\usepackage{breakurl}
\usepackage{bm}
\usepackage{color}
\usepackage{tikz}
\usepackage{todonotes}
\usepackage[labelformat=simple]{subcaption}

\usepackage[utf8]{inputenc}

\usetikzlibrary{snakes}
\usetikzlibrary{decorations}
\usetikzlibrary{trees}
\usetikzlibrary{decorations.pathmorphing}
\usetikzlibrary{decorations.markings}
\usetikzlibrary{external}
\usetikzlibrary{intersections}
\usetikzlibrary{shapes,arrows}
\usetikzlibrary{arrows.meta}
\usetikzlibrary{calc}
\usetikzlibrary{shapes.misc}
\usetikzlibrary{decorations.text}
\usetikzlibrary{backgrounds}
\usetikzlibrary{fadings}
\usetikzlibrary{tikzmark,calc,arrows,shapes,decorations.pathreplacing}

\makeatletter
\def\simgt{\mathrel{\lower2.5pt\vbox{\lineskip=0pt\baselineskip=0pt
           \hbox{$>$}\hbox{$\sim$}}}}
\def\simlt{\mathrel{\lower2.5pt\vbox{\lineskip=0pt\baselineskip=0pt
           \hbox{$<$}\hbox{$\sim$}}}}

\def\fig#1{Fig.~\ref{#1}}
\def\sect#1{Sec.~\ref{#1}}

\makeatother

\def\eqn#1{Eq.~\eqref{#1}}
\def\spa#1.#2{\left\langle#1\,#2\right\rangle}
\def\spb#1.#2{\left[#1\,#2\right]}
\def\sand#1.#2.#3{%
\left\langle#1{\vphantom1}\right|{#2}\left|#3\right]}%
\def\sandmp#1.#2.#3{%
\left\langle#1{\vphantom1}\right|{#2}\left|#3\right]}%
\def\sandpm#1.#2.#3{%
\left[#1{\vphantom1}\right|{#2}\left|#3\right\rangle}%
\def\sandmm#1.#2.#3{%
\left\langle#1{\vphantom1}\right|{#2}\left|#3\right\rangle}%
\def\sandpp#1.#2.#3{%
\left[#1{\vphantom1}\right|{#2}\left|#3\right]}%

\def\nn{\nonumber}

\newcommand{\be}{\begin{equation}}
\newcommand{\ee}{\end{equation}}

\newcommand{\Eq}[1]{Eq.~\eqref{#1}}

\newcommand{\Sec}[1]{Sec.~\ref{#1}}

\newcommand{\vect}{\boldsymbol}

\renewcommand{\imath}{\mathrm{i}}

\newcommand{\eikSum}{\mathcal{E}}

\def\topbotatom#1{\hbox{\hbox to 0pt{$#1\bot$\hss}$#1\top$}}

\definecolor{myred}{rgb}{0.8,0,0}
\definecolor{myblue}{rgb}{0,0.28,0.69}

\tikzset{
	graviton/.style={decorate,line width=0.1mm, decoration={snake,amplitude=.5mm, segment length=2mm}},
	photon/.style={decorate, decoration={snake,amplitude=.5mm, segment length=2mm}, draw=red},
	photonBG/.style={decorate, decoration={snake,amplitude=.5mm, segment length=2mm}, draw=white,line width= 1mm},
	scalar/.style={postaction={decorate},
	},
	massive/.style={postaction={decorate},
		line width=0.75mm,
	},
   massiveRed/.style={postaction={decorate},
		line width=0.6mm,
       myred
	},
   massiveBlue/.style={postaction={decorate},
		line width=0.6mm,
       myblue
	},
	massless/.style={postaction={decorate},
	},
	masslessOverlap/.style={
		decoration={show path construction, lineto code={
				\draw[white,line width=1.5mm,line cap=round] ($(\tikzinputsegmentfirst)!0.2!(\tikzinputsegmentlast)$) to ($(\tikzinputsegmentfirst)!0.8!(\tikzinputsegmentlast)$);
				\draw (\tikzinputsegmentfirst) to (\tikzinputsegmentlast);
		}},
		decorate
	},
	masslessWithDot/.style={postaction={decorate},
		decoration={
			markings,
			mark=at position 0.5 with {\fill circle (2pt);}}
	},
	masslessWithArrow/.style={postaction={decorate},
		decoration={
			markings,
			mark=at position 0.5 with {\arrow{latex}}}
	},
    masslessWithArrowNearEnd/.style={postaction={decorate},
		decoration={
			markings,
			mark=at position 0.75 with {\arrow{latex}}}
	},
    massiveWithDot/.style={postaction={decorate},
		line width=0.5mm,
		decoration={
			markings,
			mark=at position 0.5 with {\fill circle (2pt);}}
	},
	massiveLinWithDot/.style={postaction={decorate},
		double,
		thick,
		fill=white,
		decoration={
			markings,
			mark=at position 0.5 with {\fill[black] circle (2pt);}}
	},	
	massiveWithArrow/.style={postaction={decorate},
		line width=0.75mm,
		decoration={
			markings,
			mark=at position 0.5 with {\arrow{latex}}}
	},
	massiveWithArrow2/.style={postaction={decorate},
		line width=0.75mm,
		decoration={
			markings,
			mark=at position 0.9 with {\arrow{latex}}}
	},
	massiveLin/.style={postaction={decorate},
		double,
		thick,
		fill=white
	},
	massiveLinRed/.style={postaction={decorate},
		myred,
        double,
		thick,
		fill=white
	},
 	massiveLinBlue/.style={postaction={decorate},
        myblue,
		double,
		thick,
		fill=white
	},
    massiveLinStatic/.style={postaction={decorate},
		double,
		thick,
		fill=white
	},
 	static/.style={postaction={draw,densely dashed,black,thin},
		line width = 0.1cm,white
	},
    massivePhi/.style={postaction={decorate},
		line width=0.75mm,
		dashed
	},
	masslessPhi/.style={postaction={decorate},
		dashed
	},
	unitaryCut/.style={postaction={draw,densely dashed,black,thin},
		line width = 0.2cm,white
	},
	gluon/.style={decorate, draw=magenta,
		decoration={coil,amplitude=4pt, segment length=5pt}},
	partial ellipse/.style args={#1:#2:#3}{
		insert path={+ (#1:#3) arc (#1:#2:#3)}
	},
	cross/.style={cross out, draw=black, minimum size=2*(#1-\pgflinewidth), inner sep=0pt, outer sep=0pt},
	branchCut/.style={postaction={decorate},
		snake=zigzag,
		decoration = {snake=zigzag,segment length = 2mm, amplitude = 2mm}	
	},
	cross/.default={1pt},
	massiveWithResidue/.style={postaction={decorate},
		decoration={
			markings,
			mark=at position 0.5 with {\draw node[cross,red] {};}}
	},
	massiveLinWithResidue/.style={postaction={decorate},
		double,
		thick,
		fill=white,
		decoration={
			markings,
			mark=at position 0.5 with {\draw node[cross=4pt,red] {};}}	
	}
}

\newcommand{\treetGraviton}{
\includegraphics[scale=1.15]{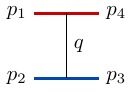}
}

\newcommand{\PlanarQuadLadder}{
		\includegraphics[]{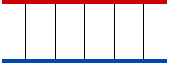}
}

\newcommand{\WindowFourLoop}{
		\includegraphics[]{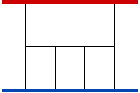}
}

\newcommand{\XWindowFourLoop}{
		\includegraphics[]{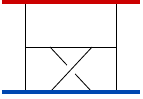}
}

\newcommand{\BNDIntA}{
\includegraphics[]{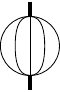}
}

\newcommand{\BNDIntB}{
\includegraphics[]{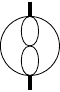}
}

\newcommand{\BNDIntC}{
\includegraphics[]{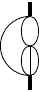}
}

\newcommand{\BNDIntD}{
\includegraphics[]{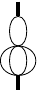}
}

\newcommand{\BNDIntE}{
\includegraphics[]{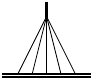}
}

\newcommand{\BNDIntF}{
\includegraphics[]{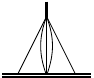}
}

\newcommand{\BNDIntG}{
\includegraphics[]{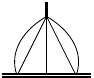}
}

\newcommand{\BNDIntH}{
\includegraphics[]{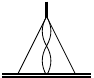}
}

\newcommand{\planarizationGA}{
\includegraphics[]{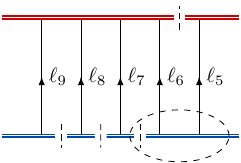}

}

\newcommand{\planarizationGB}{
\includegraphics[]{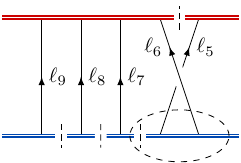}

}

\newcommand{\planarizationCut}{
\includegraphics[]{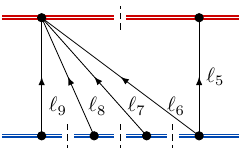}

}

\newcommand{\BNDIntEllipticAFull}{\includegraphics{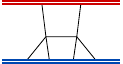}}

\newcommand{\BNDIntEllipticA}{

\includegraphics[]{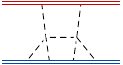}
}

\newcommand{\BNDIntADInLBL}{\includegraphics{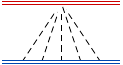}}
\newcommand{\BNDIntADInKZH}{\includegraphics{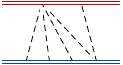}}
\newcommand{\BNDIntADInKWV}{\includegraphics{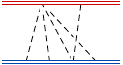}}
\newcommand{\BNDIntADInLAZ}{\includegraphics{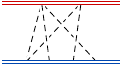}}
\newcommand{\BNDIntADInLAJ}{\includegraphics{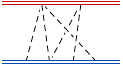}}
\newcommand{\BNDIntADInLAW}{\includegraphics{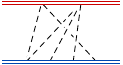}}
\newcommand{\BNDIntADJnKZB}{\includegraphics{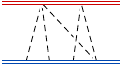}}
\newcommand{\BNDIntADKnLAH}{\includegraphics{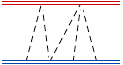}}
\newcommand{\BNDIntADPnKZD}{\includegraphics{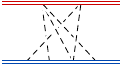}}
\newcommand{\BNDIntASMnKZH}{\includegraphics{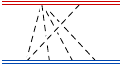}}
\newcommand{\BNDIntASMnKWV}{\includegraphics{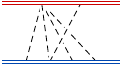}}
\newcommand{\BNDIntASMnLAZ}{\includegraphics{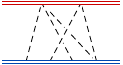}}
\newcommand{\BNDIntASMnLAJ}{\includegraphics{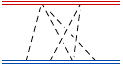}}
\newcommand{\BNDIntASMnLAW}{\includegraphics{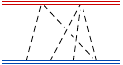}}
\newcommand{\BNDIntASNnKZB}{\includegraphics{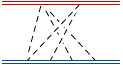}}
\newcommand{\BNDIntASOnLAH}{\includegraphics{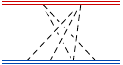}}
\newcommand{\BNDIntASTnKZD}{\includegraphics{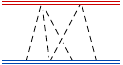}}
\newcommand{\BNDIntAXGnKZI}{\includegraphics{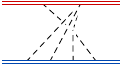}}
\newcommand{\BNDIntAGVnKIF}{\includegraphics{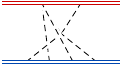}}
\newcommand{\BNDIntAGVnKID}{\includegraphics{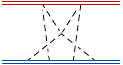}}
\newcommand{\BNDIntAGUnKIF}{\includegraphics{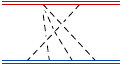}}
\newcommand{\BNDIntCInKHK}{\includegraphics{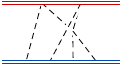}}
\newcommand{\BNDIntAINnKIF}{\includegraphics{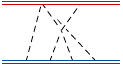}}
\newcommand{\BNDIntAGUnKID}{\includegraphics{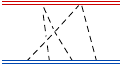}}
\newcommand{\BNDIntCPnKGU}{\includegraphics{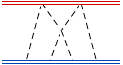}}
\newcommand{\BNDIntHOnIUQ}{\includegraphics{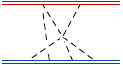}}
\newcommand{\BNDIntHUnIUQ}{\includegraphics{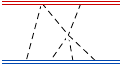}}
\newcommand{\BNDIntNDnEBW}{\includegraphics{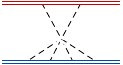}}
\newcommand{\BNDIntAICnKZI}{\includegraphics{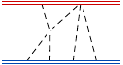}}
\newcommand{\BNDIntAVZnKID}{\includegraphics{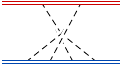}}
\newcommand{\BNDIntAVYnKIF}{\includegraphics{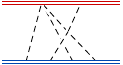}}
\newcommand{\BNDIntRMnKHK}{\includegraphics{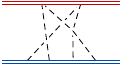}}
\newcommand{\BNDIntAVYnKID}{\includegraphics{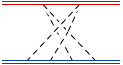}}
\newcommand{\BNDIntRTnKGU}{\includegraphics{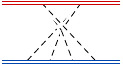}}
\newcommand{\BNDIntWSnIUQ}{\includegraphics{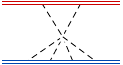}}
\newcommand{\BNDIntWYnIUQ}{\includegraphics{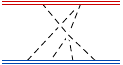}}
\newcommand{\BNDIntADInLBX}{\includegraphics{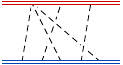}}
\newcommand{\BNDIntADInLBV}{\includegraphics{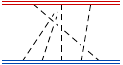}}
\newcommand{\BNDIntADInLBR}{\includegraphics{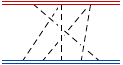}}
\newcommand{\BNDIntAnKZN}{\includegraphics{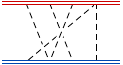}}
\newcommand{\BNDIntADInLBU}{\includegraphics{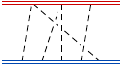}}
\newcommand{\BNDIntADInLBQ}{\includegraphics{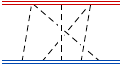}}
\newcommand{\BNDIntAnLAS}{\includegraphics{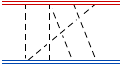}}
\newcommand{\BNDIntADInLBO}{\includegraphics{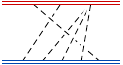}}
\newcommand{\BNDIntAnLBJ}{\includegraphics{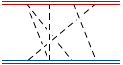}}
\newcommand{\BNDIntAnLAT}{\includegraphics{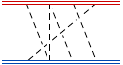}}
\newcommand{\BNDIntAnKXB}{\includegraphics{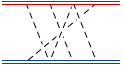}}
\newcommand{\BNDIntAnLBI}{\includegraphics{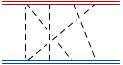}}
\newcommand{\BNDIntAnKZM}{\includegraphics{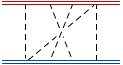}}
\newcommand{\BNDIntAnLBG}{\includegraphics{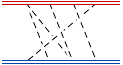}}
\newcommand{\BNDIntAnLBC}{\includegraphics{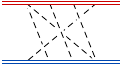}}
\newcommand{\BNDIntADJnLBV}{\includegraphics{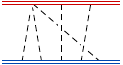}}
\newcommand{\BNDIntADJnLBR}{\includegraphics{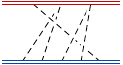}}
\newcommand{\BNDIntBnKZN}{\includegraphics{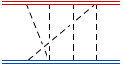}}
\newcommand{\BNDIntBnKXB}{\includegraphics{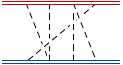}}
\newcommand{\BNDIntADJnLBO}{\includegraphics{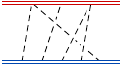}}
\newcommand{\BNDIntBnLBC}{\includegraphics{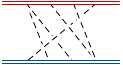}}
\newcommand{\BNDIntBnLBG}{\includegraphics{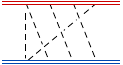}}
\newcommand{\BNDIntBnKZK}{\includegraphics{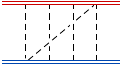}}
\newcommand{\BNDIntADKnLBX}{\includegraphics{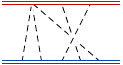}}
\newcommand{\BNDIntADKnLBV}{\includegraphics{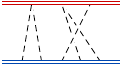}}
\newcommand{\BNDIntADKnLBR}{\includegraphics{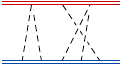}}
\newcommand{\BNDIntCnKZN}{\includegraphics{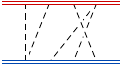}}
\newcommand{\BNDIntCnKXB}{\includegraphics{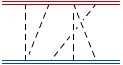}}
\newcommand{\BNDIntADKnLBU}{\includegraphics{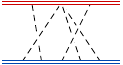}}
\newcommand{\BNDIntADKnLBQ}{\includegraphics{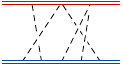}}
\newcommand{\BNDIntCnLAS}{\includegraphics{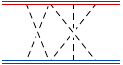}}
\newcommand{\BNDIntCnKXA}{\includegraphics{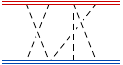}}
\newcommand{\BNDIntADKnLBO}{\includegraphics{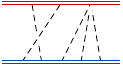}}
\newcommand{\BNDIntCnLAT}{\includegraphics{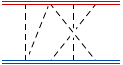}}
\newcommand{\BNDIntCnKZM}{\includegraphics{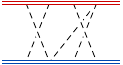}}
\newcommand{\BNDIntCnKWY}{\includegraphics{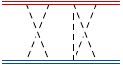}}
\newcommand{\BNDIntCnKZG}{\includegraphics{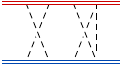}}
\newcommand{\BNDIntADLnLBV}{\includegraphics{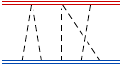}}
\newcommand{\BNDIntADLnLBR}{\includegraphics{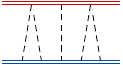}}
\newcommand{\BNDIntDnKZN}{\includegraphics{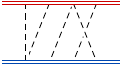}}
\newcommand{\BNDIntDnKXB}{\includegraphics{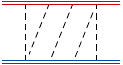}}
\newcommand{\BNDIntADLnLBU}{\includegraphics{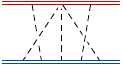}}
\newcommand{\BNDIntDnLAT}{\includegraphics{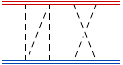}}
\newcommand{\BNDIntDnKZM}{\includegraphics{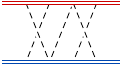}}
\newcommand{\BNDIntADMnLBU}{\includegraphics{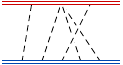}}
\newcommand{\BNDIntADMnLBQ}{\includegraphics{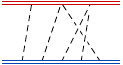}}
\newcommand{\BNDIntEnKXA}{\includegraphics{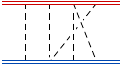}}
\newcommand{\BNDIntADMnLBO}{\includegraphics{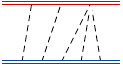}}
\newcommand{\BNDIntEnKZM}{\includegraphics{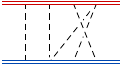}}
\newcommand{\BNDIntEnKWY}{\includegraphics{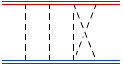}}
\newcommand{\BNDIntEnKZG}{\includegraphics{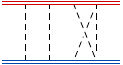}}
\newcommand{\BNDIntADNnLBU}{\includegraphics{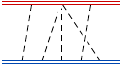}}
\newcommand{\BNDIntADNnLBQ}{\includegraphics{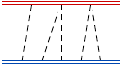}}
\newcommand{\BNDIntFnKXA}{\includegraphics{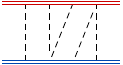}}
\newcommand{\BNDIntFnKZM}{\includegraphics{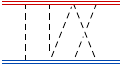}}
\newcommand{\BNDIntFnKWY}{\includegraphics{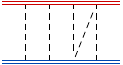}}
\newcommand{\BNDIntADOnLBU}{\includegraphics{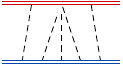}}
\newcommand{\BNDIntGnKZM}{\includegraphics{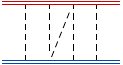}}
\newcommand{\BNDIntADPnLBX}{\includegraphics{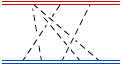}}
\newcommand{\BNDIntADPnLBV}{\includegraphics{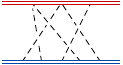}}
\newcommand{\BNDIntADPnLBR}{\includegraphics{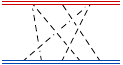}}
\newcommand{\BNDIntADPnLBQ}{\includegraphics{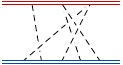}}
\newcommand{\BNDIntHnLBI}{\includegraphics{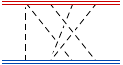}}
\newcommand{\BNDIntHnKZK}{\includegraphics{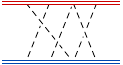}}
\newcommand{\BNDIntHnKWU}{\includegraphics{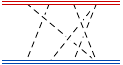}}
\newcommand{\BNDIntHnKYY}{\includegraphics{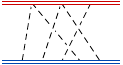}}
\newcommand{\BNDIntHnKWM}{\includegraphics{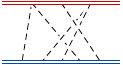}}
\newcommand{\BNDIntHnKUQ}{\includegraphics{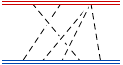}}
\newcommand{\BNDIntHnLBJ}{\includegraphics{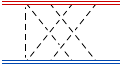}}
\newcommand{\BNDIntHnKZN}{\includegraphics{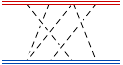}}
\newcommand{\BNDIntHnKXB}{\includegraphics{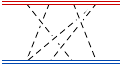}}
\newcommand{\BNDIntHnKZM}{\includegraphics{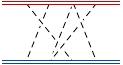}}
\newcommand{\BNDIntHnLBG}{\includegraphics{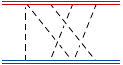}}
\newcommand{\BNDIntHnLBC}{\includegraphics{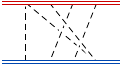}}
\newcommand{\BNDIntHnKZG}{\includegraphics{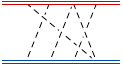}}
\newcommand{\BNDIntADQnLBV}{\includegraphics{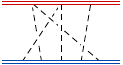}}
\newcommand{\BNDIntADQnLBR}{\includegraphics{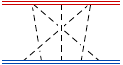}}
\newcommand{\BNDIntInKXB}{\includegraphics{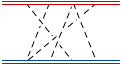}}
\newcommand{\BNDIntInLAT}{\includegraphics{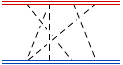}}
\newcommand{\BNDIntJnKWM}{\includegraphics{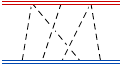}}
\newcommand{\BNDIntJnLBG}{\includegraphics{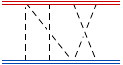}}
\newcommand{\BNDIntJnKWY}{\includegraphics{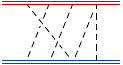}}
\newcommand{\BNDIntJnLBC}{\includegraphics{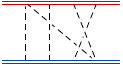}}
\newcommand{\BNDIntADSnLBV}{\includegraphics{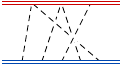}}
\newcommand{\BNDIntKnKUQ}{\includegraphics{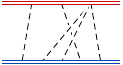}}
\newcommand{\BNDIntKnKXA}{\includegraphics{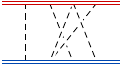}}
\newcommand{\BNDIntKnKZK}{\includegraphics{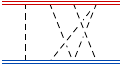}}
\newcommand{\BNDIntKnKWY}{\includegraphics{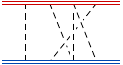}}
\newcommand{\BNDIntLnKZK}{\includegraphics{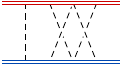}}
\newcommand{\BNDIntADUnLBV}{\includegraphics{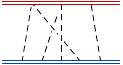}}
\newcommand{\BNDIntMnLAT}{\includegraphics{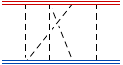}}
\newcommand{\BNDIntMnKZN}{\includegraphics{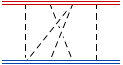}}
\newcommand{\BNDIntMnLAS}{\includegraphics{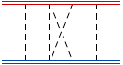}}
\newcommand{\BNDIntNnLAS}{\includegraphics{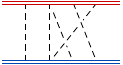}}
\newcommand{\BNDIntADWnLBV}{\includegraphics{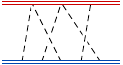}}
\newcommand{\BNDIntOnKXB}{\includegraphics{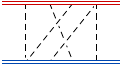}}
\newcommand{\BNDIntADWnLBO}{\includegraphics{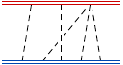}}
\newcommand{\BNDIntOnKVW}{\includegraphics{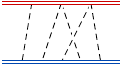}}
\newcommand{\BNDIntOnLAT}{\includegraphics{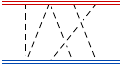}}
\newcommand{\BNDIntOnKWY}{\includegraphics{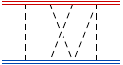}}
\newcommand{\BNDIntOnKWU}{\includegraphics{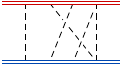}}
\newcommand{\BNDIntPnLBG}{\includegraphics{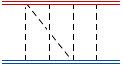}}
\newcommand{\BNDIntRnLBC}{\includegraphics{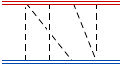}}
\newcommand{\BNDIntRnKWM}{\includegraphics{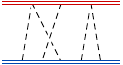}}
\newcommand{\BNDIntASMnLBX}{\includegraphics{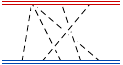}}
\newcommand{\BNDIntASMnLBV}{\includegraphics{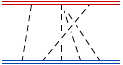}}
\newcommand{\BNDIntASMnLBR}{\includegraphics{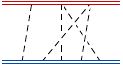}}
\newcommand{\BNDIntOEnKZN}{\includegraphics{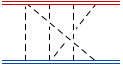}}
\newcommand{\BNDIntASMnLBU}{\includegraphics{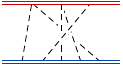}}
\newcommand{\BNDIntOEnLAS}{\includegraphics{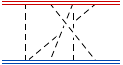}}
\newcommand{\BNDIntASMnLBO}{\includegraphics{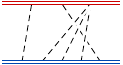}}
\newcommand{\BNDIntOEnLBJ}{\includegraphics{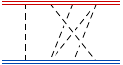}}
\newcommand{\BNDIntOEnLAT}{\includegraphics{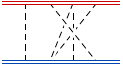}}
\newcommand{\BNDIntOEnKXB}{\includegraphics{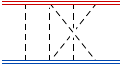}}
\newcommand{\BNDIntOEnLBI}{\includegraphics{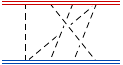}}
\newcommand{\BNDIntOEnKZM}{\includegraphics{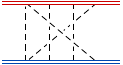}}
\newcommand{\BNDIntOEnLBG}{\includegraphics{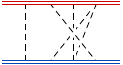}}
\newcommand{\BNDIntOEnLBC}{\includegraphics{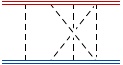}}
\newcommand{\BNDIntASNnLBV}{\includegraphics{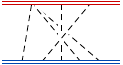}}
\newcommand{\BNDIntASNnLBR}{\includegraphics{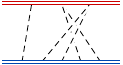}}
\newcommand{\BNDIntOFnKZN}{\includegraphics{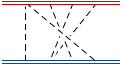}}
\newcommand{\BNDIntOFnKXB}{\includegraphics{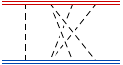}}
\newcommand{\BNDIntASNnLBO}{\includegraphics{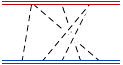}}
\newcommand{\BNDIntOFnLBC}{\includegraphics{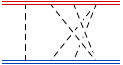}}
\newcommand{\BNDIntOFnLBG}{\includegraphics{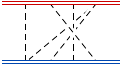}}
\newcommand{\BNDIntOFnKZK}{\includegraphics{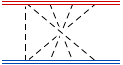}}
\newcommand{\BNDIntASOnLBX}{\includegraphics{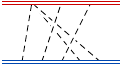}}
\newcommand{\BNDIntASOnLBV}{\includegraphics{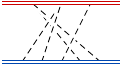}}
\newcommand{\BNDIntASOnLBR}{\includegraphics{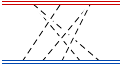}}
\newcommand{\BNDIntOGnKZN}{\includegraphics{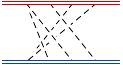}}
\newcommand{\BNDIntOGnKXB}{\includegraphics{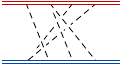}}
\newcommand{\BNDIntASOnLBU}{\includegraphics{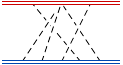}}
\newcommand{\BNDIntASOnLBQ}{\includegraphics{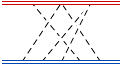}}
\newcommand{\BNDIntOGnLAS}{\includegraphics{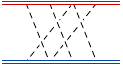}}
\newcommand{\BNDIntOGnKXA}{\includegraphics{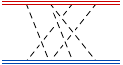}}
\newcommand{\BNDIntASOnLBO}{\includegraphics{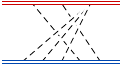}}
\newcommand{\BNDIntOGnLAT}{\includegraphics{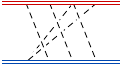}}
\newcommand{\BNDIntOGnKZM}{\includegraphics{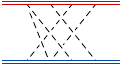}}
\newcommand{\BNDIntOGnKWY}{\includegraphics{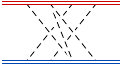}}
\newcommand{\BNDIntOGnKZG}{\includegraphics{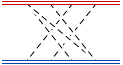}}
\newcommand{\BNDIntASPnLBV}{\includegraphics{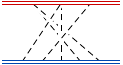}}
\newcommand{\BNDIntASPnLBR}{\includegraphics{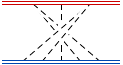}}
\newcommand{\BNDIntOHnKZN}{\includegraphics{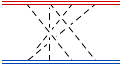}}
\newcommand{\BNDIntOHnKXB}{\includegraphics{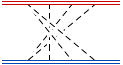}}
\newcommand{\BNDIntASPnLBU}{\includegraphics{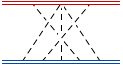}}
\newcommand{\BNDIntOHnLAT}{\includegraphics{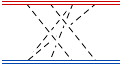}}
\newcommand{\BNDIntOHnKZM}{\includegraphics{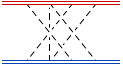}}
\newcommand{\BNDIntASQnLBU}{\includegraphics{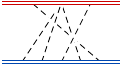}}
\newcommand{\BNDIntASQnLBQ}{\includegraphics{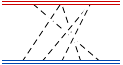}}
\newcommand{\BNDIntOInKXA}{\includegraphics{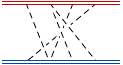}}
\newcommand{\BNDIntASQnLBO}{\includegraphics{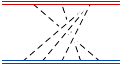}}
\newcommand{\BNDIntOInKZM}{\includegraphics{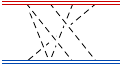}}
\newcommand{\BNDIntOInKWY}{\includegraphics{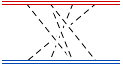}}
\newcommand{\BNDIntOInKZG}{\includegraphics{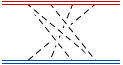}}
\newcommand{\BNDIntASRnLBU}{\includegraphics{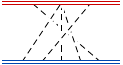}}
\newcommand{\BNDIntASRnLBQ}{\includegraphics{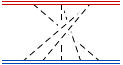}}
\newcommand{\BNDIntOJnKXA}{\includegraphics{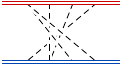}}
\newcommand{\BNDIntOJnKZM}{\includegraphics{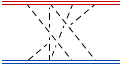}}
\newcommand{\BNDIntOJnKWY}{\includegraphics{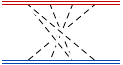}}
\newcommand{\BNDIntASSnLBU}{\includegraphics{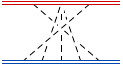}}
\newcommand{\BNDIntOKnKZM}{\includegraphics{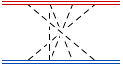}}
\newcommand{\BNDIntASTnLBX}{\includegraphics{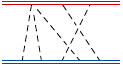}}
\newcommand{\BNDIntASTnLBR}{\includegraphics{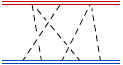}}
\newcommand{\BNDIntASTnLBQ}{\includegraphics{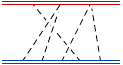}}
\newcommand{\BNDIntOLnLBI}{\includegraphics{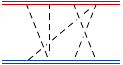}}
\newcommand{\BNDIntOLnKZK}{\includegraphics{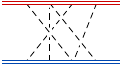}}
\newcommand{\BNDIntOLnKWU}{\includegraphics{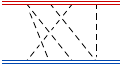}}
\newcommand{\BNDIntOLnKYY}{\includegraphics{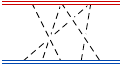}}
\newcommand{\BNDIntOLnKWM}{\includegraphics{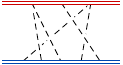}}
\newcommand{\BNDIntOLnKUQ}{\includegraphics{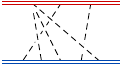}}
\newcommand{\BNDIntOLnLBJ}{\includegraphics{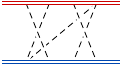}}
\newcommand{\BNDIntOLnKZN}{\includegraphics{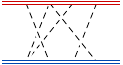}}
\newcommand{\BNDIntOLnKXB}{\includegraphics{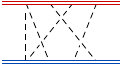}}
\newcommand{\BNDIntOLnKZM}{\includegraphics{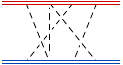}}
\newcommand{\BNDIntOLnLBG}{\includegraphics{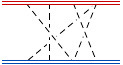}}
\newcommand{\BNDIntOLnLBC}{\includegraphics{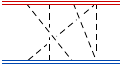}}
\newcommand{\BNDIntOLnKZG}{\includegraphics{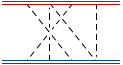}}
\newcommand{\BNDIntASUnLBV}{\includegraphics{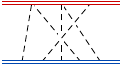}}
\newcommand{\BNDIntASUnLBR}{\includegraphics{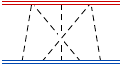}}
\newcommand{\BNDIntOMnKXB}{\includegraphics{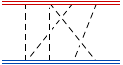}}
\newcommand{\BNDIntOMnLAT}{\includegraphics{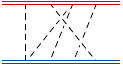}}
\newcommand{\BNDIntONnKWM}{\includegraphics{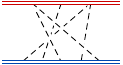}}
\newcommand{\BNDIntONnLBG}{\includegraphics{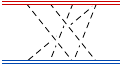}}
\newcommand{\BNDIntONnKWY}{\includegraphics{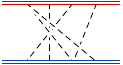}}
\newcommand{\BNDIntONnLBC}{\includegraphics{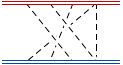}}
\newcommand{\BNDIntOOnKUQ}{\includegraphics{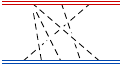}}
\newcommand{\BNDIntOOnKXA}{\includegraphics{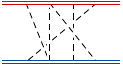}}
\newcommand{\BNDIntOOnKZK}{\includegraphics{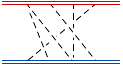}}
\newcommand{\BNDIntOOnKWY}{\includegraphics{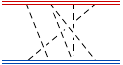}}
\newcommand{\BNDIntOPnKZK}{\includegraphics{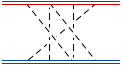}}
\newcommand{\BNDIntASYnLBV}{\includegraphics{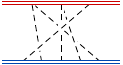}}
\newcommand{\BNDIntOQnLAT}{\includegraphics{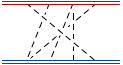}}
\newcommand{\BNDIntOQnKZN}{\includegraphics{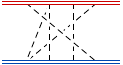}}
\newcommand{\BNDIntOQnLAS}{\includegraphics{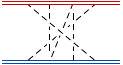}}
\newcommand{\BNDIntORnLAS}{\includegraphics{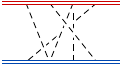}}
\newcommand{\BNDIntATAnLBV}{\includegraphics{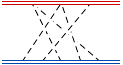}}
\newcommand{\BNDIntOSnKXB}{\includegraphics{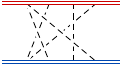}}
\newcommand{\BNDIntATAnLBO}{\includegraphics{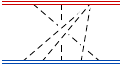}}
\newcommand{\BNDIntOSnKVW}{\includegraphics{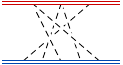}}
\newcommand{\BNDIntOSnLAT}{\includegraphics{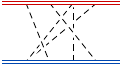}}
\newcommand{\BNDIntOSnKWY}{\includegraphics{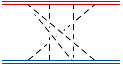}}
\newcommand{\BNDIntOSnKWU}{\includegraphics{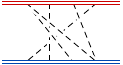}}
\newcommand{\BNDIntOTnLBG}{\includegraphics{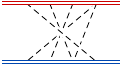}}
\newcommand{\BNDIntOVnLBC}{\includegraphics{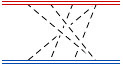}}
\newcommand{\BNDIntOVnKWM}{\includegraphics{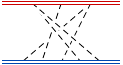}}
\newcommand{\BNDIntCFnKSE}{\includegraphics{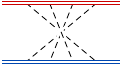}}
\newcommand{\BNDIntCGnKSE}{\includegraphics{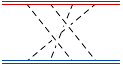}}
\newcommand{\BNDIntCHnKII}{\includegraphics{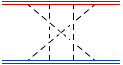}}
\newcommand{\BNDIntCInKII}{\includegraphics{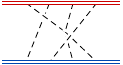}}
\newcommand{\BNDIntCJnKII}{\includegraphics{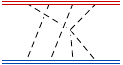}}
\newcommand{\BNDIntCKnKII}{\includegraphics{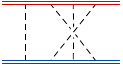}}
\newcommand{\BNDIntCLnKII}{\includegraphics{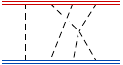}}
\newcommand{\BNDIntCMnKII}{\includegraphics{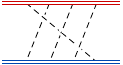}}
\newcommand{\BNDIntCNnKII}{\includegraphics{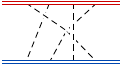}}
\newcommand{\BNDIntCOnKII}{\includegraphics{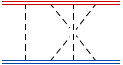}}
\newcommand{\BNDIntCPnKII}{\includegraphics{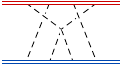}}
\newcommand{\BNDIntCQnKII}{\includegraphics{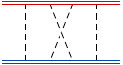}}
\newcommand{\BNDIntCRnKII}{\includegraphics{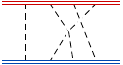}}
\newcommand{\BNDIntRJnKSE}{\includegraphics{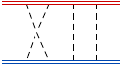}}
\newcommand{\BNDIntRKnKSE}{\includegraphics{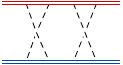}}
\newcommand{\BNDIntRLnKII}{\includegraphics{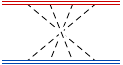}}
\newcommand{\BNDIntRMnKII}{\includegraphics{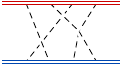}}
\newcommand{\BNDIntRNnKII}{\includegraphics{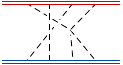}}
\newcommand{\BNDIntROnKII}{\includegraphics{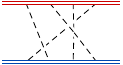}}
\newcommand{\BNDIntRPnKII}{\includegraphics{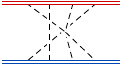}}
\newcommand{\BNDIntRQnKII}{\includegraphics{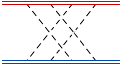}}
\newcommand{\BNDIntRRnKII}{\includegraphics{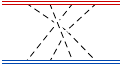}}
\newcommand{\BNDIntRSnKII}{\includegraphics{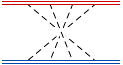}}
\newcommand{\BNDIntRTnKII}{\includegraphics{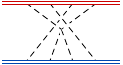}}
\newcommand{\BNDIntRUnKII}{\includegraphics{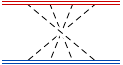}}
\newcommand{\BNDIntRVnKII}{\includegraphics{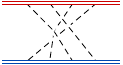}}
\newcommand{\BNDIntAnLCA}{\includegraphics{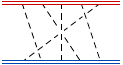}}
\newcommand{\BNDIntBnLCA}{\includegraphics{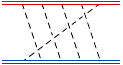}}
\newcommand{\BNDIntCnLCA}{\includegraphics{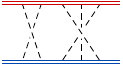}}
\newcommand{\BNDIntDnLCA}{\includegraphics{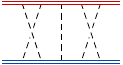}}
\newcommand{\BNDIntEnLCA}{\includegraphics{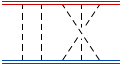}}
\newcommand{\BNDIntFnLCA}{\includegraphics{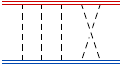}}
\newcommand{\BNDIntGnLCA}{\includegraphics{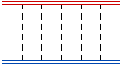}}
\newcommand{\BNDIntHnLCA}{\includegraphics{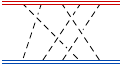}}
\newcommand{\BNDIntInLCA}{\includegraphics{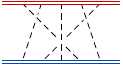}}
\newcommand{\BNDIntJnLCA}{\includegraphics{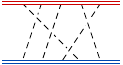}}
\newcommand{\BNDIntKnLCA}{\includegraphics{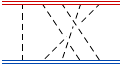}}
\newcommand{\BNDIntLnLCA}{\includegraphics{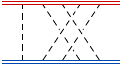}}
\newcommand{\BNDIntMnLCA}{\includegraphics{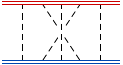}}
\newcommand{\BNDIntNnLCA}{\includegraphics{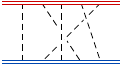}}
\newcommand{\BNDIntOnLCA}{\includegraphics{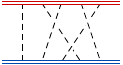}}
\newcommand{\BNDIntPnLCA}{\includegraphics{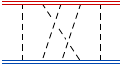}}
\newcommand{\BNDIntQnLCA}{\includegraphics{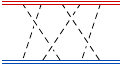}}
\newcommand{\BNDIntRnLCA}{\includegraphics{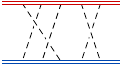}}
\newcommand{\BNDIntSnLCA}{\includegraphics{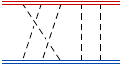}}
\newcommand{\BNDIntTnLCA}{\includegraphics{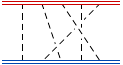}}
\newcommand{\BNDIntUnLCA}{\includegraphics{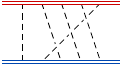}}
\newcommand{\BNDIntVnLCA}{\includegraphics{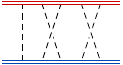}}
\newcommand{\BNDIntWnLCA}{\includegraphics{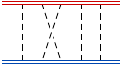}}
\newcommand{\BNDIntOEnLCA}{\includegraphics{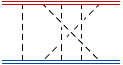}}
\newcommand{\BNDIntOFnLCA}{\includegraphics{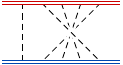}}
\newcommand{\BNDIntOGnLCA}{\includegraphics{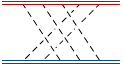}}
\newcommand{\BNDIntOHnLCA}{\includegraphics{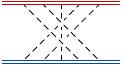}}
\newcommand{\BNDIntOInLCA}{\includegraphics{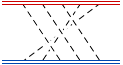}}
\newcommand{\BNDIntOJnLCA}{\includegraphics{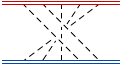}}
\newcommand{\BNDIntOKnLCA}{\includegraphics{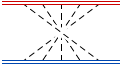}}
\newcommand{\BNDIntOLnLCA}{\includegraphics{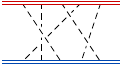}}
\newcommand{\BNDIntONnLCA}{\includegraphics{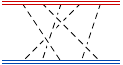}}
\newcommand{\BNDIntOOnLCA}{\includegraphics{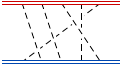}}
\newcommand{\BNDIntOPnLCA}{\includegraphics{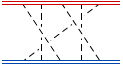}}
\newcommand{\BNDIntOQnLCA}{\includegraphics{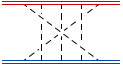}}
\newcommand{\BNDIntORnLCA}{\includegraphics{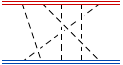}}
\newcommand{\BNDIntOSnLCA}{\includegraphics{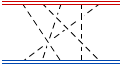}}
\newcommand{\BNDIntOTnLCA}{\includegraphics{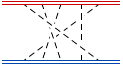}}
\newcommand{\BNDIntOUnLCA}{\includegraphics{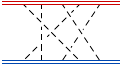}}
\newcommand{\BNDIntOVnLCA}{\includegraphics{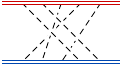}}
\newcommand{\BNDIntOWnLCA}{\includegraphics{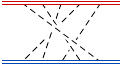}}
\newcommand{\BNDIntOXnLCA}{\includegraphics{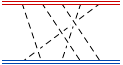}}
\newcommand{\BNDIntOYnLCA}{\includegraphics{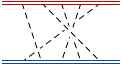}}
\newcommand{\BNDIntOZnLCA}{\includegraphics{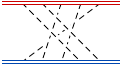}}
\newcommand{\BNDIntPAnLCA}{\includegraphics{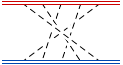}}

\newif\ifincludefigures
\includefigurestrue     

\begin{document}

\title{Amplitudes, Supersymmetric Black Hole Scattering at $\mathcal{O}(G^5)$, and Loop Integration}

\author[1]{Zvi Bern,}
\affiliation[1]{
Mani L. Bhaumik Institute for Theoretical Physics,
University of California at Los Angeles, \newline
Los Angeles, CA 90095, USA}
\emailAdd{bern@physics.ucla.edu}

\author[1]{Enrico Herrmann,}
\emailAdd{eh10@g.ucla.edu}

\author[2]{Radu Roiban,}
\affiliation[2]{Institute for Gravitation and the Cosmos,
Pennsylvania State University,\newline 
University Park, PA 16802, USA}
\emailAdd{radu@phys.psu.edu}

\author[1]{Michael~S.~Ruf,}
\emailAdd{mruf@physics.ucla.edu}

\author[4,5]{Alexander V. Smirnov,}
\affiliation[4]{Research Computing Center, Moscow State University, 119991 Moscow, Russia}
\affiliation[5]{Moscow Center for Fundamental and Applied Mathematics, 119992 Moscow, Russia}
\emailAdd{asmirnov80@gmail.com}

\author[5,6]{Vladimir A. Smirnov,}
\affiliation[6]{Skobeltsyn Institute of Nuclear Physics of Moscow State University, 119991, Moscow, Russia}
\emailAdd{smirnov@theory.sinp.msu.ru}

\author[7]{Mao Zeng}
\affiliation[7]{Higgs Centre for Theoretical Physics, University of Edinburgh, Edinburgh, EH9 3FD, UK}
\emailAdd{mao.zeng@ed.ac.uk}

\abstract{
We compute the potential-graviton contribution to the scattering amplitude, the radial action, and the scattering angle of two extremal black holes in ${\cal N}=8$ supergravity at the fifth post-Minkowskian order and to next-to-leading order in a large mass expansion (first self-force order).
Properties of classical unitarity cuts allow us to focus on the integration-by-parts reduction of planar integrals, while nonplanar integrals at this order are obtained from the planar ones by straightforward manipulations.
We present the solution to the differential equations for all master integrals  necessary to evaluate the classical scattering amplitudes of massive scalar particles at this order in all gravitational theories, in particular in ${\cal N}=8$ supergravity, and in general relativity. 
Despite the appearance of higher-weight generalized polylogarithms and elliptic functions in the solution to the differential equation for master integrals, the final supergravity answer is remarkably simple and contains only (harmonic) polylogarithmic functions up to weight 2.
The systematic analysis of elliptic integrals discussed here, as well as the particular organization of boundary integrals in ${\cal N}=8$ observables are independent of supersymmetry and may have wider applications, including to aspects of collider physics. 
}

\maketitle


\newpage

%
\newpage
\section{Introduction}

The detection of gravitational waves~\cite{LIGOScientific:2016aoc, LIGOScientific:2017vwq} marks a major scientific milestone, opening a remarkable new window on the universe.  
With the arrival of the next generation gravitational-wave detectors~\cite{Punturo:2010zz, LISA:2017pwj, Reitze:2019iox} in the coming years, vast increases in sensitivity and frequency range are expected.  
Yet, achieving precise theoretical modeling across the entire parameter spectrum poses formidable challenges.   This endeavor necessitates substantial advancements across various complementary approaches, encompassing numerical relativity~\cite{Pretorius:2005gq, Campanelli:2005dd, Baker:2005vv, Damour:2014afa}, the self-force (SF) program~\cite{Mino:1996nk, Quinn:1996am, Poisson:2011nh, Barack:2018yvs}, effective field theory (EFT)~\cite{Goldberger:2004jt}, the post-Newtonian (PN)~\cite{Droste:1916, Droste:1917, Einstein:1938yz, Ohta:1973je, Blanchet:2013haa} and the post-Minkowskian (PM)~\cite{Bertotti:1956pxu, Kerr:1959zlt, Bertotti:1960wuq, Westpfahl:1979gu, Portilla:1980uz, Bel:1981be} methods. 
These advancements would then need to be combined into accurate waveform models, such as those based on the effective-one-body (EOB) approach~\cite{Buonanno:1998gg, Buonanno:2000ef}.

Fuelled by applications to gravitational-wave physics, gravitational scattering has witnessed a renewed interest as it provides a clean theoretical probe of two-body dynamics. In this classical setup, the states representing the compact astrophysical objects are widely separated throughout the collision process so that the background spacetime is asymptotically Minkowskian and the process can be treated perturbatively. This provides straightforward definitions of physical observables, making it easier to compare results obtained via different methods. The PM framework is the natural language for gravitational scattering, as it maintains Lorentz invariance throughout the calculation, treats interactions as a perturbative series in Newton's constant $G$, and accounts for all orders in velocity.

The connection between quantum-field-theory (QFT) scattering and general relativity corrections to Newton's potential has long been appreciated~\cite{Iwasaki:1971vb, Iwasaki:1971iy, Gupta:1979br, Donoghue:1994dn, Bjerrum-Bohr:2002gqz, Bjerrum-Bohr:2013bxa, Neill:2013wsa, Damour:2016gwp, Damour:2017zjx}.   Scattering amplitudes offer a radically different approach that starts not from the Einstein-Hilbert action but instead from Feynman's observation that Minkowski space gravity is mediated by gravitons, i.e. massless spin-2 particles~\cite{Feynman:1963ax}. Moreover, the double-copy construction allows gravitational scattering amplitudes to be constructed to all orders in Newton's constant starting from corresponding gauge-theory scattering amplitudes~\cite{Kawai:1985xq, Bern:2008qj, Bern:2010ue, Bern:2019prr}. While the starting point is completely different, the results obtained in an amplitudes-based PM expansion should be identical to those derived from Einstein field equations.

On the way to evaluating the PM expansion of scattering observables in general relativity, the study of simpler theories such as electrodynamics or ${\cal N}=8$ supergravity offers a preview of the subtleties of the desired calculations while avoiding some of its complexities. Indeed, graviton dominance~\cite{tHooft:1987vrq, Amati:1990xe, Muzinich:1987in, Bellini:1992eb, Parra-Martinez:2020dzs, DiVecchia:2020ymx} made ${\cal N}=8$ supergravity an ideal testing ground for identifying graviton modes responsible for the improved high-energy behavior~\cite{DiVecchia:2020ymx} of the two-loop scattering angle subsequently found in general relativity~\cite{Damour:2020tta}.
Moreover, its extended supersymmetry, absence of one-loop triangle subgraph, and hidden leading-order integrability~\cite{Caron-Huot:2018ape} lead to simpler expressions, which nevertheless capture most of the relevant classical nonsupersymmetric physics.

The application of quantum-field-theory techniques to gravitational-wave physics fundamentally relies on the separation of scales of conservative and radiative processes, and on the distinction between the scale of the binary constituents and their mutual minimal separation. Thus, effective field theory (see, e.g.,~Ref.~\cite{Manohar:2018aog}) is a natural framework for the problem of gravitational-wave emission and was originally implemented in worldline formulations~(see, e.g., Refs.~\cite{Goldberger:2004jt, Porto:2006bt, Levi:2015msa, Kalin:2020fhe, Dlapa:2021npj, Jakobsen:2021zvh, Jakobsen:2022psy, Jakobsen:2023oow, Driesse:2024xad}). The same basic ideas underlie the recent application of quantum scattering amplitudes to the gravitational two-body problem in general relativity~\cite{Neill:2013wsa, Bjerrum-Bohr:2013bxa, Damour:2016gwp, Damour:2017zjx, Cheung:2018wkq}  (see Refs.~\cite{Iwasaki:1971vb, Iwasaki:1971iy, Donoghue:1994dn, Bjerrum-Bohr:2002gqz} for related early work). Significant strides in this approach have been fueled and sustained by methods originally developed for quantum scattering processes for collider physics, including generalized unitarity~\cite{Bern:1994zx, Bern:1994cg, Bern:1997sc, Britto:2004nc, Bern:2004cz, Bern:2007ct}, the double-copy relations between gauge and gravity theories~\cite{Kawai:1985xq, Bern:2008qj, Bern:2010ue, Bern:2019prr},  and by powerful integration methods~\cite{Chetyrkin:1981qh, Laporta:2000dsw, Maierhofer:2017gsa, Smirnov:2008iw, Smirnov:2019qkx, Kotikov:1990kg, Bern:1993kr, Remiddi:1997ny, Gehrmann:1999as, Smirnov:2020quc, Usovitsch:2020jrk, Henn:2013pwa, Henn:2013woa, Beneke:1997zp, vonManteuffel:2014ixa, Peraro:2016wsq, Klappert:2019emp, Peraro:2019svx, Laurentis:2019bjh, DeLaurentis:2022otd, Magerya:2022hvj, Belitsky:2023qho, LewinPolylogs, Goncharov:1998kja, Ablinger:2011te, Bourjaily:2022bwx, Frellesvig:2023bbf, Klemm:2024wtd}. 

Several amplitudes-based approaches to extracting classical observables from scattering amplitudes have been developed:
the EFT matching of two-body Hamiltonians ~\cite{Cheung:2018wkq, Bern:2019nnu, Bern:2019crd}, 
the observables-based (KMOC) formalism~\cite{Kosower:2018adc}, 
the eikonal approach~\cite{DiVecchia:2021bdo}, 
the amplitude~radial-action relation~\cite{Bern:2021dqo, Kol:2021jjc, Bern:2021yeh}, 
and an exponential representation of the S-matrix~\cite{Damgaard:2023ttc}. 
The expansion in a small mass ratio (or the self-force expansion) within the PM approach, employing aspects of quantum field theory in curved space, has also been established~\cite{Cheung:2023lnj, Kosmopoulos:2023bwc, Adamo:2023cfp}. More generally, the organization of observables in terms of mass dependence can be used in conjunction with any of the other strategies for extracting classical physics. In this paper, we employ this organization as a gauge-invariant separation of simpler and more complicated parts of the calculation at a fixed PM order.

An existing quantum integrand, as it is the case for ${\cal N}=8$ supergravity~\cite{Bern:2007hh, Bern:2009kd, Bern:2017ucb}, is a suitable starting point of the flat-space amplitudes-based approach to classical observables. The existence of such an integrand, however, is not required, and its parts that contribute to the classical limit can be found by directly constructing the spanning set of generalized unitarity cuts (set of cuts which probe all relevant combinations of propagators) that capture classical physics~\cite{Bern:2019crd} and then merging them into the classical amplitude.
The absence of global canonical (dual) variables~\cite{Drummond:2006rz, Alday:2007hr, Drummond:2008vq, Arkani-Hamed:2010zjl} for amplitudes beyond the planar limit can make cut merging a challenging process involving e.g.~matching a large global ansatz onto the (classical) generalized cuts.
As detailed in Ref.~\cite{Bern:2024vqs}, it is, however,  possible to {\em choose} a global basis of nonplanar integrands and actively map all cuts to the chosen basis. If desired, the generalized cuts can then be merged by simply reading off the coefficient of each basis element from the generalized cuts. The agreement of coefficients of basis integrands that are shared between different cuts yields nontrivial consistency conditions. 
However, the cut merging step is not strictly necessary for the calculation of classical observables which can be obtained from the unitarity cuts themselves.

While this unitarity-cut-based approach is beneficial for Einstein gravity, the relevant integrand for $\mathcal{N} = 8$ supergravity is readily available by dimensionally reducing the known quantum integrand~\cite{Bern:2009kd}, thus rendering the use of cut integrands unnecessary in our setup. 
The use of a global basis is, nevertheless, advantageous even though \emph{an} integrand is already available: the mapping to a global basis removes spurious singularities and manifests its gauge-invariance properties~\cite{Bern:2024vqs}, therefore yielding a cleaner starting point for the integration step.
Moreover, as we will discuss in \sect{subsec:optimizingIBP}, based on ideas of Refs.~\cite{Larsen:2015ped, Zhang:2016kfo, Bohm:2018bdy}, spanning sets of generalized cuts are an extremely useful ingredient in the efficient implementation of integration-by-parts (IBP) reduction of Feynman integrals.

All current approaches to PM observables lead to the same types of integrals which share many similarities to the ones encountered in the evaluation of perturbative quantum scattering amplitudes for collider physics applications. Thus, a common key part is dealing with the integrals.
The primary differences are that the classically-relevant integrals have a lower power count, exhibit a reduction in the number of independent kinematic invariants, and have linearized (eikonal) matter propagators instead of the usual quadratic Feynman propagators.
Their evaluation employs powerful methods including the reduction of integrals to a basis of master integrals~\cite{Chetyrkin:1981qh, Laporta:2000dsw, Maierhofer:2017gsa, Smirnov:2008iw,  Smirnov:2019qkx}, differential equations for determining analytic results for master integrals~\cite{Kotikov:1990kg, Bern:1993kr, Remiddi:1997ny, Gehrmann:1999as}, improved bases of master integrals both for facilitating the IBP reduction~\cite{Smirnov:2020quc, Usovitsch:2020jrk} and for solving the differential equations~\cite{Henn:2013pwa}, the method of regions for selecting particular contributions in an asymptotic expansion~\cite{Beneke:1997zp}, the use of finite prime fields to vastly speed up solving integration-by-parts relations~\cite{vonManteuffel:2014ixa, Peraro:2016wsq, Klappert:2019emp, Peraro:2019svx, Laurentis:2019bjh, DeLaurentis:2022otd, Magerya:2022hvj, Belitsky:2023qho} and the mathematics of iterated integrals, including generalized polylogarithms, elliptic integrals, and more recently Calabi-Yau integrals (see e.g. Refs.~\cite{LewinPolylogs, Goncharov:1998kja, Ablinger:2011te, Bourjaily:2022bwx, Frellesvig:2023bbf, Klemm:2024wtd}).  Underlying these methods is dimensional regularization~\cite{tHooft:1972tcz}, which has the effect of setting to zero scaleless integrals and ensuring the vanishing of IBP boundary terms. Many of these ideas are implemented in publicly available and private codes.

A combination of the above methods have pushed the state of the art to $\mathcal{O}(G^4)$ or the fourth PM order~\cite{Bern:2019nnu, Bern:2019crd, Bern:2021dqo, Bern:2021yeh, Dlapa:2021npj, Dlapa:2021vgp, Manohar:2022dea, Dlapa:2022lmu, Bjerrum-Bohr:2022ows}. While both worldline and scattering amplitude based approaches are effective at constructing integrands at $\mathcal{O}(G^5)$, the four-loop integral evaluation remains the primary bottleneck. The essential difficulty is that the IBP algorithms used to reduce the integrals become unwieldy and require careful tuning to bring the problem within reach~\cite{Bern:2023ccb, Driesse:2024xad}.  It is helpful to attack the problem in stages. 
Two useful subdivisions are to (1) first carry out the analogous computation in simpler theories such as electrodynamics and ${\cal N}=8$ supergravity and (2) expand in the mass ratio of the two massive objects, corresponding to the self-force expansion. 
Following the former path, in Ref.~\cite{Bern:2023ccb}, we obtained the potential-mode contributions to the electrodynamics analog of the 5PM scattering angle.
The essential difference between the electromagnetic and gravitational cases is that gravitons self-interact while photons do not, reducing the number of contributing diagram topologies.  On the other hand, for the diagram topologies that overlap with the gravitational case, the complexity of the integrals is identical.\footnote{In the quantum theory, the higher tensor power in the numerators in gravity implies that the integrals are more complicated to reduce than those of the corresponding gauge-theory ones, but these more complicated integrals yield only quantum contributions and are irrelevant in the classical limit.}  
Expanding in the mass ratio subdivides the problem into well-defined pieces organized in a similar fashion as the self-force expansion.  The 0th-self-force (0SF) piece is the simplest and corresponds to the probe limit, where one of the masses is much lighter than the other. The resulting physics is that of a probe particle moving along geodesics of a fixed background metric, for which closed form results are available to all orders in $G$.  The next term in the mass expansion, which we denote as the 1SF part, is the next simplest, with the remaining 2SF piece containing the most complicated integrals.  
Recently, the 1SF conservative contributions to the 5PM gravitational impulses were computed in general relativity in Ref.~\cite{Driesse:2024xad} using the WQFT framework of~\cite{Mogull:2020sak}; these contributions were defined by integrals with zero and two radiation modes, retarded matter propagators and Feynman graviton propagators.
In this paper, we find the potential-graviton contribution to the radial action and the scattering angle of extremal black holes in ${\cal N}=8$ supergravity through amplitudes methods. 

In carrying out the computations of this paper, we make extensive use of \texttt{FIRE6}~\cite{Smirnov:2019qkx, Smirnov:2023yhb}, as well as a program written in \texttt{Mathematica} for determining a set of improved master integrals~\cite{Smirnov:2020quc, Usovitsch:2020jrk}. We also found \texttt{LiteRed}~\cite{Lee:2013mka} to be useful for identifying symmetry relations that reduce that number of master integrals. We made extensive modifications to the \texttt{FIRE6} code to increase its speed and reduce its memory consumption.  Additionally, we developed a private code, based on finite-field methods~\cite{vonManteuffel:2014ixa, Peraro:2016wsq, Klappert:2019emp, Peraro:2019svx, Laurentis:2019bjh, DeLaurentis:2022otd, Magerya:2022hvj, Belitsky:2023qho}, to  deal with integrals that are difficult
to handle with \texttt{FIRE}. In contrast to lower orders, at 5PM order it is extremely advantageous to exploit relations between integrals that arise because of the classical limit. For example, in this limit, each loop is required to contain at least one matter propagator~\cite{Cheung:2018wkq, Bern:2019nnu}, and thus, certain matter lines can be considered cut. Using this observation, we demonstrate in \sect{subsec:Planarization} that the integral reduction of all nonplanar integrals contributing to the 1SF radial action and scattering angle can be obtained through straightforward manipulations of the integral reduction of planar integrals. This observation can lead to remarkable simplification of the reduction of the ${\cal N}=8$ integrand. 

In this paper, we present a detailed description of the four-loop master integrals that appear in the 1SF potential-region calculations in gravitational theories. Differential equations, constructed using IBP reduction, determine the master integrals in terms of polylogarithms and (iterated) elliptic integrals up to boundary values, in our case, the values of the first few terms in the low-velocity expansion of master integrals. 
Individually, the boundary integrals are not well-defined and their evaluation would require an additional regulator. 
Remarkably, the boundary integrals enter the classical amplitude and observables only in well-defined special combinations, referred to as {\em eikonal sums}, for which such an additional regulator is not necessary. A multivariate generalization of the Sokhotskia Plemelj theorem~\cite{Saotome:2012vy, Akhoury:2013yua} reduces the evaluation of the loop energy integrals to the evaluation of integrals over Dirac delta functions. 
The systematic analysis of elliptic integrals developed here as well as the organization of boundary conditions in terms of eikonal sums may have wider applications, including to questions in collider-physics.

The rest of our paper is organized as follows.
In \sect{sec:integrand}, we discuss the four-massive scalar integrand in ${\cal N}=8$ supergravity and also outline the use of generalized cuts as a means to bypass the need for such an integrand in classical calculations. 
In \sect{sec:IBP}, we discuss the reduction of the relevant integrals to a basis of master integrals and the optimizations we included to carry it out. 
Most importantly, we discuss a strategy that allows us to build the IBP reduction of 1SF nonplanar integrals solely in terms of the IBP reduction of planar ones. IBP reduction on a spanning set of generalized unitarity cuts plays an essential role.  
In \sect{sec:DE}, we discuss the differential equations for the master integrals, the steps required to bring them to the canonical form, and their solution in terms of polylogarithmic and elliptic iterated integrals. We also discuss the boundary conditions and the separation and subtraction of classical iteration terms. We provide the values of integrals in a computer-readable compressed file~\texttt{masterIntegralValues.m}.
In \sect{sec:Neq8assembly}, we assemble the classical amplitude, the radial action, and the scattering angle for a collision of two extremal black holes in $\mathcal{N} = 8$ supergravity. The result is remarkably simple and, similarly to the general relativity one, it exhibits no elliptic integrals, being given solely in terms of classical polylogarithms of weight two. 
Our result passes all available consistency conditions: the cancellation of all spurious divergences, the comparison of the probe limit with known results, and the relation between the low-velocity expansion of the 5PM scattering angle and lower-order data.  Finally, supplementary material contains our conventions (\texttt{conventions.m}), the values of the master integrals up to transcendental weight three (\texttt{masterIntegralValues.m}), and the results for the potential-graviton contribution to the classical amplitude through 1SF order (\texttt{amplitude$\_$angle$\_$N8.m}).

%
\section{Structure of the integrand}
\label{sec:integrand}
%

The principles of effective field theory~\cite{Manohar:2018aog} imply that at distances much larger than their size, black holes and other compact astrophysical objects can be treated as point particles~\cite{Goldberger:2004jt}. 
The quantum field theory approach to PN or PM approximations builds on this observation and can be systematically corrected to account for the objects' sizes. In the classical limit and in the regime in which these expansions are valid it gives results that are identical to those obtained by solving the Einstein equations. 

Our calculations start with quantum scattering amplitudes, whose well-understood structure we outline here. Classical results can then be extracted via several approaches~\cite{Cheung:2018wkq, Kosower:2018adc, Bern:2019nnu, Bern:2019crd, Bern:2021dqo, DiVecchia:2021bdo, Bjerrum-Bohr:2021wwt, Damgaard:2023ttc}. 
The classical contributions are found in the limit where gravitons are soft, in the sense described in \sect{subsec:classical_limit} below. This soft region can be further refined to potential- and radiation-mode contributions following the method of regions~\cite{Beneke:1997zp}. 
In this section we do not distinguish between planar and nonplanar terms in the amplitude. However, by four loops, it becomes highly advantageous to avoid a direct evaluation of nonplanar integrals; we discuss this in~\sect{subsec:Planarization}. 

We work entirely in the context of scattering kinematics, setting aside the important question of analytic continuation to the bound regime~\cite{Bini:2017wfr, Cho:2021arx}, the details of which are still to be clarified at 4PM order and beyond~\cite{Bini:2017wfr, Cho:2021arx, Dlapa:2024cje}. 

\subsection{Quantum amplitudes} 
\label{AmplitudeSubsection}

\begin{figure}[tb!]
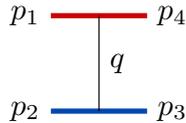

\begin{center}
\treetGraviton
\end{center}
\vskip -.32 cm  
 \caption{The leading order scattering process as a diagram and the associated momenta in an all-outgoing convention. The thick (red and blue) lines represent massive particles, and the thin line represents gravitons or other massless bosons.}
\label{LOFeynmanFigure}
\end{figure}

Computing quantum scattering amplitudes of point-like particles is a well-developed subject, either using   Feynman diagrams or via more modern methods based on generalized unitarity and factorization.  
The basic kinematics of the four-point scattering process can be read from the leading-order diagram shown in \fig{LOFeynmanFigure}, where we consider all momenta to be outgoing. In this diagram, the thick lines represent massive black holes or other compact astrophysical objects, while the thin lines are the massless bosons of the theory under consideration; for Einstein gravity, this would be the gravitons, while in $\mathcal{N}=8$ supergravity this also includes massless scalars and graviphotons. The momentum transfer $q = p_1 + p_4$ plays an important role in the identification of the classical contributions.

Here, we focus on $\mathcal N = 8$ supergravity as a stepping stone between our previous calculation in electrodynamics~\cite{Bern:2023ccb} and Einstein gravity, which captures most of the features of the latter, such as they share the same master integrals.
Ref.~\cite{Caron-Huot:2018ape} sets up the problem of the dynamics of a pair of extremal black holes in $\mathcal N = 8$ supergravity as a potentially solvable model of gravitational dynamics.  This model has also been used to understand the effects of radiative contributions to the high-energy limit~\cite{DiVecchia:2020ymx}, and for developing improved integration methods~\cite{Parra-Martinez:2020dzs} in the classical limit.  

Massless higher-dimensional four-point integrands  for $\mathcal N = 8$ supergravity were constructed through five-loop order using generalized unitarity method~\cite{Bern:1994zx, Bern:1994cg, Bern:1997sc, Britto:2004nc, Bern:2004cz, Bern:2007ct} in conjunction with the double copy~\cite{Kawai:1985xq, Bern:2008qj, Bern:2010ue, Bern:2019prr} with the goal of studying the ultraviolet properties of supergravity theories~\cite{Bern:1998ug, Bern:2009kd, Bern:2014sna}. 
Integrands of massive scalar amplitudes follow by dimensional reduction, see Ref.~\cite{Parra-Martinez:2020dzs} and below.
The existence of quantum integrands is, however, not a prerequisite for extracting classical observables. In general, such calculations make use of the generalized cuts that capture classical physics~\cite{Bern:2019nnu, Bern:2019crd}. 
Since such cuts also feature in our integration method, we therefore begin by discussing some of their relevant aspects.

At any order, the integrands of scattering amplitudes are rational functions with poles specified by the contributing Feynman graphs and residues -- or generalized cuts -- ensuing from the factorization properties of Feynman diagrams.
Specifying a particular singularity in the form of a collection of on-shell propagators, the corresponding generalized cut is given by the product of the corresponding tree amplitudes summed over all possible states on the on-shell propagators. 
Knowledge of all generalized $D$-dimensional cuts uniquely specifies the integrand.

\begin{figure}
\centering
\begin{subfigure}[b]{0.3\textwidth}
    \includegraphics{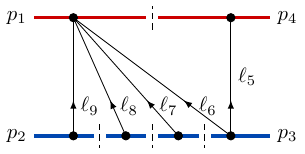}
    \caption{}
    \label{subfig:cutA}
\end{subfigure}\hskip 1.4 cm
\begin{subfigure}[b]{0.3\textwidth}
    \includegraphics{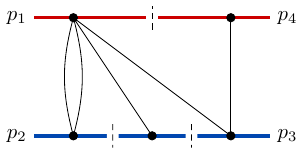}
    \caption{}
    \label{subfig:cutB}
\end{subfigure}
    \caption{Example of generalized cuts of the quantum amplitude. All graphs in cut \subref*{subfig:cutA} contribute in the classical limit. Graphs with support on only cut \subref*{subfig:cutB} (graphs with a graviton bubble) contain purely quantum information; they may be removed by introducing a further cut either on the lower (blue) matter line to obtain cut~\subref*{subfig:cutA} or on the upper (red) matter line.}
    \label{fig:examplesofcuts}
\end{figure}

A complementary interpretation of a generalized cut is that it is an operation that removes from the amplitude's integrand all terms that do not contain a specified set of propagators. 
This makes cuts an ideal tool to remove early in the construction of an integrand all terms 
that do not contribute to e.g. the classical amplitude~\cite{Bern:2019nnu, Bern:2019crd}. Indeed, since the separation of scales in the classical limit requires matter particles to be widely separated, we need to consider only cuts that do not allow matter lines to intersect.
Moreover, since in the classical limit, the scale of each loop integral must be set by the velocities of the matter particles, we need to consider only cuts that contain at least one matter line per loop. See \fig{fig:examplesofcuts} for examples. The residue, together with the on-shell conditions for the cut in Fig.~\ref{subfig:cutA}, is  
\begin{align}
{\cal C}_{(a)} &= 
\delta((p_2+\ell_9)^2-m_2^2)
\delta((p_2+\ell_{89})^2-m_2^2)
\delta((p_2+\ell_{789})^2-m_2^2)
\nonumber
\\
&\times \delta((p_4-\ell_5)^2-m_1^2)
\prod_{i=5}^9\delta(\ell_i^2)
M_0(p_1, -\ell_9, -\ell_8,-\ell_7,-\ell_6)
M_0(p_4, -\ell_5)
\nonumber
\\
&\times 
M_0(p_2, \ell_9)
M_0(p_2+\ell_9, \ell_8)
M_0(p_2+\ell_{89}, \ell_7)
M_0(p_2+\ell_{879}, \ell_6, \ell_5)\,,
\label{eq:cutexample}
\end{align}
where $\ell_{i\dots k} = \ell_i + \dots +\ell_k$ and $M_0$ denotes various tree-amplitudes in all-outgoing momentum conventions. The classical expansion discussed in \sect{subsec:classical_limit} applied to generalized cuts introduces additional features, such as the appearance of derivatives of the delta functions imposing on-shell conditions, which we discuss in \sect{subsec:Planarization}. 

Apart from determining classical and quantum amplitude integrands, 
generalized cuts are, as we discuss in \sect{subsec:optimizingIBP}, 
also a means to speed up and parallelize the IBP reduction.
Carrying out the IBP reduction on a generalized cut selects the subset of master integrals that have the cut propagators. Complete reduction formulae follow then by merging the reductions on the elements of a spanning set of generalized cuts, that is a set of generalized cuts that touch at least once every possible master integral.

\begin{figure}
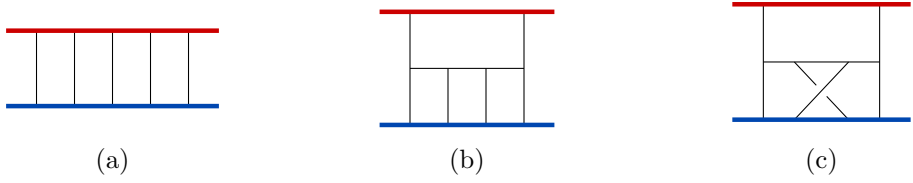

\centering
\begin{subfigure}[b]{0.3\textwidth}
\centering
    \raisebox{.25 cm}{\PlanarQuadLadder}
    \caption{}
    \label{subfig:DiagA}
\end{subfigure}
\begin{subfigure}[b]{0.3\textwidth}
\centering
    \WindowFourLoop
    \caption{}
    \label{subfig:DiagB}
\end{subfigure}
\begin{subfigure}[b]{0.3\textwidth}
\centering
    \XWindowFourLoop
    \caption{}
    \label{subfig:DiagC}
\end{subfigure}
\caption{Sample four-loop diagrams for a four-point scattering amplitude. The thick solid lines represent massive particles, and the thin lines represent gravitons or other massless bosons  }
\label{fig:FourLoopFeynmanFigure}
\end{figure}

At any order, integrands are typically organized into diagrams (see Fig.~\ref{fig:FourLoopFeynmanFigure} for four-loop examples), since this format is compatible with standard IBP reduction and modern integration methods.  At $L$ loops, the amplitude has the form,
\begin{equation}
M_L=  G^{L+1} \sum_i \int 
\prod_{j=1}^L \frac{\mathrm{d}^D \ell_j}{(2 \pi)^D} \,
\frac{N_i(p_e,\ell_j)}{\prod\limits_{\alpha_i} (k_{\alpha_i}^2 -m_{\alpha_i}^2 + \imath \varepsilon)}\,,
\end{equation}
where the sum runs over the diagrams labeled by the index~$i$, $\ell_j$ denote the independent loop momenta, $p_e$ the external momenta, and $\imath \varepsilon$ is Feynman's prescription reflecting micro-causality of the QFT.  The label  $\alpha_i$ in the denominator runs over the internal propagators of diagram $i$, and the masses $m_{\alpha_i}$ vanish except for the massive particles representing the black holes.  The numerators $N_i$ are functions of loop and external momenta. For spinning external particles, the numerators also depend on spinors and/or polarization vectors. Each loop corresponds to an additional power of Newton's constant as reflected in the overall power of $G$.   

Starting from the available four-point four-loop $D \le 10$ massless integrand~\cite{Bern:2009kd, Bern:2012uf}, we obtain the four-loop massive scalar $\mathcal N = 8$ integrand via the same construction as used in Ref.~\cite{Parra-Martinez:2020dzs} at two loops.  
This uses dimensional reduction from $D=6$ dimensions to obtain scattering amplitudes with two distinct masses describing scattering of two distinct extremal black holes. The amplitudes of this theory have a particular property that the integrand itself has no explicit dependence on the dimension, which leaves the number of bosonic and fermionic states unchanged in the dimensional reduction, though the type of particles changes.  We start from the massless amplitudes given in Ref.~\cite{Bern:2009kd} in terms of 50 cubic diagrams and their kinematic numerators.  
The dimensional reduction procedure is straightforward: we start from $D=6$ with external scalars, reduce the momenta to four dimensions and keep the desired Kaluza-Klein modes for the external scalar lines (see Ref.~\cite{Parra-Martinez:2020dzs} for further details),
\begin{equation}
k_1 = \begin{pmatrix}p_1\\0 \\ m_1\end{pmatrix}\,,  \quad
k_2 = 
\begin{pmatrix}p_2\\ m_2 \sin \phi  \\ m_2 \cos \phi\end{pmatrix}\,,  \quad  
k_3 = \begin{pmatrix}p_3\\ -m_2 \sin \phi  \\ -m_2 \cos \phi\end{pmatrix}\,,  \quad
k_4 = \begin{pmatrix}p_4\\0 \\ -m_1\end{pmatrix}\,, 
\end{equation}
where the $k_i$ are massless external momenta in $D=6$ dimensions, the $p_i$ are massive external momenta in $D=4$ dimensions, and $\phi$ is a BPS angle.  The Lorentz invariants become 
\begin{equation}
 (k_1+k_2)^2 \rightarrow s - |m_1 + m_2 e^{\imath\phi} |^2 \,, \hskip .6 cm 
 (k_1+k_4)^2\rightarrow t \,, \hskip .6 cm 
 (k_1+k_3)^2\rightarrow u - |m_1 + m_2 e^{\imath\phi} |^2 \,,
\end{equation}
where $s = (p_1+p_2)^2>(m_1+m_2)^2$, $t = q^2= (p_1+p_4)^2<0$, and $u = (p_1+p_3)^2<0$ in physical scattering kinematics in all-outgoing momentum conventions.
When performing the reduction, it is crucial to account for all possible routes that massive scalar lines can take through a diagram. A single diagram in the massless amplitude generically turns into multiple diagrams in the massive case.  Since the integrand can be straightforwardly derived from the massless results through dimensional reduction, which was spelled out in Ref.~\cite{Parra-Martinez:2020dzs}, we refrain from giving further details here. The net result is a quantum integrand organized around cubic diagrams, some of which are illustrated in Fig.~\ref{fig:FourLoopFeynmanFigure}. This integrand can then be expanded in the classical limit and fed into the machinery for evaluating integrals.
 
\subsection{The classical limit}
\label{subsec:classical_limit}

The correspondence principle identifies the classical limit/expansion of a quantum integrand or generalized cuts as the limit of large quantum numbers.
Thus, in this limit, the angular momenta must be large, $J \gg 1$  in $\hbar=1$ units (see, e.g., Refs.~\cite{Cheung:2018wkq,Kosower:2018adc,Bern:2019crd} for further details). In terms of two-particle momentum invariants, this corresponds to the hierarchy of scales,
\begin{equation}
s,u,m_1^2,m_2^2 \, \sim \, J^2 |t| \gg |t| = |q^2|\,.
\end{equation}
The typical momenta of exchanged gravitons split into two regimes,
\begin{equation}
\text{hard:}\quad\ell\gg |q| \,, \hskip 2 cm \text{soft:}\quad      \ell\sim |q| \,,
\label{eq:hard_soft}
\end{equation}
where classical physics is contained in the soft region and we can systematically expand loop integrands in this limit via the method of regions \cite{Beneke:1997zp}. 
Upon integration, at $L$ loops, the classical part of the amplitude is proportional to $|q|^{L-2} \ln |q|$ and $|q|^{L-2}$ for even and odd $L$, respectively, corresponding to a classical long-range interaction potential of the form $(G/r)^{L-1}$ in position space. In the amplitude, the necessary appearance of a $\ln|{ q}|$ implies that the four-loop amplitudes must contain terms that are proportional to $|{q}|^{2 - 8 \epsilon} /\epsilon$ and hence singular in the dimensional regularization parameter $\epsilon = (4-D)/2$, analogous to the situation at two loops~\cite{Bern:2019nnu, Bern:2019crd, Parra-Martinez:2020dzs}.\footnote{In position space, the singular terms are analytic in $q^2$ and therefore correspond to head-on collisions of the two particles, and this does not contribute to classical long-distance physics.} 

Scattering amplitudes exhibit terms with stronger than classical scaling for soft graviton momenta. This feature can be understood intuitively by recalling that the $n$-th order scattering amplitude for potential scattering in nonrelativistic quantum mechanics includes terms with $n$ factors of the leading-order potential. 
These terms typically referred to as {\em iteration terms}, {\em superclassical terms}, or {\em classically singular terms}, must be consistently removed in order to identify the classical amplitude, which at $(L+1)$PM order appears at $L^{{\rm th}}$ order in the soft expansion.

The soft expansion can be further subdivided into potential and radiation subregions~\cite{Beneke:1997zp}, formally by using a characteristic velocity $v$ of the incoming bodies:\footnote{At small velocities (i.e. in the PN expansion) this separates gravitons into off shell and close to on shell. For relativistic velocities, this scaling no longer realizes this separation. From this perspective, we {\em define} the PM expansion as the resummation of the PN expansion to all orders in the velocity.}
\begin{equation}
\label{eq:ClassicalRegions}
\null \hskip -.35 cm
\text{potential:}\quad \ell\sim(v,\bm 1)|q|\,, \hskip 2 cm 
\text{radiation:}\quad \ell\sim(v,\bm v)|q| \, .
\end{equation}
Here, the first entry refers to the graviton's energy component and the second entry to the typical scale of the spatial momentum components. Physically, the two regions encode the physics of exchanged gravitons that are either off or close to their mass shell, respectively.

Here, we focus on the contribution where all graviton loop momenta $\ell_i$ are in the potential region. As in general relativity, the potential region does not account for all conservative effects, which, starting at 4PM~\cite{Manohar:2006nz, Porto:2017dgs, Bern:2021yeh}, also require the inclusion of an even number of radiation modes together with at least one potential mode.
                      
In taking the classical limit, it is very useful to choose variables carefully. Starting from the $p_i$ in \fig{LOFeynmanFigure}, we introduce special variables~\cite{Parra-Martinez:2020dzs},
\begin{align}
\label{eq:soft_vars}
\hskip - .3 cm 
p_1 = - \overline{m}_1\, u_1  + \frac{q}{2} \,, \hskip .6 cm  
p_2 = - \overline{m}_2\, u_2 - \frac{q}{2} \,, \hskip .6 cm  
p_3 =   \overline{m}_2\, u_2 -  \frac{q}{2} \,, \hskip .6 cm      
p_4 =   \overline{m}_1\, u_1 +  \frac{q}{2} \,,     
\hskip .2 cm 
\end{align}
so that,
\begin{align}
u^2_1 = u^2_2 = 1,\, \hskip 1 cm
u_1 \cdot q = u_2 \cdot q = 0,\, \hskip 1 cm  
\overline{m}_i^2 = m_i^2 \Bigl(1 - \frac{q^2}{4 m_i^2} \Bigr) \,.
\end{align}
This choice greatly simplifies the integration after the soft expansion; the dependence on the particle masses $\overline{m}_i$ factors out of the integral, and the dependence on the only remaining dimensionful variable, $q^2$, can be scaled out by a change of integration variables or fixed by dimensional analysis. 
Therefore, all integrals are effectively functions of the single kinematic variable,
\begin{align}
\label{defYvsXvsSigma}
y  = u_1 \cdot u_2 \,.
\end{align}
Up to terms of $\mathcal{O}(q^2)$, $y$ is proportional to the dot product of the incoming momenta,
\begin{align}
\label{eq:sigmaDef}
y = \sigma + \mathcal O (q^2)\,,
\qquad
\text{where}\quad 
\sigma = \frac{p_1{\cdot} p_2}{m_1\, m_2} \,.
\end{align}

Starting from the dimensionally reduced quantum integrand constructed in \sect{AmplitudeSubsection}, we insert the soft parametrization of the external kinematics defined in Eq.~(\ref{eq:soft_vars}) and series expand the integrand in the classical region where $\ell_j \sim |q|$ to the relevant classical order. 
Here we restrict ourselves to the potential-region contributions, which allows us to discard a number of diagram topologies that have no support in this subregion~\cite{Bern:2019crd}. There are 394 relevant
cubic diagram topologies up to crossing. Further truncating to 1SF reduces the number of diagrams to 374 plus crossings. Most of these are nonplanar and, as we discuss in \sect{subsec:Planarization}, a drastically smaller subset of simpler integrals has to be considered explicitly. We provide the relevant diagram information in the supplementary material in computer-readable form in the file~\texttt{conventions.m}. 

Due to the simplifications appearing in maximal supergravity (for example, the absence of explicit one-loop triangle sub-topologies in the massless quantum integrand), before any further processing, we only find 215 non-zero numerators. Since the quantum integrand was originally constructed with the aim of studying UV properties of maximal supergravity, it does not have manifest Feynman diagram power-counting in the loop momenta, artificially enhancing the small-momentum scaling. In the classical limit, this feature translates to the presence of spurious classically-singular terms in diagrams that should be classical from a Feynman diagram analysis. To remedy this problem, we have devised a novel mapping procedure (to be discussed elsewhere~\cite{Bern:2024vqs}) to bring the classical integrand to a global diagrammatic basis that leads to the manifest cancellation of all spurious superclassical terms prior to integration. 

In a cut-based approach to classical observables, the soft expansion introduces additional features. While the expansion of the tree-level amplitude factors, such as those in the example in Eq.~\eqref{eq:cutexample}, is straightforward and follows the discussion above, the expansion of the delta functions imposing the cut conditions requires some care. 
We may expand the delta functions in terms of their derivatives, e.g.,
\begin{equation}
\delta((p_2+\ell_9)^2-m_2^2) =\delta(2 p_2\cdot \ell_9) 
= \delta(-2\overline{m}_2 u_2\cdot \ell_9 + q\cdot \ell_9)
=\sum_{n=0}^\infty \frac{1}{n!}\delta^{(n)}(-2\overline{m}_2 u_2\cdot \ell_9 ) 
(q\cdot \ell_9)^n  ,
\label{eq:delta_fct}
\end{equation}
where in the first step we set $\ell_9^2 = 0$ because of $\delta(\ell_9^2)$ in Eq.~\eqref{eq:cutexample}.\footnote{More generally, this expansion may generate inverse graviton propagators; such terms can be ignored if they lead to disallowed contact terms between the two matter lines. }
Alternatively, we may also take the perspective that the delta function enforcing the on-shell condition simply sets to zero all contributions in which the corresponding propagator is absent. 
Thus, we may replace the delta functions with the corresponding propagator and expand the resulting integral as if it belonged to an off-shell integrand, e.g.
\begin{align}
\frac{1}{(p_2+\ell_9)^2-m_2^2+\imath\varepsilon} = 
\frac{1}{-2\overline{m}_2 u_2\cdot \ell_9 + q\cdot \ell_9+\imath\varepsilon}
=
\sum_{n=0}^\infty\frac{(-q\cdot \ell_9)^n}{(-2\overline{m}_2 u_2\cdot \ell_9 +\imath\varepsilon)^{n+1}} \,,
\label{eq:propagator_expansion}
\end{align}
and then discard all terms that contain non-positive powers of the corresponding linearized propagator, which may be generated by numerator terms.   
There is a one-to-one correspondence between the terms of Eqs.~\eqref{eq:delta_fct} and \eqref{eq:propagator_expansion} arising from 
\begin{equation}
\frac{1}{x+\imath\varepsilon} - \frac{1}{x-\imath\varepsilon} = -2\pi \imath \delta(x)\,,
\qquad
\frac{1}{(x+\imath\varepsilon)^{n+1}} - \frac{1}{(x-\imath\varepsilon)^{n+1}} = -2\pi \imath\frac{(-1)^n}{n!} \delta^{(n)}(x) \,.
\label{eq:delta_derivatives}
\end{equation}
Each of these two perspectives has advantages in different applications, e.g. the latter for the direct application of IBP reduction to cut integrals and the former, as we will see in \sect{subsec:Planarization}, to the planarization of IBP reduction.    
When assembling the classical amplitude and radial action of ${\cal N}=8$ supergravity in \sect{sec:Neq8assembly}, we will encounter eikonal sums of boundary integrals. A multivariate generalization~\cite{Saotome:2012vy, Akhoury:2013yua} of Eq.~\eqref{eq:delta_derivatives},
\begin{equation}
\label{eq:Eikonal}
    \delta(\omega_1+\dots \omega_n)\sum_{\text{perms}\, \omega_i}
    \frac{1}{\omega_1+\imath\varepsilon}\cdots 
    \frac{1}{\omega_1+\cdots + \omega_{n-1}+\imath\varepsilon}
    =(-2\pi\imath)^n \delta(\omega_1)\cdots\delta(\omega_n)
    \,,
\end{equation}
will be useful in their evaluation, as it localizes the energy integrals in all loops and renders them well-defined without an additional regulator. See \sect{subsec:boundaryconditions} and \sect{sec:Neq8assembly} for details.

\subsection{Classical amplitudes, the radial action, and classical observables}

An efficient method for extracting physical observables from the amplitude utilizes the amplitude~radial-action relation~\cite{Bern:2021dqo}, which asserts that, in the classical limit, the amplitude can be expressed as a suitably-defined exponential function of the radial action\footnote{While the relation between the elastic scattering matrix and the radial action is similar to that between the scattering matrix and the eikonal, there are important practical differences. First, in the radial action formulation, there is no prefactor of the exponential that separates exponentiating and non-exponentiating terms. Second, the precise relation to the scattering angle differes: on the one hand the angle is the derivative of the radial action with respect to the angular momentum and on the other, it is the tangent of half the angle which is related to the derivative of the eikonal with respect to the impact parameter. The key conceptual difference is the admixture of quantum contributions in the two quantities: the radial action is defined to contain no terms scaling quantum mechanically in the barred variables, defined in \Eq{eq:soft_vars}, while the eikonal contains such terms. Such quantum terms at lower orders interfere with classically-singular terms and alter the classical terms at higher order.},
%
\begin{equation}
\label{aarelation}
        \imath \mathcal M(\bm q) = \int_J (e^{\imath I_r(J)} - 1) \,.
\end{equation}
The classical radial action~\cite{Landau:1975pou}, $I_r(J)=\int p_r\, \mathrm{d}r$ is defined as an integral of the radial momentum, $p_r$, over the scattering trajectory. The radial action is a function of the total angular momentum $J=|\vect{p}|  |\vect{b}|$ of the $2\to2$ scattering process of two massive particles with momentum $\vect{p}$ and asymptotic impact parameter $b=|\vect{b}|$ in the center-of-mass reference frame.  The Fourier conjugate variable of $\bm b$ is the momentum transfer~$\vect{q}$,
\begin{equation}
\tilde I_r (\bm q) = \int_{J} I_r(J) \coloneqq \mu^{-2\epsilon}  4E\, |\vect{p}| \int {\mathrm{d}^{D-2}\bm b}\, e^{\imath\bm q\cdot \bm b}\, I_r(J) \,,
\label{eq:FT_J}
\end{equation}
where we work in dimensional regularization $D=4-2\epsilon$ and $\mu$ is the dimensional regularization scale.
In terms of the kinematic variables introduced above, we have
\begin{align}
\label{eq:com_E_p}
   E  = \sqrt{s} =  \sqrt{2 m_1 m_2 \sigma + m^2_1 + m^2_2}
   \,, \hskip 1 cm 
   |\vect{p}| = \frac{m_1 m_2\sqrt{\sigma^2-1}}{E}\,.
\end{align}
Expanding perturbatively the exponential in \eqn{aarelation}, we find {\em classical iteration} terms of the form ($n \ge 2$)
\begin{align}
&\int_{J} \frac{(\imath I_r(J))^n }{n!}
= \imath\int_{\bm \ell}  \frac{\tilde I_r(\bm{\ell}_1) \dots \tilde I_r(\bm{\ell}_n)}{Z_1 \dots Z_{n-1}} 
\,,
\label{eq:product} 
\\[5pt]
&
Z_j = - 4 E\,|\vect{p}| \, \big((\bm \ell_{1}+\bm \ell_2+\dots +\bm\ell_{j})\cdot \hat{\bm z}+\imath\varepsilon \big)\,, \nn
\end{align}
where $\hat{\vect{z}}= {\bm p}/|\bm p|$ denotes the unit vector in the direction of the three-momentum in the  center-of-mass reference frame, and the integration measure is
\begin{align}
\int_{\bm \ell} \; \coloneqq \; \mu^{2(n-1) \epsilon} \int \Biggl( \prod_{i=1}^n \,\frac{\mathrm{d}^{D-1}\bm \ell_i}{(2\pi)^{D-1}} \Biggr) (2\pi)^{D-1} \delta^{(D-1)}\Biggl(\sum_{j=1}^n\bm\ell_j-\bm q\Biggr)\,.
\end{align}
The iteration terms are not particularly interesting in the classical limit as they are singular and contain information only from earlier orders of perturbation theory. More explicitly, for the first two orders, the amplitude~radial-action relation reads\footnote{We label the classical amplitude by the PM order, while in \sect{AmplitudeSubsection} we labeled quantum amplitudes by their loop order.}
\begin{align}
	{\cal M}_1 &= \tilde{I}_{r,1}\,, \qquad  
	{\cal M}_2 = \tilde{I}_{r,2} + \int_{\bm \ell}\frac{\tilde{I}_{r,1} \tilde{I}_{r,1}}{Z_1} \,, 
\end{align}
where we have introduced the expansions
\begin{equation}
    \mathcal{M}=\sum_{k=1}\mathcal{M}_k\,,\hskip 1 cm  \tilde{I}_r=\sum_{k=1}\tilde{I}_{r,k}\,.
 \end{equation}
Here, the powers of Newton's constant are implicit,
\begin{equation}
    \mathcal{M}_k=\mathcal{O}(G^k)\,, \hskip 1 cm 
    \tilde{I}_{r,k}=\mathcal{O}(G^k)\,.
\end{equation}
Note that we keep only the leading classical expansion in the radial action. At the fifth order, we have 
 \begin{align}
 {\cal M}_5 ={}& \tilde{I}_{r,5} + \mathcal{M}_5^{{\rm it.}}\label{eq:amp_action_5}\\
\mathcal{M}_5^{{\rm it.}}={}&
  \int_{\bm \ell}\frac{\tilde{I}_{r,1}^5}{Z_1 Z_2 Z_3 Z_4}
  + \int_{\bm \ell}\frac{\tilde{I}_{r,1} \tilde{I}_{r,1} \tilde{I}_{r,1} \tilde{I}_{r,2}}{Z_1 Z_2 Z_3} 
  + \int_{\bm \ell}\frac{\tilde{I}_{r,1} \tilde{I}_{r,2}   \tilde{I}_{r,2}}{Z_1 Z_2} 
  + \int_{\bm \ell}\frac{\tilde{I}_{r,1} \tilde{I}_{r,1}   \tilde{I}_{r,3}}{Z_1 Z_2}  
  \nonumber \\
  &\phantom{= \tilde{I}_{r,5}(\bm q)\ }
  + \int_{\bm \ell}\frac{\tilde{I}_{r,1} \tilde{I}_{r,4}}{Z_1}
  + \int_{\bm \ell}\frac{\tilde{I}_{r,2} \tilde{I}_{r,3}}{Z_1} \,,
\end{align} 
where the sum over permutations of distinct $\tilde{I}_{r,n}$ is implicit; for instance,
\begin{equation}
\tilde{I}_{r,1} \tilde{I}_{r,4} \coloneqq \tilde{I}_{r,1}(\bm \ell_1) \tilde{I}_{r,4}(\bm \ell_2)+\tilde{I}_{r,4}(\bm \ell_1) \tilde{I}_{r,1}(\bm \ell_2) \, ,
\end{equation}
and $\mathcal{M}_5^{{\rm it.}}$ contains all the classically-singular terms in the soft region. 
The radial action (and, therefore, the classical limit of the amplitude) then determines the scattering angle,
\begin{equation}
\label{eq:angle_Irad}
\chi= - \frac{\partial I_r(J)}{\partial J} \,.
\end{equation}
The scattering angle has a PM expansion
\begin{align}
\label{eq:angle_PM}
\chi=\sum_{k=1}\chi_k\,,\hskip 1.5 cm  \chi_k=\mathcal{O}(G^k)\,.
\end{align}
As we discuss in \sect{sec:DE}, we do not have to explicitly subtract the iteration terms, since they can be automatically separated by the boundary conditions in the differential equations for the master integrals, thereby effectively subtracting them.

In general, it is helpful to organize the PM results in a series expansion in the mass ratio, corresponding to the self-force (SF) expansion.   As noted in Ref.~\cite{Damour:2019lcq},  1SF, corresponding to the first correction beyond the probe mass limit, determine the impulse through 4PM, with 5PM being the first order where 2SF terms are present.  This mass dependence represents the {\em good mass polynomiality rule} of Ref.~\cite{Damour:2019lcq} and is manifest in the soft expansion of the amplitudes where it can be understood as a reflection of Lorentz invariance. In particular, the radial action takes the form,
\begin{align}
\label{eq:5PM_amp_SF_org}
    \tilde{I}_{r,5} = {}&
       m_1^2 m_2^2 (m_1^4+m_2^4) \tilde{I}_{r,5}^{0 {\rm SF}} 
    +  m_1^3 m_2^3 (m_1^2+m_2^2) \tilde{I}_{r,5}^{1 {\rm SF}} 
    +  m_1^4 m_2^4 \,\tilde{I}_{r,5}^{2 {\rm SF}} 
    \\[3pt]
    = {}&
       M^8\nu^2\left[ \tilde{I}_{r,5}^{0 {\rm SF}} 
    +\nu \left(\tilde{I}_{r,5}^{1 {\rm SF}}-4\tilde{I}_{r,5}^{0 {\rm SF}}\right) 
    +\nu^2\left(  \,\tilde{I}_{r,5}^{2 {\rm SF}}-2\tilde{I}_{r,5}^{1 {\rm SF}}+2\tilde{I}_{r,5}^{0 {\rm SF}}\right)\right]
\,,\nonumber
\end{align}
where 
\begin{equation}
M=m_1+m_2\,, 
\qquad
\text{and}
\quad
\nu=\frac{m_1 m_2}{M^2}\,,
\end{equation}
are the total mass and the symmetric mass ratio, respectively. In this expression, we have also factored the explicit mass dependence from the terms corresponding to the radial action.  In this paper, we discuss only 0SF- and 1SF-type terms at 5PM, leaving the more complicated 2SF terms for future work. The additional complication stems primarily from the fact that the IBP system and class of functions are more involved.

%
\section{Integral reduction}
\label{sec:IBP}
%

Once the integrand has been constructed, the next step is to perform the integration.  Following standard procedure, we first carry out integration-by-parts reduction~\cite{Chetyrkin:1981qh, Tkachov:1981wb} to reduce the complete set of integrals to a relatively small number of simpler integrals known as \emph{master integrals}.   IBP identities derive from the fact that in dimensional regularization~\cite{tHooft:1972tcz}, total derivatives integrate to zero.  This generates linear relations between Feynman integrals, which can then be solved in terms of 
the master integrals.   This method is a crucial tool in most modern multi-loop amplitude calculations. The most automated algorithmic approaches~\cite{Maierhofer:2017gsa, Smirnov:2008iw, Smirnov:2019qkx} rely on the Laporta algorithm~\cite{Laporta:2000dsw} and its refinements. As the number of loops or kinematic variables increases, the required linear systems can become challenging to solve. 

\subsection{Optimizing IBP reduction}
\label{subsec:optimizingIBP}

The algorithm needs multiple optimizations in order to carry out IBP reduction at 5PM.  Typically, the algorithm performs much better with an \emph{improved basis} of master integrals chosen so that the coefficients of each master integral have denominators with a factorized dependence on the kinematic variables and the spacetime dimension~\cite{Smirnov:2020quc, Usovitsch:2020jrk}. This greatly simplifies the resulting integral reduction tables and improves the reconstruction of analytic IBP reduction tables using finite-field numerical methods~\cite{vonManteuffel:2014ixa, Peraro:2016wsq, Klappert:2019emp, Peraro:2019svx, Laurentis:2019bjh, DeLaurentis:2022otd, Magerya:2022hvj, Belitsky:2023qho}. Additionally, it greatly facilitates finding simpler differential equations for the chosen master integrals.

Another very useful optimization is to use a spanning set of generalized cuts as a means to reduce memory usage and parallelize the IBP reduction of an integral family~\cite{Larsen:2015ped, Zhang:2016kfo, Bohm:2018bdy}. Following the discussion in \sect{AmplitudeSubsection}, the reduction of integrals on a generalized cut writes them as a linear combination of master integrals that are nonvanishing on this cut, i.e., the subset of master integrals that contain only positive powers of the propagators that are cut. Reducing the integrals on a spanning set of cuts and merging the results produces the complete IBP tables.

Some integrals involved in our calculation are reduced using a development version of \texttt{FIRE6} in conjunction with symmetry rules generated by \texttt{LiteRed}~\cite{Lee:2013mka}. The new version of \texttt{FIRE7} (private version) builds upon the latest public release~\cite{Smirnov:2023yhb} and offers further speedups. The most significant improvements in \texttt{FIRE} compared to the public version 6.5 are:
\begin{itemize}
    \item  the ability to work only with integrals of a given parity under simultaneous transformation $u_i\to -u_i$, either odd or even, as explained in the context of heavy-quark effective theory in Ref.~\cite{Grozin:2006xm},
    \item  a tool that allows the identification of integrals across different integral families,
    \item  improved IBP presolving without substituting the specific power, or index, of the propagator,
    \item  reconstruction algorithms implemented in C++,
    \item implementation of the Zippel algorithm~\cite{ZIPPEL1990375, ZippelProceedings} and a ``balanced Zippel'' algorithm (to be described in a separate paper) for efficient modular reconstruction,
    \item the capability to use other modular reduction programs and automatic reconstruction of results using the Message Passing Interface (MPI).
\end{itemize}
A separate private program is tuned to handle integrals that are still
difficult to evaluate using \texttt{FIRE}. This private program
specializes in applying finite-field
techniques~\cite{vonManteuffel:2014ixa, Peraro:2016wsq,
  Klappert:2019emp, Peraro:2019svx, Laurentis:2019bjh,
  DeLaurentis:2022otd, Magerya:2022hvj, Belitsky:2023qho} to IBP
reduction.  A carefully trimmed selection of seed integrals, whose
total derivatives give IBP identities, minimizes the size of the
linear system to be solved. A similar observation was made in a
related calculation in Ref.~\cite{Driesse:2024xad}. We independently
arrived at a minimal selection of seed integrals by taking inspiration
from the syzygy equation method for IBP reduction~\cite{Gluza:2010ws,
  Schabinger:2011dz, Ita:2015tya, Larsen:2015ped, Abreu:2017xsl,
  Abreu:2017hqn, Bohm:2018bdy, Bendle:2019csk, Wu:2023upw}, which
suggests that Feynman integrals without doubled propagators can often
be reduced by IBP equations generated from seed integrals without
doubled propagators.\footnote{While we do not use syzygy equations to
  find special IBP equations without doubled propagators in
  intermediate steps, such special IBP equations would all be linear
  combinations of simpler IBP equations generated from seed integrals
  without doubled propagators.}  When we reduce integrals involving
one or more doubled propagators, we adjust the seed integrals to have
a similar number of doubled propagators. Additionally, we implement a
heuristic algorithm~\cite{dumas2002computing} for ``pivoting'',
i.e.~dynamic reordering of equations and variables, to preserve the
sparsity of the linear system during Gaussian elimination. The benefit
of pivoting in the context of IBP reduction has been previously
discussed in Ref.~\cite{Bendle:2019csk}. Another speedup comes from
caching certain steps in the Gaussian elimination process
\cite{duff2017direct, Magerya:2022hvj}, as the steps are the same with
different numerical values for the dimension and kinematic parameters
and therefore can be reused. The method of spanning cuts as applied to
IBP reduction~\cite{Larsen:2015ped} is used to divide the jobs into
smaller pieces that can be run in parallel. In the future, we will
explore implementing the algorithms of the private program in FIRE.

Analytic IBP reduction results are then reconstructed from numerous numeric runs, scanning different values of the spacetime dimension $D$ and the kinematic variable $y$, as well as several 64-bit prime moduli. The number of numerical $(D,y)$ pairs typically ranges from a few hundred to about a thousand, and up to five prime moduli are used, depending on the integral being reduced. The factorization of denominators into kinematic and dimensional parts, when using an improved master basis, allows the denominators to be obtained through Thiele reconstruction. This involves reconstructing the rational dependence on $D$ at a random numerical value of $y$, and then reconstructing the rational dependence on $y$ at a random numerical value of $D$. This process leaves only a numerator polynomial to be reconstructed. The polynomial reconstruction is performed by proposing an ansatz and solving linear constraints for the parameters involved. Another special property of our integrals is that all IBP coefficients are either even or odd under the transformation
$u_1 \cdot u_2 = y \rightarrow -y$. This property, which is a direct consequence of the parity of the IBP relations under $u_1\to -u_1$, is exploited to reduce the number of finite-field numerical probes needed in the polynomial reconstruction.

\subsection{Planarization on unitarity cuts}\label{subsec:Planarization}

\begin{figure}
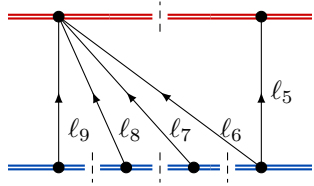

    \centering
    \planarizationCut
    \caption{A sample generalized cut.  The (red and blue) doubled lines represent eikonal matter lines, with the dashed lines cutting them representing that the propagators are replaced with on-shell delta functions.  The thin solid lines represent massless propagators, with the arrows indicating the direction of momentum flow.  The vertices in this diagram represent on-shell tree amplitudes.}
    \label{fig:planarization_eg_cut}
\end{figure}

Having discussed general aspects of the IBP reduction, 
we now present a four-loop example to illustrate how to avoid nontrivial computations of nonplanar diagram topologies by favoring planar ones in a cut-based construction of classical observables.
Equivalently, we use this method to produce IBP reduction tables for non-planar integrals starting from IBP reduction tables of planar ones.
Our method employs a set of spanning cuts and merges the results from each cut at the end.

\begin{figure}
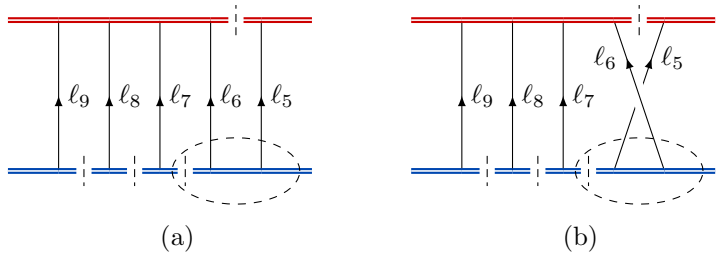

\centering
\begin{subfigure}[b]{0.3\textwidth}
    \planarizationGA
    \caption{}
    \label{subfig:planarizationGA}
\end{subfigure}\qquad
\begin{subfigure}[b]{0.3\textwidth}
    \planarizationGB
    \caption{}
    \label{subfig:planarizationGB}
\end{subfigure}
\caption{Planar ladder and nonplanar ladder graphs. The cut lines correspond to the cut in Figure~\ref{fig:planarization_eg_cut}. The part of the diagrams relevant for the discussion in the main text is circled with a dashed line.}
\label{fig:planarization}
\end{figure}

For concreteness, we look at the cut in \fig{fig:planarization_eg_cut}. Among numerous other diagram topologies, consider the diagrams depicted in \fig{fig:planarization}.
These diagrams contribute to the cut in \fig{fig:planarization_eg_cut}, where we highlighted the part of the diagrams where we apply our planarization procedure. On support of the cut in \fig{fig:planarization_eg_cut}, we can relate the propagators of the nonplanar diagram in \fig{fig:planarization} to those in the planar diagram in \fig{fig:planarization}, but with a different $\imath \varepsilon$ prescription
\begin{align}
\label{eq:eg_diag_prop_relations_on_cut}
 \delta(\ell_{56}\cdot u_2) \, \frac{1}{2 \ell_6\cdot u_2 + \imath \varepsilon} 
 = - \delta(\ell_{56}\cdot u_2)  \,\frac{1}{2 \ell_5\cdot u_2 - \imath \varepsilon}\,,
\end{align}
where $\ell_{56} = \ell_5 + \ell_6$. The relation \eqref{eq:eg_diag_prop_relations_on_cut} allows us to map integrals from the nonplanar topology to the planar one. Crucially, all IBP relations are insensitive to the sign of the $\imath \varepsilon$, so our mapping procedure is as follows: On the cut, use identity \eqref{eq:eg_diag_prop_relations_on_cut} to convert nonplanar integrals to the planar one, perform the planar IBP reduction, and convert the reduced result back to the nonplanar labels which restore the proper $\imath \varepsilon$ prescription.  In practice, this produces little overhead because all the collected planar integrals can be reduced together.

There is, however, one subtlety that we need to discuss. Our classically expanded integrands generically contain matter propagators raised to higher powers resulting from the classical expansion; equivalently, in classically-expanded cuts one encounters derivatives of delta functions. 
As discussed at the end of \sect{subsec:classical_limit}, see Eqs.~\eqref{eq:delta_fct}--\eqref{eq:delta_derivatives}, there is a one-to-one map between derivatives of delta functions and (linear combinations of) propagators raised to higher powers. 
It is convenient to keep the derivatives of delta functions as such for our current purpose of planarizing cut integrals.
In these cases, the delta function identity \eqref{eq:eg_diag_prop_relations_on_cut} is not sufficient. Indeed,
\begin{align}
\label{eq:eq_diag_prop_relations_on_cut}
 \delta^{(n)}(\ell_{56}\cdot u_2)  \frac{1}{2 \ell_6\cdot u_2 + \imath \varepsilon} 
 \neq - \delta^{(n)}(\ell_{56}\cdot u_2)  \frac{1}{2 \ell_5\cdot u_2 - \imath \varepsilon}\,, \qquad n\neq 0\,,
\end{align}
which can be verified by integrating the left- and right-hand sides above against a test function. 

Instead, we proceed as follows.  We find simple special integration-by-parts relations that allow us to shift the raised indices away from the matter lines whose delta function support we wish to employ to planarize diagrams. In the example above, this can be accomplished via the special IBP relation generated by 
\begin{align}
  \check{u}_2 \cdot \partial_{\ell_5} \,, \hskip 1 cm 
  \hbox{with} \quad \check{u}^\mu_2 = \frac{y u^\mu_1 - u^\mu_2}{y^2-1}\,,
\end{align}
which allows us to remove all integrals where the propagator $1/(2\ell_{56}\cdot u_2)$ is raised to higher powers. The special vector $\check{u}^\mu_2$ is constructed so that 
\begin{align}
 \check{u}_2 \cdot u_2 = 1\,, \hskip 1 cm 
 \check{u}_2 \cdot u_1 = 0\,, \hskip 1cm 
 \check{u}_2 \cdot q = 0\,.
\end{align}
This choice is advantageous since it guarantees that matter propagators of the other particle (involving $u_1$) that contain $\ell_5$ are unaffected. 

The IBP reduction of most integrals can be obtained 
from the reduction of planar integrals following the procedure described above. The reduction of the few remaining nonplanar integrals at 1SF order follows similarly after additional partial fractioning~\cite{Driesse:2024xad}.

The main result of this discussion can be summarized as follows:
judiciously applying cut propagator and partial fraction identities
allows us to export all integration-by-parts results\footnote{We
  remark that the export of IBP tables is straightforward, assuming
  diagram symmetries have not been already solved (by
  e.g. \texttt{LiteRed}~\cite{Lee:2013mka}).} from planar diagram
families to nonplanar ones. This observation is very powerful because,
based on experience, traditional IBP programs tend to have performance
issues for nonplanar diagram topologies. 
For the 1SF problem under
consideration in this work, we have explicitly checked that our
mapping procedure agrees with the direct application of IBP reduction
in nonplanar families. As expected, use of both the support of the
delta functions on the spanning cuts, as well as the partial fraction
identity~\cite{Driesse:2024xad}, effectively planarizes the QFT integrand 
in a similar form as obtained by worldline methods.
%

%
\section{Differential Equations}
\label{sec:DE}
%

A standard method for evaluating master integrals in modern quantum field theory computations is by constructing differential equations~\cite{Kotikov:1990kg, Bern:1993kr, Remiddi:1997ny, Gehrmann:1999as, Henn:2013pwa}.   These differential equations are obtained by differentiating a master integral with respect to external kinematic variables and then reducing the resulting integrals back to the basis of master integrals, leading to a closed system of first-order (partial) differential equations. The master integrals discussed in this work depend on a single kinematic variable, $y$, so that the problem reduces to the solution of a system of first-order ordinary differential equations. As explained, the integration-by-parts relations split into a parity-even and a parity-odd sector under the simultaneous transformation $u_i\to -u_i$, which directly implies a similar split for the master integrals. Here we only consider the parity-even sector necessary for the evaluation of the conservative scattering angle. To this end, we first compute differential equations for a set of 2379 master integrals $\vec{I}$ organized into 374 families plus their $y\to -y$ images, which are relevant for scattering through 1SF corresponding to the different diagram topologies with only cubic vertices of our problem. 
If we focus on the classical integrals and ignore classically divergent iterations that are predicted by lower-order information, all relevant integrals are given via planarization identities by the single family depicted in Fig.~\ref{subfig:DiagB}, which only has 71 master integrals.

The full list of 2379 master integrals satisfy 
\begin{equation}
\partial_x \vec{I}=A(x,\epsilon)\vec{I}\,,\hskip 1 cm 
y=\frac{1+x^2}{2x}\,,
\end{equation}
where $A(x, \epsilon)$ is rational in the variables  $x,\epsilon$. The variable $x$ is introduced in order to rationalize the square-root $\sqrt{y^2-1}=\frac{1-x^2}{2 x}$ that appears in the problem. The physical scattering region $1<y$ can be mapped onto the variable range $0<x<1$ in the $x$ parametrization. As mentioned in \Sec{sec:IBP}, we streamline the computation by selecting master integrals such that the denominators in $A$ do not contain mixed dependence on both $\epsilon$ and $x$~\cite{Smirnov:2020quc, Usovitsch:2020jrk}. Additionally, it is beneficial to rearrange the system into an $\epsilon$-factorized form~\cite{Henn:2013pwa,Adams:2018yfj}:
\begin{equation}
\partial_x \vec{J}=\epsilon\, B(x)\vec{J}\,, 
\hskip 1 cm 
\vec{J}=T(\epsilon,x)\vec{I}\,.
\label{eq:CanoncialDE}
\end{equation}
In cases where the Feynman integrals evaluate to generalized polylogarithms, the transformation matrix $T(\epsilon,x)$ usually depends rationally on the kinematics and $\epsilon$. However, starting at three loops, we find elliptic functions, and the transformation $T$ is no longer rational and depends on 
\begin{equation}
\Psi(x)\coloneqq\frac{K(1-x^2)}{\pi} \,, \quad 
\end{equation}
and its derivative, where $K$ is the complete integral of the first kind
\begin{equation}
    K(z)\coloneqq \int^{1}_0 \frac{\mathrm{d}t}{\sqrt{1-t^2}\sqrt{1-z t^2}}\,.
\end{equation}
In our four-loop problem, the matrix $B$ takes the form 
\begin{equation}\label{eq:DEExpandedKernels}
B(x)=\sum_{i=1}^{10} b_i \, k_i(x) \,,
\end{equation}
where $b_i$ are matrices with entries in the rational numbers and $k_i$ are taken from the alphabet
\begin{align}
\mathbb{W}=&\left\{\frac{1}{x},\, \frac{1}{x+1},\, \frac{1}{x-1},\, \frac{x}{x^2+1},\, \frac{1}{x \left(1-x^2\right) \Psi
   (x)^2}, \, \frac{\Psi (x)^2}{x}, \, x \Psi (x)^2,\right.\nonumber\\
   &\qquad\left.\frac{x \Psi (x)^2}{1-x^2}, \, \frac{\left(1+x^2\right) \Psi
   (x)^2}{1-x^2}, \, \frac{\left(1+x^2\right) \left(x^2-2 x-1\right) \left(x^2+2 x-1\right) \Psi (x)^4}{x
   \left(1-x^2\right)}\right\} .
\label{eq:alphabet}   
\end{align}
A feature of the matrices $b_i$, $i=5,\dots 10$ is that they are nilpotent.
Note that the kernels in Eq.~\eqref{eq:alphabet} have at most logarithmic singularities at $x=0,\pm 1, \pm \mathrm{i}$ and $\infty$. The solution to the differential equation~\eqref{eq:CanoncialDE} is then straightforward to obtain in terms of iterated integrals\footnote{The choice of base-point at $x=1$ is natural here; relations to different representations with base-point $x=0$ follow straightforwardly.}
\begin{equation}
\Gamma_{i_1,\dots i_n}(x)= \int_{1}^{x}\mathrm{d}x_n\, k_{i_n}(x_n)\cdots\int_{1}^{x_2}\mathrm{d}x_1\, k_{i_1}(x_1)\,.
\label{eq:iteratedIntegrals}  
\end{equation}
Since the integrals are regular at $x=1$, we find that the last index is fixed such that $i_n\neq 3$ and as such, the integrals of the form $\Gamma_{i_1,\dots i_n}$ are well-defined without reference to endpoint regularization.

We note that the $\Gamma_{i_1,\dots i_n}$ do not satisfy any linear relations. Shuffle identities can be used to express some functions as products of simpler ones, as in the polylogarithmic case (see e.~g.~Ref.~\cite{Duhr:2014woa}).
The functions contain important subclasses of harmonic polylogarithms $i_r\in \{1,2,3\}$ and for $i_r\in \{1,2,3,4\}$ a special class of cyclotomic polylogarithms introduced in  Ref.~\cite{Ablinger:2011te} that was discussed in Ref.~\cite{Bern:2023ccb}. We note that all iterated integrals with elliptic kernels $i_r>4$ cancel in the final result for the potential-region amplitude in classical maximal supergravity. This is analogous to the cancellation of iterated elliptic integrals at the 4PM order~\cite{Bern:2021dqo, Dlapa:2021npj} and the cancellation at 5PM order in general relativity~\cite{Driesse:2024xad}.
Note that starting from Eq.~\eqref{eq:iteratedIntegrals}, it is straightforward to obtain series expansions around $x=1$ by iterative integration of series expansions. In this way, we can produce expansions, e.g., to 1000 orders in $(x-1)$, which allows for a precise numerical evaluation. This approach is mostly insensitive to the explicit form of the kernels in Eq.~\eqref{eq:alphabet}.

\begin{figure}
    \centering
    \begin{equation}
    \begin{pmatrix}
    D_{1,1}(\epsilon,x) & 0 &0\\
    D_{2,1}(\epsilon,x) & D_{2,2}(\epsilon,x)&0\\
    \vdots & \vdots&\ddots\\
    \end{pmatrix}\stackrel{T}{\to} \begin{pmatrix}
    \epsilon D_{1,1}^\prime(x) & 0 &0\\
    D_{2,1}^\prime(\epsilon,x) & \epsilon D_{2,2}^\prime(x)&0\\
    \vdots & \vdots&\ddots\\
    \end{pmatrix}\stackrel{T^\prime}{\to} \begin{pmatrix}
    \epsilon D_{1,1}^{\prime\prime}(x) & 0 &0\\
    \epsilon D_{2,1}^{\prime\prime}(x) & \epsilon D_{2,2}^{\prime\prime}(x)&0\\
    \vdots & \vdots&\ddots\\
    \end{pmatrix}\nonumber
\end{equation}
    \caption{A canonical differential equation is found in stages. First, the diagonal blocks $D_{i, i}$ are brought to canonical form through the transformation $T$. Subsequently, the transformation $T^\prime$ is constructed, bringing the off-diagonal elements $D_{i,j}$ $i\neq j$ to canonical form.}
    \label{fig:CononicalizeDE}
\end{figure}

The most involved step in the computation is determining a basis $\vec{J}$ that produces the $\epsilon$-factorized form~\eqref{eq:CanoncialDE}.
Fortunately, this step can be done in an algorithmic way. 
It is a particular feature of the differential equations and IBP reduction that taking a derivative leads to integrals with the same denominator factors (with some of them potentially canceled). This makes it natural to structure the computation around sectors, i.e., classes of integrals that share the same denominators.
In this setup, the differential equations are block-triangular, with the diagonal blocks determined by the master integrals in a given sector and the off-diagonal blocks relating integrals with more propagators to integrals with fewer propagators.
The problem of finding an epsilon-factorized form is naturally separated into two tasks: finding epsilon forms for the diagonal blocks and then simplifying the off-diagonal parts of the system. This procedure is summarized in \fig{fig:CononicalizeDE}.
Typically, the first step is more involved. There are two classes of sectors. The first class can be brought to $\epsilon$-factorized form~\eqref{eq:CanoncialDE} using rational transformation, while the second class involves transformations depending on $\Psi$ and $\partial_x\Psi$. It is natural to call the first class polylogarithmic and the second elliptic. For the polylogarithmic sectors, we use a private implementation of the algorithm described by Lee in Ref.~\cite{Lee:2014ioa}. (See Refs.~\cite{Prausa:2017ltv, Gituliar:2017vzm, Lee:2020zfb} for various implementations in the literature.) Given the size of the systems, we make use of finite-field methods~\cite{vonManteuffel:2014ixa, Peraro:2016wsq, Klappert:2019emp, Peraro:2019svx, Laurentis:2019bjh, DeLaurentis:2022otd,  Magerya:2022hvj, Belitsky:2023qho} through the package \texttt{FiniteFlow}~\cite{Peraro:2019svx}.  

For the parity-even integrals, there is a single elliptic sector up to permutations~\cite{Klemm:2024wtd}. We first derive the Picard-Fuchs equation for the maximal cut (see e.g.~Ref.~\cite{Muller-Stach:2012tgj}) of the scalar integral
\begin{align}\label{eq:ScalarElliptic}
\vcenter{\hbox{\BNDIntEllipticAFull}}={}&\int\prod_{i=1}^4\frac{\mathrm{d}^{D}\ell_i}{(2\pi)^D}\frac{1}{\ell_1^2 \ell_2^2 \ell_3^2 \left(\ell_{12}-\ell_4\right)^2 \ell_4^2 \left(\ell_{123}+\hat{q}\right)^2}\\
    \qquad{}& \null \times \frac{1}{ \left(\ell_4+\hat{q}\right){}^2 \left(-2 u_1\cdot\ell_4 \right)\left(2 u_2\cdot \ell_1 \right) \left(2u_2\cdot \ell_{12} \right) \left(2 u_2\cdot\ell_{123}\right)}\,,\nonumber
\end{align}
where we define $\hat{q}=q/\sqrt{-q^2}$ and make the loop momenta dimensionless. It turns out that the Picard-Fuchs equation for this four-loop example is exactly the same differential equation that is satisfied by the elliptic three-loop integral~\cite{Ruf:2021egk}. In particular, after imposing regularity in the static limit ($x\to 1$), we find 
\begin{equation}
\vcenter{\hbox{\BNDIntEllipticAFull}}= c\,x\Psi^2+\mathcal{O}(\epsilon)\,,
\label{eq:SolEp0}
\end{equation}
where $c$ is an integration constant to be determined by evaluating the integral in the static limit.

The scalar integral divided by the factor $x\Psi^2$ thus provides us with a good candidate for an integral with uniform transcendental weight. We then proceed to get a canonical basis for the sector by the \texttt{INITIAL} algorithm~\cite{Dlapa:2020cwj, Dlapa:2022wdu}. For a recently proposed algorithm involving elliptic integrals, see also \cite{Gorges:2023zgv}.

The remaining task is to bring the off-diagonal terms into the form $\eqref{eq:CanoncialDE}$. This is again algorithmic and is carried out with the algorithm described in Ref.~\cite{Lee:2014ioa} with relatively minor modifications owing to the presence of the elliptic kernels. Once derivatives of $\Psi$ are removed by integrating total derivatives, all remaining transformations can be found by so-called balance transformations~\cite{Lee:2014ioa, BARKATOU20091017,10.1145/1277548.1277550}. 
This process can be expedited by first finding simpler expressions for the sector integrals (see, e.g., Ref.~\cite{Dlapa:2023hsl}).

\subsection{Boundary conditions}\label{subsec:boundaryconditions}

In order to solve the differential equations for the integrals, we need to provide boundary conditions. These are naturally computed in the static limit $x\to 1$. To obtain the boundary conditions, we expand in the potential region $\ell=(\omega,\boldsymbol{\ell})\sim (v,1)$, following the discussion in Ref.~\cite{Parra-Martinez:2020dzs}. We choose the explicit frame,
\begin{equation}
\hskip -.2cm
  u_1=(y,0,0,v)\,,
  \quad  
  u_2=(1,0,0,0)\,,
  \quad 
  q=(0,\vect{q}_{x},\vect{q}_y,0)\,,
  \quad 
  y=\frac{1+x^2}{2x}\,, 
  \quad
  v=\frac{1-x^2}{2x}\,.
\hskip -.2cm  
\end{equation}
In the potential region, the eikonal propagators are homogeneous, while the propagators of the massless particles are formally expanded in the limit $v\ll 1$,
\begin{equation}
    \frac{1}{\ell^2}=\frac{1}{\omega^2-\boldsymbol{\ell}^2}=-\frac{1}{\boldsymbol{\ell}^2}\sum_{k=0}^\infty\frac{\omega^{2k}}{(\boldsymbol{\ell}^2)^{ k}}\,.
\end{equation}
To illustrate the general method, consider the scalar integral introduced in Eq.~\eqref{eq:ScalarElliptic}. Expanding in $v$ we find,
\begin{align} 
\label{eq:expandedIntegral}
\vcenter{\hbox{\BNDIntEllipticAFull}}
& =\vcenter{\hbox{\BNDIntEllipticA}}+\mathcal{O}(v^2)\,.
\end{align}
Here and below, we use dashed lines to represent static propagators. We will sometimes refer to the static integrals as \emph{boundary integrals}. In particular, in \Eq{eq:expandedIntegral}, we have
\begin{align} 
\vcenter{\hbox{\BNDIntEllipticA}} &=\int\prod_{i=1}^4\frac{\mathrm{d}\omega_i}{2\pi}\frac{\mathrm{d}^{D-1}\boldsymbol{\ell}_i}{(2\pi)^{D-1}}\frac{1}{\boldsymbol{\ell}_1^2 \boldsymbol{\ell}_2^2 \boldsymbol{\ell}_3^2 \left(\boldsymbol{\ell}_{12}-\boldsymbol{\ell}_4\right){}^2 \boldsymbol{\ell}_4^2 \left(\boldsymbol{\ell}_{123}+\hat{\boldsymbol{q}}\right){}^2}\\
    {}&\qquad\quad \times \frac{1}{ \left(\boldsymbol{\ell}_4+\hat{\boldsymbol{q}}\right){}^2 \left(-2 (\omega_4-\hat{\boldsymbol{z}}\cdot\boldsymbol{\ell}_4) \right) 
    2\omega_1\, 2\omega_{12} \, 2\omega_{123}}\nonumber\,,
\end{align}
where $\omega_{i\ldots j} = \omega_i + \cdots + \omega_j$ and $\boldsymbol{\ell}_{i\ldots j} = \boldsymbol{\ell}_{i}+ \cdots +\boldsymbol{\ell}_{j}$. In this expression, we have rescaled $\omega\to v\omega$ to manifest the integral's independence of $v$. The terms of order $v^2$ in Eq.~\eqref{eq:expandedIntegral} include numerators in the form of polynomials in $\omega_i$, and in general, the propagators are raised to some powers. The exact form of the series is not important to the discussion here.  In order to simplify the integrals, it is useful to apply IBP reduction for the problem with static propagators. This involves a set of non-standard propagators with some of the propagators being purely spatial and others involving the energy component.\footnote{The individual integrals are not well defined without additional regularization, but in the sum over diagrams, the denominators always appear in eikonal sums that result in well-defined delta functions---see \eqn{eq:Eikonal}.} The result is 
\begin{equation}
\vcenter{\hbox{\BNDIntEllipticAFull}}=\vcenter{\hbox{\BNDIntAICnKZI}}\times\left[\frac{1{-}10 \epsilon}{2 \epsilon }{+}\mathcal{O}(v^2)\right]{+}\vcenter{\hbox{\BNDIntNDnEBW}}\times\left[\frac{(4 \epsilon {-}1) (6 \epsilon {-}1)}{8 \epsilon ^2}{+}\mathcal{O}(v^2)\right] .
\end{equation}
Note that no master integral is associated with the 11-propagator sector. This is a more generic feature. While the total number of sectors is 905, we find only 312 master integrals, all of which are scalar integrals without numerator insertion.

Applying this process to all master integrals determines the boundary conditions of the system of differential equations. We note that the regularity of the integrals in the static limit imposes additional constraints that are useful as consistency checks. 
In terms of the differential equation, this means that the boundary vector $\vec{J}^{(0)}=\left(J_{1}^{(0)},\dots, J_{2379}^{(0)}\right)$ satisfies,
\begin{align}
B(x)\vec{J}^{(0)}& =\frac{1}{x-1}\left(b_3-2b_5-\frac{1}{8}b_8-\frac{1}{4}b_9+\frac{1}{4}b_{10}\right)\vec{J}^{(0)}+\mathcal{O}((1-x)^0) \nonumber \\
& =\mathcal{O}((1-x)^0)\,,\label{eq:regularity}
\end{align}
where the matrices $b_i$ are introduced in Eq.~\eqref{eq:DEExpandedKernels}.  We checked explicitly that the boundary vector $\vec{J}^{(0)}$ is compatible with regularity in the static limit.

In the canonical basis, the integrals exhibit at most logarithmic singularities, which means they only need to be expanded to the order of $v^0$. However, since a canonical integral is typically a sum of integrals in the inverse propagator basis, we must expand individual integrals to higher orders. The order of this expansion depends significantly on the selection of master integrals in the inverse propagator basis. For the problem at hand, we find it generally advantageous to select integrals with numerator insertions rather than doubled propagators.

Since we will have the same boundary master integrals at all orders in the expansion, we can conclude that, more generally, the result takes the form
\begin{equation}
     \vcenter{\hbox{\BNDIntEllipticAFull}}=
     \vcenter{\hbox{\BNDIntAICnKZI}}\times I_1(x,\epsilon)
     \ + \ 
     \vcenter{\hbox{\BNDIntNDnEBW}}\times I_2(x,\epsilon)\,.
\end{equation}
Furthermore, the boundary integrals are independent of $x$, so that the coefficient functions $I_1$ and $I_2$ have to satisfy the same differential equation as the full integral. Organizing the solution to the differential equation around the evaluation of the coefficient functions of each of the boundary integrals is very natural, and for example, as we will discuss later, the individual coefficient functions can be spanned by functions with different analytic and transcendental properties.  This allows us, for example, to disentangle the contribution from the elliptic sectors and polylogarithmic ones~\cite{Ruf:2021egk, Bern:2022jvn}. In practice, we typically do not have to evaluate each of the coefficient functions separately, as they will be related. Another feature of this organization is that the boundary vectors are exact in $\epsilon$
\begin{equation}
\label{eq:FactorizingBND}     
\vec{J}^{(0)}=
r_1(\epsilon)\vec{c}_1\vcenter{\hbox{\BNDIntAICnKZI}}
+r_2(\epsilon)\vec{c}_2\vcenter{\hbox{\BNDIntNDnEBW}}
+\cdots\,,
\end{equation}
where $\vec{c}_i$ are vectors of rational numbers and $r_i(\epsilon)$ are rational functions of the dimensional regularization parameter $\epsilon$, and the ellipsis stands for the other boundary master integrals, labeled by $i=1, \dots, 312$. For example, we find
\begin{equation}
   r_1(\epsilon)=  \frac{1{-}10 \epsilon}{ 2\epsilon }\,,\hskip 1 cm  r_2(\epsilon)=  \frac{(4 \epsilon {-}1) (6 \epsilon {-}1)}{8 \epsilon ^2}\,.
\end{equation}
The property \eqref{eq:FactorizingBND} provides a non-trivial check of our computation of boundary vectors. It is interesting to note that the $\vec{c}_i$ are linearly independent and provide a basis for the space of all vectors satisfying the regularity condition \eqref{eq:regularity}.
 
The organization in terms of boundary integrals makes it possible to derive solutions to the differential equations that are exact in $\epsilon$ but truncated in the weight of the transcendental functions. Since the maximal weight is fixed at a given loop order (e.g., transcendental weight 3 for the present computation), this provides a good way to determine an integral table that is sufficient for any computation at that loop order.
  
It is also important to note that we have not specified an $\imath\varepsilon$-prescription above. By organizing the computation in terms of boundary integrals, this approach can be applied as well to problems involving delta function (cut) propagators.

We have carried out extensive checks on the results.
The integration is naturally organized in terms of families defined by cubic diagrams. We have checked that the resulting integrals on the overlap agree; in particular, we have checked the consistency with the QED-type integrals results derived in Ref.~\cite{Bern:2023ccb}. Although this implies some unnecessary over-computation, the fact that overlapping integrals agree is highly nontrivial.

We conclude this section by pointing out that while we have independently obtained differential equations for nonplanar integrals, the procedure discussed in \sect{subsec:Planarization} for IBP relations naturally carries over to the differential equations. In particular, canonical bases found for planar integrals provide canonical bases for nonplanar families. It is striking that this process drastically reduces the number of master integrals that have to be computed to 71. We found that verifying this relation at the level of the differential equations' solutions provides a valuable consistency check.
Similar considerations might also be relevant to effective field theories of QCD in the context of collider physics.

The solutions to the differential equations up to transcendental weight 3 are provided in a computer-readable compressed format in \texttt{masterIntegralValues.m} included in the supplementary material to this manuscript.

%
\section{The $\mathcal{N} = 8$ supergravity 5PM amplitude and scattering angle}
\label{sec:Neq8assembly}
%

Using the classical integrals derived from canonical differential equations in the previous section, we compute the  $\mathcal{N} = 8$ supergravity potential-region amplitude and the associated scattering angle.  Upon dimensionally reducing the massless integrand (see \sect{AmplitudeSubsection}) and performing the soft expansion, the integrand mapping and subsequent IBP reduction (see \sect{sec:IBP}), we obtain the even in $|{\vect{q}}|$ amplitude in terms of the global master integrals for which we have solved the differential equations.

In order to organize the calculation, it turns out to be convenient to arrange the different contributions according to the boundary conditions. This procedure has the advantage that we can directly confirm the appearance of eikonal sums of boundary integrals. 
The eikonal sums \eqref{eq:Eikonal} generate cuts on the eikonal matter lines.
While the individual integrals are not well defined without further regularization, after summing the combination of integrals is well defined.  The amplitude is expanded in $|{\vect{q}}|$,
\begin{equation}
\label{eq:amp_sterman_organized}
    \mathcal{M}_5=\left(\frac{|{\vect{q}}|^2}{\mu^2}\right)^{-4\epsilon}\left[
    \sum_{k=-2}^2|{\vect{q}}|^{k}\sum_{n=1}^{N_k}
    f_{k,n}(\sigma,\epsilon) \, \eikSum_{k,n}(\epsilon)
    +\mathcal{O}(|{\vect{q}}|^3)\right]\,,
\end{equation}
where $f_{k,n}(\sigma,\epsilon)$ are transcendental functions and $\eikSum_{k,n}(\epsilon)$ are eikonal sums of static eikonal integrals. The classical part of the amplitude, i.e. the radial action $\tilde{I}_{r,5}$, is obtained from the $k=2$ term in the sum. Since we restrict our attention to the parity-even sector under simultaneous flip of $u_i\to -u_i$, we only have access to the even powers in the $|q|$ expansion. The number $N_k$ counts the independent eikonal sums and depends on the power, $k$, of $|{\vect{q}}|$. For our case, we find that the functions $f_{k,n}(\sigma,\epsilon)$ are finite in $\epsilon$. Therefore, the logarithm $\ln(|{\vect{q}}|)$ required to obtain classical physics is generated whenever the static integrals have poles in $\epsilon$. We remark that we could simply drop the classically singular iteration terms $k=-2,-1,0,1$ in \Eq{eq:amp_sterman_organized} if we are only interested in the genuine classical contribution that is relevant to extract classical observables at $\mathcal{O}(G^5)$. Here, however, we keep classical iterations, which are far simpler computationally, to serve as nontrivial checks.

\subsection{Iteration contributions}

For the leading $k=-2$ term in \Eq{eq:amp_sterman_organized}, there is a single eikonal sum $\eikSum_{-2,1}$
\begin{align}
\label{eq:sterman_m2_1}
 \eikSum_{-2,1}   ={}
&1{\times}\!\!\vcenter{\hbox{\BNDIntGnLCA}}{+}1{\times}\!\!\vcenter{\hbox{\BNDIntOKnLCA}}{+}1{\times}\!\!\vcenter{\hbox{\BNDIntDnLCA}}{+}1{\times}\!\!\vcenter{\hbox{\BNDIntMnLCA}}{+}1{\times}\!\!\vcenter{\hbox{\BNDIntOHnLCA}}{+}\nonumber\\[3pt] 
&1{\times}\!\!\vcenter{\hbox{\BNDIntOQnLCA}}{+}2{\times}\!\!\vcenter{\hbox{\BNDIntAnLCA}}{+}2{\times}\!\!\vcenter{\hbox{\BNDIntBnLCA}}{+}2{\times}\!\!\vcenter{\hbox{\BNDIntCnLCA}}{+}2{\times}\!\!\vcenter{\hbox{\BNDIntEnLCA}}{+}\nonumber\\[3pt] 
&2{\times}\!\!\vcenter{\hbox{\BNDIntFnLCA}}{+}2{\times}\!\!\vcenter{\hbox{\BNDIntInLCA}}{+}2{\times}\!\!\vcenter{\hbox{\BNDIntLnLCA}}{+}2{\times}\!\!\vcenter{\hbox{\BNDIntPnLCA}}{+}2{\times}\!\!\vcenter{\hbox{\BNDIntQnLCA}}{+}\nonumber\\[3pt] 
&2{\times}\!\!\vcenter{\hbox{\BNDIntVnLCA}}{+}2{\times}\!\!\vcenter{\hbox{\BNDIntWnLCA}}{+}2{\times}\!\!\vcenter{\hbox{\BNDIntOEnLCA}}{+}2{\times}\!\!\vcenter{\hbox{\BNDIntOFnLCA}}{+}2{\times}\!\!\vcenter{\hbox{\BNDIntOGnLCA}}{+}\nonumber\\[3pt] 
&2{\times}\!\!\vcenter{\hbox{\BNDIntOInLCA}}{+}2{\times}\!\!\vcenter{\hbox{\BNDIntOJnLCA}}{+}2{\times}\!\!\vcenter{\hbox{\BNDIntOPnLCA}}{+}2{\times}\!\!\vcenter{\hbox{\BNDIntOTnLCA}}{+}2{\times}\!\!\vcenter{\hbox{\BNDIntOUnLCA}}{+}\nonumber\\[3pt] 
&2{\times}\!\!\vcenter{\hbox{\BNDIntOZnLCA}}{+}2{\times}\!\!\vcenter{\hbox{\BNDIntPAnLCA}}{+}4{\times}\!\!\vcenter{\hbox{\BNDIntHnLCA}}{+}4{\times}\!\!\vcenter{\hbox{\BNDIntJnLCA}}{+}4{\times}\!\!\vcenter{\hbox{\BNDIntKnLCA}}{+}\nonumber\\[3pt] 
&4{\times}\!\!\vcenter{\hbox{\BNDIntNnLCA}}{+}4{\times}\!\!\vcenter{\hbox{\BNDIntOnLCA}}{+}4{\times}\!\!\vcenter{\hbox{\BNDIntRnLCA}}{+}4{\times}\!\!\vcenter{\hbox{\BNDIntSnLCA}}{+}4{\times}\!\!\vcenter{\hbox{\BNDIntTnLCA}}{+}\nonumber\\[3pt] 
&4{\times}\!\!\vcenter{\hbox{\BNDIntUnLCA}}{+}4{\times}\!\!\vcenter{\hbox{\BNDIntOLnLCA}}{+}4{\times}\!\!\vcenter{\hbox{\BNDIntONnLCA}}{+}4{\times}\!\!\vcenter{\hbox{\BNDIntOOnLCA}}{+}4{\times}\!\!\vcenter{\hbox{\BNDIntORnLCA}}{+}\nonumber\\[3pt] 
&4{\times}\!\!\vcenter{\hbox{\BNDIntOSnLCA}}{+}4{\times}\!\!\vcenter{\hbox{\BNDIntOVnLCA}}{+}4{\times}\!\!\vcenter{\hbox{\BNDIntOWnLCA}}{+}4{\times}\!\!\vcenter{\hbox{\BNDIntOXnLCA}}{+}4{\times}\!\!\vcenter{\hbox{\BNDIntOYnLCA}}\,.
\end{align}
In \eqn{eq:sterman_m2_1} and similar expressions below, the integrals that appear in a sum are obtained from the first term by permuting the vertices on the massive lines. The various numeric factors in front of different integrals account for relative differences in diagram symmetries.

The coefficient of integral combination (\ref{eq:sterman_m2_1}) in the amplitude (\ref{eq:amp_sterman_organized}) is proportional to the fifth power of the tree-level radial action,
\begin{equation}
\label{eq:fFactor_m2_1}
    f_{-2,1}= \frac{(8 \pi G\,  4 m^2_1m^2_2 c_{\phi }^2)^5}{\left(m_1m_2 \sqrt{\sigma^2 - 1}\right)^4} \propto \tilde{I}_{r,0}^5 \,,
\end{equation}
where $c_{\phi } = \cos \phi - \sigma$ is a recurring quantity involving the angle between the two BPS vectors defining the external states.
Using the eikonal identity \eqref{eq:Eikonal}, we find
\begin{align}
\label{eq:stermanValue_m2_1}
     \hspace{-0.25cm}\frac{5!}{120}\eikSum_{-2,1} {=}&\left(\frac{\imath}{2}\right)^4\!\!\!\!\vcenter{\hbox{\BNDIntE}}\!\!\!\!\\
     {=}&\frac{1}{16}\int\prod_{i=1}^4\frac{\mathrm{d}^{3-2\epsilon}\boldsymbol{\ell}_i}{(2\pi)^{3-2\epsilon}}\frac{1}{\boldsymbol{\ell}_1^2\boldsymbol{\ell}_2^2\boldsymbol{\ell}_3^2\boldsymbol{\ell}_4^2(\boldsymbol{\ell}_{1234}+\hat{\vect{q}})^2}\frac{1}{(2\hat{\vect{z}}\cdot\vect{\ell}_1)(2\hat{\vect{z}}\cdot\vect{\ell}_{12})(2\hat{\vect{z}}\cdot\vect{\ell}_{123})(2\hat{\vect{z}}\cdot \vect{\ell}_{1234})}\nonumber\,.
\end{align}
The combinatorial factor $\frac{5!}{120} = 1$ associated to $\eikSum_{-2,1}$ in Eq.~\eqref{eq:stermanValue_m2_1} is obtained from the fact that we permute the five vertices on the bottom matter line and we divide out by the number of integrals (120) appearing in the eikonal sum in Eq.~\eqref{eq:sterman_m2_1}.
Combining Eqs.~\eqref{eq:fFactor_m2_1} and \eqref{eq:stermanValue_m2_1} we find that the iteration term at $\mathcal{O}(|\vect{q}|^{-2})$ is consistent with the prediction from the amplitude-action relation~\eqref{eq:amp_action_5}.

For the purposes of this work, we only focus on the even powers in the $|{\vect{q}}|$ expansion and, therefore, skip the classically-singular terms at $|{\vect{q}}|^{\pm 1}$. At order $|{\vect{q}}|^0$ we find five possible different eikonal sums
\begin{align}
    \eikSum_{0,1}={}&\vcenter{\hbox{\BNDIntADOnLBU}}+\dots\,,
    \qquad
    \eikSum_{0,2}=\vcenter{\hbox{\BNDIntADLnLBR}}+\dots\,,\nonumber
    \\
    \eikSum_{0,3}={}&\vcenter{\hbox{\BNDIntDnKXB}}+\dots\,,
    \qquad
    \eikSum_{0,4}=\vcenter{\hbox{\BNDIntGnKZM}}+\dots \,,\nonumber 
    \\
    \eikSum_{0,5}={}& \vcenter{\hbox{\BNDIntCQnKII}}+\dots \,.
\end{align}
The full expressions include 32, 47, 42, 84, and 26 terms, respectively. They are explicitly given in the appendix in Eqs.~\eqref{eq:Eikonal_01}--\eqref{eq:Eikonal_05}. The first and the fourth sums are of the form (two-loop)$\times$(tree)${}^2$ and evaluate to
\begin{align}
\label{eq:stermanValue_0_1_4}
     \frac{4!}{60}\eikSum_{0,1}=  \frac{4!}{144}\eikSum_{0,4}
     & =\left(\frac{\imath}{2}\right)^4\vcenter{\hbox{\BNDIntF}}\,.
\end{align}
Similarly, the fifth sum evaluates to 
\begin{align}
\label{eq:stermanValue_0_5}
     \frac{4!}{72}\eikSum_{0,5}=\left(\frac{\imath}{2}\right)^4 
     \hspace{-.2cm}
     \vcenter{\hbox{\BNDIntH}}\,.
\end{align}
The two remaining sums are of the type (one-loop)$^2\times$(tree)${}$ and are given by
\begin{align}
\label{eq:stermanValue_0_2_3}
     \frac{4!}{90}\eikSum_{0,2}= \frac{4!}{72} \eikSum_{0,3}
     & =\left(\frac{\imath}{2}\right)^4\vcenter{\hbox{\BNDIntG}}\,.
\end{align}
The integrals appearing on the right hand side of Eqs.~\eqref{eq:stermanValue_0_1_4}--\eqref{eq:stermanValue_0_2_3}  are, explicitly,
\begin{align}
\vcenter{\hbox{\BNDIntF}}
     ={}&\left(\frac{1}{16 \pi^2 \epsilon}+\mathcal{O}(\epsilon^0)\right)\int\prod_{i=1}^2\frac{\mathrm{d}^{3-2\epsilon}\boldsymbol{\ell}_i}{(2\pi)^{3-2\epsilon}}\frac{1}{\boldsymbol{\ell}_1^2
     (\boldsymbol{\ell}_2^2)^{2\epsilon} (\vect{\ell}_{12}+\hat{\vect{q}})^2}\frac{1}{(2\hat{\vect{z}}\cdot\vect{\ell}_1)(2\hat{\vect{z}}\cdot\vect{\ell}_{12})}\,,\\
\vcenter{\hbox{\BNDIntH}}
     = {}&
     \left(\frac{1}{16}+\mathcal{O}(\epsilon)\right) 
     \int\prod_{i=1}^2\frac{\mathrm{d}^{3-2\epsilon}\boldsymbol{\ell}_i}{(2\pi)^{3-2\epsilon}}
     \frac{1}{\boldsymbol{\ell}_1^2
     (\boldsymbol{\ell}_2^2)^{2\epsilon} 
     (\vect{\ell}_{12}+\hat{\vect{q}})^2}\frac{1}
     {(2\hat{\vect{z}}\cdot\vect{\ell}_1)
     (2\hat{\vect{z}}\cdot\vect{\ell}_{12})}\,,
\\
    \vcenter{\hbox{\BNDIntG}}
     ={}&\left(\frac{1}{16}+\mathcal{O}(\epsilon)\right)\int\prod_{i=1}^2\frac{\mathrm{d}^{3-2\epsilon}\boldsymbol{\ell}_i}{(2\pi)^{3-2\epsilon}}\frac{1}{ (\boldsymbol{\ell}_1^2)^{1/2+2\epsilon}\boldsymbol{\ell}_2^2\left[(\vect{\ell}_{12}{+}\hat{\vect{q}})^2\right]^{1/2+2\epsilon}}\frac{1}{(2\hat{\vect{z}}\cdot\vect{\ell}_1)(2\hat{\vect{z}}\cdot\vect{\ell}_{12})}\,.
\end{align}
In these expressions, we have performed the trivial subloop integration of bubble integrals, which leads to the fractional powers on some of the remaining propagators and gives rise to the $\epsilon$-dependent coefficients.

As predicted by the amplitude-action relation~\eqref{eq:amp_action_5}, we find in $\mathcal{N}=8$ supergravity,
\begin{align}
    f_{0,1}& = -(8 \pi  G)^5 \frac{\left(4 m_1 m_2 c^2_\phi \right)^5   \left(m^2_1+m^2_2\right)}{4\, m_1 m_2 \, (\sigma ^2-1)^3}
    \propto \tilde{I}_{r,0} \tilde{I}^{{\rm 0 SF}}_{r,2}\,,
    \\[6 pt]
    f_{0,4}& = -(8 \pi  G)^5  \frac{(m_1 m_2)^5\left(4  c^2_\phi \right)^4 }{(\sigma ^2-1)^3} \left[\frac{2}{3}c^2_\phi\,  \sigma {-} \left(\sigma^2{-}1\right)^{\frac{3}{2}}
    \operatorname{arccosh}(\sigma)\right]
    \propto \tilde{I}_{r,0}^2 \, \tilde{I}^{{\rm 1 SF}}_{r,2}\,,
\end{align}
the individual sums naturally correspond to different SF sectors of the 3PM radial action
\begin{equation}\tilde{I}_{r,2}= m_1^2m_2^2\left[(m_1^2+m_2^2) \tilde{I}^{{\rm 0 SF}}_{r,2}+m_1m_2\tilde{I}^{{\rm 1 SF}}_{r,2}\right]\,.
\end{equation}
This structure is consistent with the amplitude-action relation~\eqref{eq:amp_action_5}. As expected, 
$ f_{0,2} =  f_{0,3} =0$ due to the absence of a classical one-loop contribution $\tilde{I}_{r,1}=0$ in $\mathcal{N}=8$~\cite{Caron-Huot:2018ape}. 
In addition, we find $f_{0,5}=0$ consistent with the fact that the corresponding structure is absent at two loops. The two-loop result has been computed in Ref.~\cite{Parra-Martinez:2020dzs}.

\subsection{Classical contributions}
%
The first class of eikonal sums at order $|{\vect{q}}|^2$ comes from probe diagrams and is given by 
\begin{align}
\label{eq:sterman_2_1}
\eikSum_{2,1}= {}&1{\times}\!\!\vcenter{\hbox{\BNDIntADInLBL}}\,.
\end{align}
Furthermore, there are two classes already present in the QED,
\begin{align}
\label{eq:sterman_2_2}
    \eikSum_{2,2}= {}&1{\times}\!\!\vcenter{\hbox{\BNDIntADInKZH}}{+}1{\times}\!\!\vcenter{\hbox{\BNDIntADInKWV}}{+}1{\times}\!\!\vcenter{\hbox{\BNDIntASMnKZH}}{+}1{\times}\!\!\vcenter{\hbox{\BNDIntASMnKWV}}\,,\hphantom{{+}1{\times}\!\!\vcenter{\hbox{\BNDIntADInLBL}}}   
 \\
\label{eq:sterman_2_3}    
\eikSum_{2,3}= {}&1{\times}\!\!\vcenter{\hbox{\BNDIntADKnLAH}}{+}1{\times}\!\!\vcenter{\hbox{\BNDIntADInLAJ}}{+}1{\times}\!\!\vcenter{\hbox{\BNDIntADInLAW}}{+}1{\times}\!\!\vcenter{\hbox{\BNDIntADJnKZB}}{+}1{\times}\!\!\vcenter{\hbox{\BNDIntADInLAZ}}{+}\nonumber
\\ 
&1{\times}\!\!\vcenter{\hbox{\BNDIntADPnKZD}}{+}1{\times}\!\!\vcenter{\hbox{\BNDIntASMnLAZ}}{+}1{\times}\!\!\vcenter{\hbox{\BNDIntASMnLAJ}}{+}1{\times}\!\!\vcenter{\hbox{\BNDIntASMnLAW}}{+}1{\times}\!\!\vcenter{\hbox{\BNDIntASNnKZB}}{+}\nonumber
\\ 
&1{\times}\!\!\vcenter{\hbox{\BNDIntASOnLAH}}{+}1{\times}\!\!\vcenter{\hbox{\BNDIntASTnKZD}}\,.\hphantom{{+}1{\times}\!\!\vcenter{\hbox{\BNDIntADInLBL}}}\hphantom{{+}1{\times}\!\!\vcenter{\hbox{\BNDIntADInLBL}}}\hphantom{{+}1{\times}\!\!\vcenter{\hbox{\BNDIntADInLBL}}}
\end{align}
In gravity, while there are no eikonal sums involving diagrams with cubic interactions of massless particles, we find eikonal sums that involve quartic and quintic couplings between the massless particles
\begin{align}
\label{eq:sterman_2_4}
    \eikSum_{2,4}= {}&1{\times}\!\!\vcenter{\hbox{\BNDIntAICnKZI}}{+}1{\times}\!\!\vcenter{\hbox{\BNDIntAGVnKIF}}{+}1{\times}\!\!\vcenter{\hbox{\BNDIntAGUnKIF}}{+}1{\times}\!\!\vcenter{\hbox{\BNDIntAINnKIF}}{+}1{\times}\!\!\vcenter{\hbox{\BNDIntAXGnKZI}}{+}\nonumber\\ 
&1{\times}\!\!\vcenter{\hbox{\BNDIntAVYnKIF}},\hphantom{{+}1{\times}\!\!\vcenter{\hbox{\BNDIntADInLBL}}}\hphantom{{+}1{\times}\!\!\vcenter{\hbox{\BNDIntADInLBL}}}\hphantom{{+}1{\times}\!\!\vcenter{\hbox{\BNDIntADInLBL}}}\hphantom{{+}1{\times}\!\!\vcenter{\hbox{\BNDIntADInLBL}}}
\\
\label{eq:sterman_2_5}
   \eikSum_{2,5}= {}&1{\times}\!\!\vcenter{\hbox{\BNDIntCPnKGU}}{+}1{\times}\!\!\vcenter{\hbox{\BNDIntRTnKGU}}{+}1{\times}\!\!\vcenter{\hbox{\BNDIntAGVnKID}}{+}1{\times}\!\!\vcenter{\hbox{\BNDIntAVZnKID}}{+}2{\times}\!\!\vcenter{\hbox{\BNDIntCInKHK}}{+}\nonumber\\ 
&2{\times}\!\!\vcenter{\hbox{\BNDIntRMnKHK}}{+}2{\times}\!\!\vcenter{\hbox{\BNDIntAGUnKID}}{+}2{\times}\!\!\vcenter{\hbox{\BNDIntAVYnKID}}\,,\hphantom{{+}1{\times}\!\!\vcenter{\hbox{\BNDIntADInLBL}}}\hphantom{{+}1{\times}\!\!\vcenter{\hbox{\BNDIntADInLBL}}}
\\
\label{eq:sterman_2_6}
   \eikSum_{2,6}= {}&1{\times}\!\!\vcenter{\hbox{\BNDIntHUnIUQ}}{+}1{\times}\!\!\vcenter{\hbox{\BNDIntHOnIUQ}}{+}1{\times}\!\!\vcenter{\hbox{\BNDIntWSnIUQ}}{+}1{\times}\!\!\vcenter{\hbox{\BNDIntWYnIUQ}}\,.\hphantom{{+}1{\times}\!\!\vcenter{\hbox{\BNDIntADInLBL}}}
\end{align}
Finally, there is a class that is factorizing and involves a six-point coupling between the massless particles
\begin{align}
\label{eq:sterman_2_7}
      \eikSum_{2,7}={}&\vcenter{\hbox{\BNDIntNDnEBW}}\,.
\end{align}
Using the eikonal identity \eqref{eq:Eikonal} for the cases  $n=2,4,5$ we can evaluate the sums. The energy integrations are performed straightforwardly and the resulting integrals are three-dimensional Euclidean integrals. We find the following values 

\begin{align}
\label{eq:stermanValue_2_1_2_3}
    5!\times\eikSum_{2,1}=
    \frac{4!\times 2!}{4}\times\eikSum_{2,2}=
    \frac{4!\times 2!}{12}\times\eikSum_{2,3}={}&\left(\frac{\imath}{2}\right)^4\vcenter{\hbox{\BNDIntA}}\,,\\
\label{eq:stermanValue_2_4_5}
    \frac{2!\times 4!}{6}\times\eikSum_{2,4}
    =\frac{2!\times 4!}{12}\times\eikSum_{2,5}={}&\left(\frac{\imath}{2}\right)^4\vcenter{\hbox{\BNDIntB}}\,,\\
\label{eq:stermanValue_2_6}
    \frac{2!\times4!}{4}\times\eikSum_{2,6}={}&\left(\frac{\imath}{2}\right)^4\vcenter{\hbox{\BNDIntC}}\,,\\
\label{eq:stermanValue_2_7}
    2!\times4!\times\eikSum_{2,7}={}&\left(\frac{\imath}{2}\right)^4\vcenter{\hbox{\BNDIntD}}\,.
\end{align}
The Euclidean integrals can be evaluated in terms of iterated bubble integrals and have the values
\begin{align}
    \vcenter{\hbox{\BNDIntA}}{}&=\int\prod_{i=1}^4\frac{\mathrm{d}^{3-2\epsilon}\boldsymbol{\ell}_i}{(2\pi)^{3-2\epsilon}}\frac{1}{\boldsymbol{\ell}_1^2\boldsymbol{\ell}_2^2\boldsymbol{\ell}_3^2\boldsymbol{\ell}_4^2(\boldsymbol{\ell}_{1234}+\hat{\boldsymbol{q}})^2}={}\frac{1}{(16\pi^2)^4}\left[-\frac{16\pi^4}{3\epsilon}+\mathcal{O}(\epsilon^0)\right]\,,
    \\[3pt]
    \vcenter{\hbox{\BNDIntB}}{}&=\int\prod_{i=1}^4\frac{\mathrm{d}^{3-2\epsilon}\boldsymbol{\ell}_i}{(2\pi)^{3-2\epsilon}}\frac{1}{\boldsymbol{\ell}_1^2\boldsymbol{\ell}_2^2\boldsymbol{\ell}_3^2\boldsymbol{\ell}_4^2(\boldsymbol{\ell}_{123}+\hat{\boldsymbol{q}})^2\boldsymbol{\ell}_{234}^2}={}\frac{1}{(16\pi^2)^4}\left[\frac{8\pi^6}{\epsilon}+\mathcal{O}(\epsilon^0)\right]\,,
    \\[3pt]
    \vcenter{\hbox{\BNDIntC}}{}&=\int\prod_{i=1}^4\frac{\mathrm{d}^{3-2\epsilon}\boldsymbol{\ell}_i}{(2\pi)^{3-2\epsilon}}\frac{1}{\boldsymbol{\ell}_1^2\boldsymbol{\ell}_2^2\boldsymbol{\ell}_3^2\boldsymbol{\ell}_4^2(\boldsymbol{\ell}_{123}+\hat{\boldsymbol{q}})^2\boldsymbol{\ell}_{1234}^2}\nonumber\\
    &={}\frac{1}{(16\pi^2)^4}\left[\frac{8\pi^4}{\epsilon^2}{+}\frac{32\pi^4}{\epsilon}(4{-}\gamma_{\mathrm{E}}{+}\ln(4\pi))+\mathcal{O}(\epsilon^0)\right] ,
    \\[3pt]
        \vcenter{\hbox{\BNDIntD}}{}&=\int\prod_{i=1}^4\frac{\mathrm{d}^{3-2\epsilon}\boldsymbol{\ell}_i}{(2\pi)^{3-2\epsilon}}\frac{1}{\boldsymbol{\ell}_1^2\boldsymbol{\ell}_2^2\boldsymbol{\ell}_3^2\boldsymbol{\ell}_4^2(\boldsymbol{\ell}_{123}+\hat{\boldsymbol{q}})^2(\boldsymbol{\ell}_4+\hat{\boldsymbol{q}})^2}=\mathcal{O}(\epsilon^0)\,,
        \label{eq:iteratedBubble}
\end{align}
All diagrams within a given eikonal sum are related by permuting vertices on the matter lines. In our case, each of the graphs represents a separate boundary condition.
Crucially, it serves as a nontrivial check that the coefficients of the amplitude on eikonal-related boundary integrals must agree as functions of the kinematics up to relative numerical symmetry factors.

The organization in terms of the eikonal sums, exposes some interesting structures. First, the groups cleanly separate different mass structures in the final result. To leading order in the large mass expansion only $\eikSum_{2,1}$ contributes, at the next order $\eikSum_{2,i}$ $i=2,\dots 7$ contribute. 
At the second order of the gravitational self-force, there are additional classes of diagrams with two lines on either side not included here. Next, we note that the coefficients of the boundary integrals feature different classes of special functions. In particular, while elliptic integrals cancel from the final result, they only arise in the coefficients $f_{2,4}$ and $f_{2,7}$ and at higher order in $\epsilon$. To leading order in $\epsilon$ the contributions from $f_{2,4}$ and  $f_{2,5}$ are particularly simple and only contain rational functions. Interestingly, we find that the coefficients come with opposite signs $f_{2,4}=-f_{2,5}$, and thus, their contribution cancels in the final result.
The functions $f_{2,2}$ and $f_{2,3}$ are the most complicated and involve polylogarithms of up to weight 2. The divergence arising from the split between near-zone and far-zone, i.e. the separation of potential and radiation region, is localized in $\eikSum_{2,3}$.

\subsection{Results}

After evaluating the eikonal sums and adding all contributions, we find\footnote{We recall that, in our conventions for the SF organization of the radial action \Eq{eq:5PM_amp_SF_org}, we have factored out the explicit mass dependence $m^2_1 m^2_2 (m^4_1+m^4_2)$ and $m^3_1 m^3_2 (m^2_1+m^2_2)$ for the 0SF and 1SF contributions, respectively.}
\begin{align}
\label{eq:result0SF}
    \tilde{I}_{r,5}^{0 {\rm SF}}& =
    G^5 \pi  c_{\phi }^4\,   
    |{\vect{q}}|^2 \left(\frac{|{\vect{q}}|^2}{\bar{\mu}^2}\right)^{-4\epsilon}
    \left[-\frac{4\, c_{\phi }^6}{5 \epsilon \left(\sigma^2-1\right)^4}+ \mathcal{O}(\epsilon^0)\right]
    ,
    \\
 \label{eq:result1SF}
    \tilde{I}_{r,5}^{1 {\rm SF}}&=
    G^5 \pi  c_{\phi }^4 \,
     |{\vect{q}}|^2 \left(\frac{|{\vect{q}}|^2}{\bar{\mu}^2}\right)^{-4\epsilon}
    \left[
    \frac{\tilde{I}_{r,5}^{1 {\rm SF,div.}}}{\epsilon^2} + \frac{\tilde{I}_{r,5}^{1 {\rm SF,fin.}}}{\epsilon} + \mathcal{O}(\epsilon^0)
    \right],
\end{align}
where $ \bar \mu^ 2 =  4\pi e^{-\gamma_{\rm E}}\,\mu^2$ is the $\overline{\text{MS}}$ renormalization scale. Note that classical physics is encoded in the non-analytic $|{\vect{q}}|^2\ln|{\vect{q}}|$ term at 5PM, which is obtained by expanding in $\epsilon$. In presenting our results, however, we prefer to keep the $|\vect{q}^2|^{-4\epsilon}$ term intact, and it is implicitly understood that we need a $1/\epsilon$ accompanying this factor in order to ultimately generate the $\ln|{\vect{q}}|$. 
The appearance of the $1/\epsilon^2$ factor signals a true divergence, which arises 
because of the overlap of the potential and radiation contributions~\cite{Manohar:2006nz, Porto:2017dgs}.
The divergence is expected to cancel after tail contributions are included. This cancellation appears at three loops~\cite{Bern:2021yeh} and is generally common when separating regions~\cite{Beneke:1997zp}.

The 5PM radial action through the first order in self-force is
\begin{align}
\label{eq:m5_1sf_div}
\tilde{I}_{r,5}^{1 {\rm SF,div.}} & = 
16 c_{\phi }^2 \left\{\mathrm{F}_0 \left[\frac{\sigma ^2 c_{\phi }^2}{\left(\sigma ^2-1\right)^2}+\frac{4 \sigma  c_{\phi }}{\sigma ^2-1}+4\right]+\frac{\sigma  c_{\phi }^2}{\left(\sigma ^2-1\right)^2}+\frac{2 c_{\phi }}{\sigma ^2-1}\right\}
\,,\\
\label{eq:m5_1sf_fin}
\tilde{I}_{r,5}^{1 {\rm SF,fin.}} &= r_1
+ r_2 \, \mathrm{F}_0
+ r_3 \, \mathrm{F}_0^2
+ r_4 \,  \mathrm{F}_1
+ r_5 \, \mathrm{F}_2\,.
\end{align}
Surprisingly, we encounter only simple transcendental functions up to weight two,
\begin{align}
\label{eq:transcendental_fcts_n8_amp0}
\mathrm{F}_0 & = \frac{2 x}{1-x^2} \ln (x)
\,,\\
\mathrm{F}_1 & = \frac{2 x}{1-x^2} \left[-\operatorname{Li}_2(1-x)-\operatorname{Li}_2(-x)-\ln (x) \ln (x+1)-\frac{1}{2}\zeta_2\right]
,
\label{eq:transcendental_fcts_n8_amp1}
\\ 
\mathrm{F}_2 & = 
\frac{2 x}{1-x^2} \left[-\operatorname{Li}_2(1-x)+\operatorname{Li}_2(-x)-\frac{1}{2} \ln ^2(x)+\ln (x) \ln (x+1) +\frac{1}{2}\zeta_2\right]
.
\label{eq:transcendental_fcts_n8_amp2}
\end{align}
The corresponding rational prefactors are
\begin{align}
\hspace{-.5cm}
 r_1 & = \frac{16 c_{\phi }^3}{\sigma ^2 - 1}
 \left[
    \frac{-\sigma  \left(5 \sigma ^2-4\right) c_{\phi }^3}{5 \left(\sigma ^2 - 1\right)^3}
    -\frac{2 c_{\phi }^2}{\sigma ^2 - 1}
    +8
\right] 
 , \\
 \hspace{-.5cm}
 r_2  & = 32 c_{\phi }^2 \left[
    \frac{-\sigma ^2 c_{\phi }^4}{\left(\sigma ^2-1\right)^3}
    -\frac{4 \sigma  c_{\phi }^3}{\left(\sigma ^2-1\right)^2}
    -\frac{9 \left(2 \sigma ^2-1\right) c_{\phi }^2}{2 \left(\sigma ^2-1\right)^2}
    -\frac{8 \sigma  c_{\phi }}{\sigma ^2-1}+1\right] , \\
 \hspace{-.5cm}
 r_3  & = 
 16 c_{\phi } 
 \Bigg[
 \frac{{-} \sigma ^3 c_{\phi }^5}{\left(\sigma ^2{-}1\right)^3}
 {-}\frac{6 \sigma ^2 c_{\phi }^4}{\left(\sigma ^2{-}1\right)^2}
 {+}\frac{6 \sigma  \left(2{-}3 \sigma ^2\right) c_{\phi }^3}{\left(\sigma ^2{-}1\right)^2}
 {+}\frac{\left(8{-}32 \sigma ^2\right) c_{\phi }^2}{\sigma ^2{-}1} 
 {-} 29\, \sigma\,  c_{\phi } 
 {-} 10 \left(\sigma ^2{-}1\right) 
 \Bigg]
 \hspace{-.5cm} \\
 %
\hspace{-.5cm}
 r_4 & =32 c_{\phi }  \Bigg[
 \frac{-\sigma  c_{\phi }^3}{\left(\sigma ^2-1\right)^2}
 -\frac{2 c_{\phi }^2}{\sigma ^2-1}
 -\frac{3\, \sigma \, c_{\phi }}{\left(\sigma ^2-1\right)}
 -6 
 \Bigg] 
 \,, \\
  \hspace{-.5cm} 
 r_5  & = 64 c_{\phi } \left[
    \frac{\sigma ^2 c_{\phi }^3}{\left(\sigma ^2-1\right)^2}
    +\frac{4 \sigma  c_{\phi}^2}{\sigma ^2-1}
    +\frac{(11 \sigma^2-8)c_{\phi}}{2(\sigma^2-1)}
    +3 \, \sigma
    \right]\, .
\end{align}
Thus, the resulting classical amplitude in maximal supergravity is extremely simple, especially when compared to expectations from lower-loop calculations.  Besides finding at most transcendental functions of weight two, all elliptic integrals have canceled in the final result in a trivial manner; all elliptic contributions appear only in higher orders in the $\epsilon$ expansion, in close similarity with the analogous calculation in Einstein gravity~\cite{Driesse:2024xad}. Out of the 10 integration kernels in \Eq{eq:alphabet} that are relevant for the differential equations for the master integrals, the amplitude only contains the kernels $1/x$ and $1/(x\pm 1)$ and is therefore expressible in terms of harmonic polylogarithms~\cite{Remiddi:1999ew}. Since we only encounter weight two functions, we are able to re-write the result in terms of the classical polylogarithms in Eqs.~\eqref{eq:transcendental_fcts_n8_amp0}--\eqref{eq:transcendental_fcts_n8_amp1}.

Note that the divergent part of the four-loop potential-region amplitude \Eq{eq:m5_1sf_div} makes a nontrivial prediction~\cite{Bini:2017wfr, Blanchet:2019rjs, Bini:2020hmy} (see also~\cite{Driesse:2024xad} for the check in general relativity) for the velocity-odd part of the energy loss at three-loop order to be proportional to,
%
\begin{align}
16 c_{\phi }^2 \left[\mathrm{F}_0 \left(\frac{\sigma ^2 c_{\phi }^2}{\left(\sigma ^2-1\right)^2}+\frac{4 \sigma  c_{\phi }}{\sigma ^2-1}+4\right)+\frac{\sigma  c_{\phi }^2}{\left(\sigma ^2-1\right)^2}+\frac{2 c_{\phi }}{\sigma ^2-1}\right]\,,
\end{align}
which, surprisingly, contains only the logarithm $\ln(x)$ in $\mathrm{F}_0$ given in \eqn{eq:transcendental_fcts_n8_amp0} and appears simpler than the two-loop result in Refs.~\cite{Herrmann:2021lqe,Herrmann:2021tct}. It would be interesting to verify this three-loop prediction by an explicit computation using either the KMOC formalism~\cite{Kosower:2018adc} or the recently computed NLO (super-)gravitational waveform~\cite{Herderschee:2023fxh} (for analogous computations in GR, see \cite{Brandhuber:2023hhy,Georgoudis:2023lgf,Georgoudis:2023ozp}).

\noindent
To convert the momentum-space radial action \eqref{eq:result0SF}--\eqref{eq:result1SF} to the 5PM scattering angle, 
\begin{align}
\chi_5 = (m^4_1+m^4_2)\,  \chi_5^{0 {\rm SF}} 
        + m_1 m_2(m^2_1+m^2_2)\, \chi_5^{1 {\rm SF}}  
        + m^2_1 m^2_2\, \chi_5^{2 {\rm SF}}
\end{align}
we first Fourier-transform the former to impact parameter space, paying attention to the interference of the $1/\epsilon$ divergence with the $\epsilon$-dependent power of the momentum transfer. Then, the scattering angle follows from Eq.~\eqref{eq:angle_Irad}. These steps are equivalent to simply replacing the $|{\vect{q}}|$-dependent 
factors in Eqs.~(\ref{eq:result0SF}) and (\ref{eq:result1SF}) by 
\begin{align}
|{\vect{q}}|^2 \left(\frac{|{\vect{q}}|^2}{\bar \mu^2}\right)^{-4\epsilon} 
\longmapsto & \,
\left(\mu^2\, \tilde{b}^2\right)^{5 \epsilon} \frac{1}{b^{5}}
\frac{\sqrt{2 m_1 m_2 \sigma +m^2_1+m^2_2}}{\pi \, m^2_1 m^2_2 \left(\sigma ^2-1\right)}
\underbrace{\frac{2  (e^{-9 \, \epsilon\, \gamma_{\rm E}})\, \Gamma (3-5 \epsilon) }{\Gamma (4 \epsilon-1)} }_{=\left(-16\epsilon + 184 \epsilon^2 + \mathcal{O}(\epsilon^3)\right)} \,,
\end{align}
where we defined $\tilde{b}^2 = b^2 \pi e^{\gamma_{\rm E}}$ and $b$ is the magnitude of the impact parameter.
Explicitly, the 0SF and 1SF potential-region contributions to the 5PM scattering angle are
\begin{align}
\label{eq:result0SFAngle}
    \chi_5^{0 {\rm SF}}& =\frac{G^5}{b^{5}} 
    \left(\mu^2\, \tilde{b}^2\right)^{5 \epsilon}
    c_{\phi }^4 \,  
\frac{\sqrt{2 m_1 m_2 \sigma +m^2_1+m^2_2}}{\left(\sigma ^2-1\right)}
    \left[\frac{64\, c_{\phi }^6}{5 \left(\sigma^2-1\right)^4}\right]+ \mathcal{O}(\epsilon)
    \,,
    \\
    \chi_5^{1 {\rm SF}}&=  \frac{G^5}{b^{5}}  
    \left(\mu^2\, \tilde{b}^2\right)^{5 \epsilon}
    c_{\phi }^4\, 
\frac{\sqrt{2 m_1 m_2 \sigma +m^2_1+m^2_2}}{\left(\sigma ^2-1\right)}    
    \nonumber \\
    & \hskip 2cm \times
    \Bigg[
    -\! \frac{16}{\epsilon} \tilde{I}_{r,5}^{1 {\rm SF,div.}}
    +\left(184 
    \tilde{I}_{r,5}^{1 {\rm SF,div.}}
    -16 \tilde{I}_{r,5}^{1 {\rm SF,fin.}}
    \right)
    \Bigg]+ \mathcal{O}(\epsilon)\,.  \label{eq:result1SFAngle}
\end{align}
As already noted, we expect that the divergent term in the 1SF potential scattering angle cancels once even-in-velocity conservative radiation contributions are included.

\subsection{Consistency checks}

Unlike in general relativity, there are no available post-Newtonian results in $\mathcal{N}=8$ supergravity for comparison. Instead, we have conducted various consistency checks on our results. First, we verified the zero self-force scattering angle against the $\mathcal{O}(G^5)$ prediction from the exact geodesic motion result~\cite{Parra-Martinez:2020dzs}\footnote{Note that we organize the SF expansion in terms of mass structures, not in terms of the total mass $M = m_1 + m_2$ and the symmetric mass ratio $\nu = \frac{m_1m_2}{M^2}$, where, for example, $m^4_1 + m^4_2 = M^4 (1 - 4\nu + 2\nu^2)$ would contribute to higher orders in the $\nu$ polynomial. Therefore, to check for agreement to a certain SF order, we need to expand, say, in $m_1 \ll m_2$ to the desired order.},
\begin{align}
 \chi^{0 {\rm SF}} = 2 \operatorname{arctan}\left(
 \frac{G m_1 m_2}{J }
 \frac{2 c^2_\phi}{\sqrt{\sigma ^2-1}}
 \right).
\end{align}
We have furthermore checked the leading in velocity terms of the scattering angle that are predicted from lower-order iterations. In this, we follow the discussion in Ref.~\cite{Bern:2019crd}.
We use the fact that the scattering angle is written in terms of the radial momentum,
\begin{align}
 \chi = \pi - 2 J \int\limits_{r_{\mathrm{min}}}^\infty \frac{\mathrm{d}r}{r^2 \sqrt{p^2_r(r)}}\,.
\end{align}
Next, we expand the radial momentum in powers of the coupling $G$,
\begin{align}
 p^2_r(r) & = \vect{p}^2-\frac{J^2}{r^2}+\sum_{k=1}\left(\frac{G}{r}\right)^{k}(\mu r)^{2k\epsilon}P_k(\sigma)\,,
\end{align}
where $\vect{p}$ is the center-of-mass momentum at infinity, see \Eq{eq:com_E_p},  and the epsilonic factors are important to handle divergent quantities. We perform the radial integral order by order in a $G$ expansion and compare it to the PM expansion of the scattering angle~\eqref{eq:angle_PM}.
This yields the following expression, predicting the angle at orders $v^{-2k}$, $k=2,3,4,5$
\begin{equation}
    \chi_5=\frac{16 \chi _1\chi_4 \left[1+\epsilon\left(\frac{55}{6}-16 \ln 2\right)\right]}{3 \pi }-\frac{9 \chi _1^5}{80}+\frac{3 \chi _2 \chi _1^3}{\pi
   }-\frac{3}{2} \chi _3 \chi _1^2-\frac{12 \chi _2^2 \chi _1}{\pi ^2}+\frac{4 \chi _2 \chi _3}{\pi }+\mathcal{O}(v^{-2},\epsilon)\,.
   \label{SmallVelPredict}
\end{equation}
Note that the one-loop correction to the scattering angle is zero, $\chi_2 = 0$, in $\mathcal{N}=8$ supergravity due to the absence of one-loop triangles. We checked that the general results in Eqs.~\eqref{eq:result0SF} and \eqref{eq:result1SF} agree with this prediction\footnote{
To perform this check, we have used an unpublished result for the 4PM angle $\chi_4$~\cite{4PM-N8-unpublished} in \eqn{SmallVelPredict}.
}.

%
\section{Conclusions and Outlook}
%

$\mathcal{N}=8$ supergravity provides an outstanding model to study
interesting aspects of gravitational interactions, such as the
potential integrability of orbital motion~\cite{Caron-Huot:2018ape}
and the universality in the high-energy
limit~\cite{Parra-Martinez:2020dzs, DiVecchia:2020ymx}. Classical
calculations in this theory via scattering amplitudes also provide a
robust test of loop integration methods, as they
sample~\cite{Parra-Martinez:2020dzs} nearly all the master integral
types encountered in Einstein gravity.

While constructing amplitude integrands at high orders in the PM expansion is relatively straightforward, the most significant conceptual and technical challenges arise elsewhere. 
The most obvious difficulty is that by the fifth PM (or fourth loop) order analyzed here, the standard IBP and differential equation methods become increasingly involved. In a previous paper~\cite{Bern:2023ccb}, we computed the potential-mode contribution for the simpler case of electrodynamics, demonstrating that with careful tuning, the standard methods can be made to work at this order. 
Recently, Ref.~\cite{Driesse:2024xad} computed 1SF conservative contributions to the 5PM impulse in Einstein gravity.

In this paper, we computed the graviton potential-mode contribution to the classical four-massive-scalar four-loop amplitude in $\mathcal N = 8$ supergravity and assembled their 1SF  contributions to the radial action and scattering angle, and confirmed the self-consistency of its small velocity expansion. To this end, we used master integrals that can also be chosen in Einstein gravity.
The amplitude displays all the expected properties after integration.  Apart from being remarkably compact, it only exhibits weight-two classical polylogarithms, Eqs.~\eqref{eq:transcendental_fcts_n8_amp0}--\eqref{eq:transcendental_fcts_n8_amp2}; similar to Einstein gravity, no elliptic integrals feature in our results.
The radiation-mode contributions can be analyzed in a similar manner, which we leave for future work. Our analysis here is restricted to the 1SF-type terms, as the remaining 2SF contributions are more complex.

Our analysis explicitly demonstrates through four loops that basic structures encountered in gravitational amplitudes in the eikonal limit continue to be relevant in the classical limit. In particular, the appearance of eikonal sums of integrals~\cite{Saotome:2012vy, Akhoury:2013yua} enforces the expected property that classical amplitudes have one cut matter line in each loop~\cite{Cheung:2018wkq, Bern:2019nnu}. The corresponding constraints realized as Dirac delta functions of linearized matter propagators, enable an algebraic evaluation of the energy integrals in all loops. 

The emergence of these eikonal sums also implies that the boundary integrals necessary to define the solution to the differential equations for master integrals also appear in similar combinations, which are well defined even though the individual integrals are not. An analogous feature was already encountered at lower loops~\cite{Bern:2019crd}.
Integrals with divergences not regulated by dimensional regularization are reminiscent of effective field theories of QCD, see e.g.~\cite{Chiu:2012ir, Rothstein:2016bsq}. Here, the need for an additional regulator is removed by the explicit appearance of the eikonal sums of integrals.
It would be interesting to understand whether a similar strategy can be employed in QCD.  

For our calculations, we used an off-shell four-scalar integrand derived via Kaluza-Klein reduction~\cite{Parra-Martinez:2020dzs} from the available~\cite{Bern:2009kd} complete massless $D$-dimensional integrand of $\mathcal N = 8$ supergravity.
We explained how cuts of linear propagators---either appearing because of eikonal sums or effectively because the collapse of certain matter lines sets integrals to zero or simply as a means to reduce the complexity of integral reduction---can be used to extract the IBP reduction of nonplanar integrals from that of planar ones.
As we explained in \sect{subsec:Planarization}, the IBP reduction tables are essentially identical except for keeping track of the $\imath \varepsilon$ prescription for matter propagators. 

Generally, however, and especially in Einstein gravity, it is more convenient to forego the assembly of soft-expanded generalized cuts into gravitational integrands and instead use the cut conditions to planarize them as early as possible. 
While derivatives of delta functions that appear in the classical limit are not natural objects from the perspective of generalized cuts, simple IBP identities can be applied to remove all derivatives of delta functions, simplifying the mapping between nonplanar and planar contributions.

Because the massless four-point four-loop ${\cal N}=8$ integrand~\cite{Bern:2009kd} that we used in our calculations
was constructed for the study of its UV properties, its IR properties are different from the ones implied by Feynman diagram analysis. To correct this feature, we found it useful to map the soft-expanded integrand to a complete basis
of unique integrals specified within a choice of irreducible scalar products.
This map leads to the cancellation of all terms with spurious soft scaling, which also lead to integrals of higher-than-expected complexity.
In a cut-based construction, the mapping to unique integrand terms effectively diagonalizes the merging of generalized cuts and, up to combinatorial factors, puts them in direct correspondence with the off-shell integrand.
As we will explain in Ref.~\cite{Bern:2024vqs}, this property makes it straightforward to apply tree-level double copy to build complete gravitational integrands at any loop order. By bypassing the difficulties in finding gauge-theory integrands at high loop orders that satisfy the duality between color and kinematics~\cite{Mogull:2015adi, Bern:2017ucb}, this improved approach to cut merging clears the path for renewed progress in studying the high-energy properties of supergravity theories with various numbers of supercharges, including ${\cal N}=8$ supergravity.

An important spin-off from our work here is various improvements in the IBP codes. The off-the-shelf application of standard IBP-reduction algorithms to the integrals with 22 propagators and irreducible numerators encountered at four loops is highly inefficient. This is mitigated by careful tuning of the algorithms and new ideas to minimize memory requirements and maximize speed.  We have considerably upgraded \texttt{FIRE} and have also constructed an independent private program based on finite prime fields. \texttt{Kira} has also been recently upgraded for gravity calculations by another group \cite{Driesse:2024xad}. Not only is this of direct importance for problems in gravitational waves, but we expect that these developments will impact, in due course, collider physics and quantum scattering amplitude calculations.  We anticipate that further advancements are likely achievable in the coming years, not only from refinements of existing algorithms but also following ideas, for example, from algebraic geometry~\cite{Gluza:2010ws, Ita:2015tya, Zhang:2016kfo, Gambuti:2023eqh}, intersection theory~\cite{Mastrolia:2018uzb} or careful selection of integral relations~\cite{Guan:2019bcx}.

Besides the obvious application of the methods to the two-body dynamics in general relativity, a natural direction is to complete the $\mathcal N = 8$ supergravity conservative part to include radiative conservative and dissipative contributions, including the 2SF terms not evaluated in this paper. 
The structure of this theory is sufficiently simple to offer new insights into two-body dynamics, as at lower loops~\cite{Caron-Huot:2018ape, Parra-Martinez:2020dzs, DiVecchia:2020ymx}. In particular, it would be interesting to study the fate of graviton dominance and high-energy universality~\cite{DiVecchia:2020ymx}. At the same time, this theory is sufficiently close to general relativity so that these insights should also apply to this more general case. 
Although significant conceptual and technical challenges remain, continued refinements in the standard integration by parts and differential equation methods will be able to overcome a major obstacle to obtaining the complete two-body dynamics at 5PM.
\vspace{-.3cm}
\acknowledgments
\vspace{-.1cm}
We thank Julio Parra-Martinez, Chia-Hsien Shen, and Mikhail Solon for collaboration on the 4PM scattering angle and for permission to use the unpublished results~\cite{4PM-N8-unpublished}. We also thank Chia-Hsien Shen for comparing the dimensional reduction of the $\mathcal{N}=8$ integrand. We also thank Alessandra Buonanno, Gustav Mogull, Mikhail Solon and Fei Teng for useful discussions.
Z.B., E.H., and M.R.~are supported by the U.S. Department of Energy (DOE) under award number DE-SC0009937.
R.R.~is supported by the U.S.  Department of Energy (DOE) under award number~DE-SC00019066.
The work of A.S. and V.S. was supported by the Russian Science
Foundation under agreement No. 21-71-30003 (development of new
features of the \texttt{FIRE} program) and by the Ministry
of Education and Science of the Russian Federation as part of the
program of the Moscow Center for Fundamental and Applied Mathematics
under Agreement No. 075-15-2022-284 (constructing improved
bases of master integrals).
M.Z.'s work is supported in part by the U.K.\ Royal Society through Grant
URF\textbackslash R1\textbackslash 20109. For the purpose of open access,
the author has applied a Creative Commons Attribution (CC BY) license to any
Author Accepted Manuscript version arising from this submission.
The authors acknowledge the Texas Advanced Computing Center (TACC) at The University of Texas at Austin for providing high performance computing resources that have contributed to the research results reported within this paper (URL: http://www.tacc.utexas.edu).
In addition, this work used computational and storage services associated with the Hoffman2 Shared Cluster provided by UCLA Office of Advanced Research Computing \& Research Technology Group.
This work has made use of the resources provided by the Edinburgh Compute and Data Facility (ECDF) (http://www.ecdf.ed.ac.uk/).
We are also grateful to the Mani L. Bhaumik Institute for Theoretical Physics for support.
\vspace{-.5cm}

\newpage
\appendix

\section{Eikonal sums}

In the appendix, we list the remaining eikonal sums for the classically singular terms computed in this work. Since we restrict our attention to the parity-even sector under simultaneous flip of $u_i\to -u_i$, we have access only to the even powers in the $|q|$ expansion.

The first eikonal sum at $\mathcal{O}(q^0)$ is given by 
\begin{align}\label{eq:Eikonal_01}
    \eikSum_{0,1}={}&1{\times}\!\!\vcenter{\hbox{\BNDIntADOnLBU}}{+}1{\times}\!\!\vcenter{\hbox{\BNDIntADLnLBU}}{+}1{\times}\!\!\vcenter{\hbox{\BNDIntASPnLBU}}{+}1{\times}\!\!\vcenter{\hbox{\BNDIntASSnLBU}}{+}2{\times}\!\!\vcenter{\hbox{\BNDIntHnKUQ}}{+}\nonumber\\ 
&2{\times}\!\!\vcenter{\hbox{\BNDIntKnKUQ}}{+}2{\times}\!\!\vcenter{\hbox{\BNDIntOLnKUQ}}{+}2{\times}\!\!\vcenter{\hbox{\BNDIntOOnKUQ}}{+}2{\times}\!\!\vcenter{\hbox{\BNDIntADInLBX}}{+}2{\times}\!\!\vcenter{\hbox{\BNDIntADInLBV}}{+}\nonumber\\ 
&2{\times}\!\!\vcenter{\hbox{\BNDIntADInLBO}}{+}2{\times}\!\!\vcenter{\hbox{\BNDIntADJnLBV}}{+}2{\times}\!\!\vcenter{\hbox{\BNDIntADKnLBX}}{+}2{\times}\!\!\vcenter{\hbox{\BNDIntADKnLBU}}{+}2{\times}\!\!\vcenter{\hbox{\BNDIntADKnLBO}}{+}\nonumber\\ 
&2{\times}\!\!\vcenter{\hbox{\BNDIntADMnLBU}}{+}2{\times}\!\!\vcenter{\hbox{\BNDIntADMnLBO}}{+}2{\times}\!\!\vcenter{\hbox{\BNDIntADNnLBU}}{+}2{\times}\!\!\vcenter{\hbox{\BNDIntADPnLBX}}{+}2{\times}\!\!\vcenter{\hbox{\BNDIntADWnLBO}}{+}\nonumber\\ 
&2{\times}\!\!\vcenter{\hbox{\BNDIntASMnLBX}}{+}2{\times}\!\!\vcenter{\hbox{\BNDIntASMnLBV}}{+}2{\times}\!\!\vcenter{\hbox{\BNDIntASMnLBO}}{+}2{\times}\!\!\vcenter{\hbox{\BNDIntASNnLBV}}{+}2{\times}\!\!\vcenter{\hbox{\BNDIntASOnLBX}}{+}\nonumber\\ 
&2{\times}\!\!\vcenter{\hbox{\BNDIntASOnLBU}}{+}2{\times}\!\!\vcenter{\hbox{\BNDIntASOnLBO}}{+}2{\times}\!\!\vcenter{\hbox{\BNDIntASQnLBU}}{+}2{\times}\!\!\vcenter{\hbox{\BNDIntASQnLBO}}{+}2{\times}\!\!\vcenter{\hbox{\BNDIntASRnLBU}}{+}\nonumber\\ 
&2{\times}\!\!\vcenter{\hbox{\BNDIntASTnLBX}}{+}2{\times}\!\!\vcenter{\hbox{\BNDIntATAnLBO}}\,.\hphantom{{+}1{\times}\!\!\vcenter{\hbox{\BNDIntADInLBL}}}\hphantom{{+}1{\times}\!\!\vcenter{\hbox{\BNDIntADInLBL}}}\hphantom{{+}1{\times}\!\!\vcenter{\hbox{\BNDIntADInLBL}}}\
\end{align}
The second eikonal sum at $\mathcal{O}(q^0)$ is given by 
\begin{align}\label{eq:Eikonal_02}
    \eikSum_{0,2}=&1{\times}\!\!\vcenter{\hbox{\BNDIntADLnLBR}}{+}1{\times}\!\!\vcenter{\hbox{\BNDIntADQnLBR}}{+}1{\times}\!\!\vcenter{\hbox{\BNDIntASPnLBR}}{+}1{\times}\!\!\vcenter{\hbox{\BNDIntASUnLBR}}{+}2{\times}\!\!\vcenter{\hbox{\BNDIntHnKYY}}{+}\nonumber\\ 
&2{\times}\!\!\vcenter{\hbox{\BNDIntHnKWM}}{+}2{\times}\!\!\vcenter{\hbox{\BNDIntJnKWM}}{+}2{\times}\!\!\vcenter{\hbox{\BNDIntOnKVW}}{+}2{\times}\!\!\vcenter{\hbox{\BNDIntRnKWM}}{+}2{\times}\!\!\vcenter{\hbox{\BNDIntOLnKYY}}{+}\nonumber\\ 
&2{\times}\!\!\vcenter{\hbox{\BNDIntOLnKWM}}{+}2{\times}\!\!\vcenter{\hbox{\BNDIntONnKWM}}{+}2{\times}\!\!\vcenter{\hbox{\BNDIntOSnKVW}}{+}2{\times}\!\!\vcenter{\hbox{\BNDIntOVnKWM}}{+}2{\times}\!\!\vcenter{\hbox{\BNDIntADInLBR}}{+}\nonumber\\ 
&2{\times}\!\!\vcenter{\hbox{\BNDIntADInLBU}}{+}2{\times}\!\!\vcenter{\hbox{\BNDIntADInLBQ}}{+}2{\times}\!\!\vcenter{\hbox{\BNDIntADJnLBR}}{+}2{\times}\!\!\vcenter{\hbox{\BNDIntADJnLBO}}{+}2{\times}\!\!\vcenter{\hbox{\BNDIntADKnLBV}}{+}\nonumber\\ 
&2{\times}\!\!\vcenter{\hbox{\BNDIntADKnLBR}}{+}2{\times}\!\!\vcenter{\hbox{\BNDIntADKnLBQ}}{+}2{\times}\!\!\vcenter{\hbox{\BNDIntADLnLBV}}{+}2{\times}\!\!\vcenter{\hbox{\BNDIntADMnLBQ}}{+}2{\times}\!\!\vcenter{\hbox{\BNDIntADNnLBQ}}{+}\nonumber\\ 
&2{\times}\!\!\vcenter{\hbox{\BNDIntADPnLBV}}{+}2{\times}\!\!\vcenter{\hbox{\BNDIntADPnLBR}}{+}2{\times}\!\!\vcenter{\hbox{\BNDIntADPnLBQ}}{+}2{\times}\!\!\vcenter{\hbox{\BNDIntADQnLBV}}{+}2{\times}\!\!\vcenter{\hbox{\BNDIntADSnLBV}}{+}\nonumber\\ 
&2{\times}\!\!\vcenter{\hbox{\BNDIntADUnLBV}}{+}2{\times}\!\!\vcenter{\hbox{\BNDIntADWnLBV}}{+}2{\times}\!\!\vcenter{\hbox{\BNDIntASMnLBR}}{+}2{\times}\!\!\vcenter{\hbox{\BNDIntASMnLBU}}{+}2{\times}\!\!\vcenter{\hbox{\BNDIntASNnLBR}}{+}\nonumber\\ 
&2{\times}\!\!\vcenter{\hbox{\BNDIntASNnLBO}}{+}2{\times}\!\!\vcenter{\hbox{\BNDIntASOnLBV}}{+}2{\times}\!\!\vcenter{\hbox{\BNDIntASOnLBR}}{+}2{\times}\!\!\vcenter{\hbox{\BNDIntASOnLBQ}}{+}2{\times}\!\!\vcenter{\hbox{\BNDIntASPnLBV}}{+}\nonumber\\ 
&2{\times}\!\!\vcenter{\hbox{\BNDIntASQnLBQ}}{+}2{\times}\!\!\vcenter{\hbox{\BNDIntASRnLBQ}}{+}2{\times}\!\!\vcenter{\hbox{\BNDIntASTnLBR}}{+}2{\times}\!\!\vcenter{\hbox{\BNDIntASTnLBQ}}{+}2{\times}\!\!\vcenter{\hbox{\BNDIntASUnLBV}}{+}\nonumber\\ 
&2{\times}\!\!\vcenter{\hbox{\BNDIntASYnLBV}}{+}2{\times}\!\!\vcenter{\hbox{\BNDIntATAnLBV}}\,.\hphantom{{+}1{\times}\!\!\vcenter{\hbox{\BNDIntADInLBL}}}\hphantom{{+}1{\times}\!\!\vcenter{\hbox{\BNDIntADInLBL}}}\hphantom{{+}1{\times}\!\!\vcenter{\hbox{\BNDIntADInLBL}}}
\end{align}
The third eikonal sum at $\mathcal{O}(q^0)$ is given by 
\begin{align}\label{eq:Eikonal_03}
    \eikSum_{0,3}={}&1{\times}\!\!\vcenter{\hbox{\BNDIntDnKXB}}{+}1{\times}\!\!\vcenter{\hbox{\BNDIntBnLBC}}{+}1{\times}\!\!\vcenter{\hbox{\BNDIntCnLAS}}{+}1{\times}\!\!\vcenter{\hbox{\BNDIntBnKXB}}{+}1{\times}\!\!\vcenter{\hbox{\BNDIntLnKZK}}{+}\nonumber\\ 
&1{\times}\!\!\vcenter{\hbox{\BNDIntOnKXB}}{+}1{\times}\!\!\vcenter{\hbox{\BNDIntOFnKXB}}{+}1{\times}\!\!\vcenter{\hbox{\BNDIntOFnLBC}}{+}1{\times}\!\!\vcenter{\hbox{\BNDIntOGnLAS}}{+}1{\times}\!\!\vcenter{\hbox{\BNDIntOHnKXB}}{+}\nonumber\\ 
&1{\times}\!\!\vcenter{\hbox{\BNDIntOPnKZK}}{+}1{\times}\!\!\vcenter{\hbox{\BNDIntOSnKXB}}{+}2{\times}\!\!\vcenter{\hbox{\BNDIntAnKZN}}{+}2{\times}\!\!\vcenter{\hbox{\BNDIntAnLAS}}{+}2{\times}\!\!\vcenter{\hbox{\BNDIntBnKZN}}{+}\nonumber\\ 
&2{\times}\!\!\vcenter{\hbox{\BNDIntCnKZN}}{+}2{\times}\!\!\vcenter{\hbox{\BNDIntCnKXB}}{+}2{\times}\!\!\vcenter{\hbox{\BNDIntCnKXA}}{+}2{\times}\!\!\vcenter{\hbox{\BNDIntDnKZN}}{+}2{\times}\!\!\vcenter{\hbox{\BNDIntEnKXA}}{+}\nonumber\\ 
&2{\times}\!\!\vcenter{\hbox{\BNDIntFnKXA}}{+}2{\times}\!\!\vcenter{\hbox{\BNDIntHnLBI}}{+}2{\times}\!\!\vcenter{\hbox{\BNDIntHnKZK}}{+}2{\times}\!\!\vcenter{\hbox{\BNDIntHnKWU}}{+}2{\times}\!\!\vcenter{\hbox{\BNDIntInKXB}}{+}\nonumber\\ 
&2{\times}\!\!\vcenter{\hbox{\BNDIntMnLAT}}{+}2{\times}\!\!\vcenter{\hbox{\BNDIntRnLBC}}{+}2{\times}\!\!\vcenter{\hbox{\BNDIntOEnKZN}}{+}2{\times}\!\!\vcenter{\hbox{\BNDIntOEnLAS}}{+}2{\times}\!\!\vcenter{\hbox{\BNDIntOFnKZN}}{+}\nonumber\\ 
&2{\times}\!\!\vcenter{\hbox{\BNDIntOGnKZN}}{+}2{\times}\!\!\vcenter{\hbox{\BNDIntOGnKXB}}{+}2{\times}\!\!\vcenter{\hbox{\BNDIntOGnKXA}}{+}2{\times}\!\!\vcenter{\hbox{\BNDIntOHnKZN}}{+}2{\times}\!\!\vcenter{\hbox{\BNDIntOInKXA}}{+}\nonumber\\ 
&2{\times}\!\!\vcenter{\hbox{\BNDIntOJnKXA}}{+}2{\times}\!\!\vcenter{\hbox{\BNDIntOLnLBI}}{+}2{\times}\!\!\vcenter{\hbox{\BNDIntOLnKZK}}{+}2{\times}\!\!\vcenter{\hbox{\BNDIntOLnKWU}}{+}2{\times}\!\!\vcenter{\hbox{\BNDIntOMnKXB}}{+}\nonumber\\ 
&2{\times}\!\!\vcenter{\hbox{\BNDIntOQnLAT}}{+}2{\times}\!\!\vcenter{\hbox{\BNDIntOVnLBC}}\,.\hphantom{{+}1{\times}\!\!\vcenter{\hbox{\BNDIntADInLBL}}}\hphantom{{+}1{\times}\!\!\vcenter{\hbox{\BNDIntADInLBL}}}\hphantom{{+}1{\times}\!\!\vcenter{\hbox{\BNDIntADInLBL}}}\
\end{align}
The fourth eikonal sum at $\mathcal{O}(q^0)$ is given by 
\begin{align}\label{eq:Eikonal_04}
    \eikSum_{0,4}={}&1{\times}\!\!\vcenter{\hbox{\BNDIntGnKZM}}{+}1{\times}\!\!\vcenter{\hbox{\BNDIntAnKZM}}{+}1{\times}\!\!\vcenter{\hbox{\BNDIntAnLBC}}{+}1{\times}\!\!\vcenter{\hbox{\BNDIntBnKZK}}{+}1{\times}\!\!\vcenter{\hbox{\BNDIntCnKWY}}{+}\nonumber\\ 
&1{\times}\!\!\vcenter{\hbox{\BNDIntCnKZG}}{+}1{\times}\!\!\vcenter{\hbox{\BNDIntDnKZM}}{+}1{\times}\!\!\vcenter{\hbox{\BNDIntEnKWY}}{+}1{\times}\!\!\vcenter{\hbox{\BNDIntEnKZG}}{+}1{\times}\!\!\vcenter{\hbox{\BNDIntAnKXB}}{+}\nonumber\\ 
&1{\times}\!\!\vcenter{\hbox{\BNDIntHnLBJ}}{+}1{\times}\!\!\vcenter{\hbox{\BNDIntMnLAS}}{+}1{\times}\!\!\vcenter{\hbox{\BNDIntOEnKXB}}{+}1{\times}\!\!\vcenter{\hbox{\BNDIntOEnKZM}}{+}1{\times}\!\!\vcenter{\hbox{\BNDIntOEnLBC}}{+}\nonumber\\ 
&1{\times}\!\!\vcenter{\hbox{\BNDIntOFnKZK}}{+}1{\times}\!\!\vcenter{\hbox{\BNDIntOGnKWY}}{+}1{\times}\!\!\vcenter{\hbox{\BNDIntOGnKZG}}{+}1{\times}\!\!\vcenter{\hbox{\BNDIntOHnKZM}}{+}1{\times}\!\!\vcenter{\hbox{\BNDIntOInKWY}}{+}\nonumber\\ 
&1{\times}\!\!\vcenter{\hbox{\BNDIntOInKZG}}{+}1{\times}\!\!\vcenter{\hbox{\BNDIntOKnKZM}}{+}1{\times}\!\!\vcenter{\hbox{\BNDIntOLnLBJ}}{+}1{\times}\!\!\vcenter{\hbox{\BNDIntOQnLAS}}{+}2{\times}\!\!\vcenter{\hbox{\BNDIntAnLBJ}}{+}\nonumber\\ 
&2{\times}\!\!\vcenter{\hbox{\BNDIntAnLAT}}{+}2{\times}\!\!\vcenter{\hbox{\BNDIntAnLBI}}{+}2{\times}\!\!\vcenter{\hbox{\BNDIntAnLBG}}{+}2{\times}\!\!\vcenter{\hbox{\BNDIntBnLBG}}{+}2{\times}\!\!\vcenter{\hbox{\BNDIntCnLAT}}{+}\nonumber\\ 
&2{\times}\!\!\vcenter{\hbox{\BNDIntCnKZM}}{+}2{\times}\!\!\vcenter{\hbox{\BNDIntDnLAT}}{+}2{\times}\!\!\vcenter{\hbox{\BNDIntEnKZM}}{+}2{\times}\!\!\vcenter{\hbox{\BNDIntFnKZM}}{+}2{\times}\!\!\vcenter{\hbox{\BNDIntFnKWY}}{+}\nonumber\\ 
&2{\times}\!\!\vcenter{\hbox{\BNDIntHnKZN}}{+}2{\times}\!\!\vcenter{\hbox{\BNDIntHnKXB}}{+}2{\times}\!\!\vcenter{\hbox{\BNDIntHnKZM}}{+}2{\times}\!\!\vcenter{\hbox{\BNDIntHnLBG}}{+}2{\times}\!\!\vcenter{\hbox{\BNDIntHnLBC}}{+}\nonumber\\ 
&2{\times}\!\!\vcenter{\hbox{\BNDIntHnKZG}}{+}2{\times}\!\!\vcenter{\hbox{\BNDIntInLAT}}{+}2{\times}\!\!\vcenter{\hbox{\BNDIntJnLBG}}{+}2{\times}\!\!\vcenter{\hbox{\BNDIntJnKWY}}{+}2{\times}\!\!\vcenter{\hbox{\BNDIntJnLBC}}{+}\nonumber\\ 
&2{\times}\!\!\vcenter{\hbox{\BNDIntKnKXA}}{+}2{\times}\!\!\vcenter{\hbox{\BNDIntKnKZK}}{+}2{\times}\!\!\vcenter{\hbox{\BNDIntKnKWY}}{+}2{\times}\!\!\vcenter{\hbox{\BNDIntMnKZN}}{+}2{\times}\!\!\vcenter{\hbox{\BNDIntNnLAS}}{+}\nonumber\\ 
&2{\times}\!\!\vcenter{\hbox{\BNDIntOnLAT}}{+}2{\times}\!\!\vcenter{\hbox{\BNDIntOnKWY}}{+}2{\times}\!\!\vcenter{\hbox{\BNDIntOnKWU}}{+}2{\times}\!\!\vcenter{\hbox{\BNDIntPnLBG}}{+}2{\times}\!\!\vcenter{\hbox{\BNDIntOEnLBJ}}{+}\nonumber\\ 
&2{\times}\!\!\vcenter{\hbox{\BNDIntOEnLAT}}{+}2{\times}\!\!\vcenter{\hbox{\BNDIntOEnLBI}}{+}2{\times}\!\!\vcenter{\hbox{\BNDIntOEnLBG}}{+}2{\times}\!\!\vcenter{\hbox{\BNDIntOFnLBG}}{+}2{\times}\!\!\vcenter{\hbox{\BNDIntOGnLAT}}{+}\nonumber\\ 
&2{\times}\!\!\vcenter{\hbox{\BNDIntOGnKZM}}{+}2{\times}\!\!\vcenter{\hbox{\BNDIntOHnLAT}}{+}2{\times}\!\!\vcenter{\hbox{\BNDIntOInKZM}}{+}2{\times}\!\!\vcenter{\hbox{\BNDIntOJnKZM}}{+}2{\times}\!\!\vcenter{\hbox{\BNDIntOJnKWY}}{+}\nonumber\\ 
&2{\times}\!\!\vcenter{\hbox{\BNDIntOLnKZN}}{+}2{\times}\!\!\vcenter{\hbox{\BNDIntOLnKXB}}{+}2{\times}\!\!\vcenter{\hbox{\BNDIntOLnKZM}}{+}2{\times}\!\!\vcenter{\hbox{\BNDIntOLnLBG}}{+}2{\times}\!\!\vcenter{\hbox{\BNDIntOLnLBC}}{+}\nonumber\\ 
&2{\times}\!\!\vcenter{\hbox{\BNDIntOLnKZG}}{+}2{\times}\!\!\vcenter{\hbox{\BNDIntOMnLAT}}{+}2{\times}\!\!\vcenter{\hbox{\BNDIntONnLBG}}{+}2{\times}\!\!\vcenter{\hbox{\BNDIntONnKWY}}{+}2{\times}\!\!\vcenter{\hbox{\BNDIntONnLBC}}{+}\nonumber\\ 
&2{\times}\!\!\vcenter{\hbox{\BNDIntOOnKXA}}{+}2{\times}\!\!\vcenter{\hbox{\BNDIntOOnKZK}}{+}2{\times}\!\!\vcenter{\hbox{\BNDIntOOnKWY}}{+}2{\times}\!\!\vcenter{\hbox{\BNDIntOQnKZN}}{+}2{\times}\!\!\vcenter{\hbox{\BNDIntORnLAS}}{+}\nonumber\\ 
&2{\times}\!\!\vcenter{\hbox{\BNDIntOSnLAT}}{+}2{\times}\!\!\vcenter{\hbox{\BNDIntOSnKWY}}{+}2{\times}\!\!\vcenter{\hbox{\BNDIntOSnKWU}}{+}2{\times}\!\!\vcenter{\hbox{\BNDIntOTnLBG}}\,.\hphantom{{+}1{\times}\!\!\vcenter{\hbox{\BNDIntADInLBL}}}
\end{align}
The fifth eikonal sum at $\mathcal{O}(q^0)$ is given by 
\begin{align}
    \label{eq:Eikonal_05}
    \eikSum_{0,5}={}&1{\times}\!\!\vcenter{\hbox{\BNDIntCQnKII}}{+}1{\times}\!\!\vcenter{\hbox{\BNDIntCHnKII}}{+}1{\times}\!\!\vcenter{\hbox{\BNDIntRLnKII}}{+}1{\times}\!\!\vcenter{\hbox{\BNDIntRUnKII}}{+}2{\times}\!\!\vcenter{\hbox{\BNDIntCFnKSE}}{+}\nonumber\\ 
&2{\times}\!\!\vcenter{\hbox{\BNDIntCGnKSE}}{+}2{\times}\!\!\vcenter{\hbox{\BNDIntCMnKII}}{+}2{\times}\!\!\vcenter{\hbox{\BNDIntCOnKII}}{+}2{\times}\!\!\vcenter{\hbox{\BNDIntCPnKII}}{+}2{\times}\!\!\vcenter{\hbox{\BNDIntRJnKSE}}{+}\nonumber\\ 
&2{\times}\!\!\vcenter{\hbox{\BNDIntRKnKSE}}{+}2{\times}\!\!\vcenter{\hbox{\BNDIntRQnKII}}{+}2{\times}\!\!\vcenter{\hbox{\BNDIntRSnKII}}{+}2{\times}\!\!\vcenter{\hbox{\BNDIntRTnKII}}{+}4{\times}\!\!\vcenter{\hbox{\BNDIntCInKII}}{+}\nonumber\\ 
&4{\times}\!\!\vcenter{\hbox{\BNDIntCJnKII}}{+}4{\times}\!\!\vcenter{\hbox{\BNDIntCKnKII}}{+}4{\times}\!\!\vcenter{\hbox{\BNDIntCLnKII}}{+}4{\times}\!\!\vcenter{\hbox{\BNDIntCNnKII}}{+}4{\times}\!\!\vcenter{\hbox{\BNDIntCRnKII}}{+}\nonumber\\ 
&4{\times}\!\!\vcenter{\hbox{\BNDIntRMnKII}}{+}4{\times}\!\!\vcenter{\hbox{\BNDIntRNnKII}}{+}4{\times}\!\!\vcenter{\hbox{\BNDIntROnKII}}{+}4{\times}\!\!\vcenter{\hbox{\BNDIntRPnKII}}{+}4{\times}\!\!\vcenter{\hbox{\BNDIntRRnKII}}{+}\nonumber\\ 
&4{\times}\!\!\vcenter{\hbox{\BNDIntRVnKII}}\,.\hphantom{{+}1{\times}\!\!\vcenter{\hbox{\BNDIntADInLBL}}}\hphantom{{+}1{\times}\!\!\vcenter{\hbox{\BNDIntADInLBL}}}\hphantom{{+}1{\times}\!\!\vcenter{\hbox{\BNDIntADInLBL}}}\hphantom{{+}1{\times}\!\!\vcenter{\hbox{\BNDIntADInLBL}}}
\end{align}

\bibliographystyle{JHEP}
\bibliography{5PM_Neq8.bib}

\end{document}